\documentclass[12pt, fleqn,reqno]{article}
\usepackage{textcomp}
\newcommand{\rmb}{\textyen}
\usepackage{natbib}
\usepackage{dsfont}
\usepackage{dlfltxbcodetips}
\usepackage{amsfonts}
\usepackage{amssymb}
\usepackage{amsmath}
\usepackage{amsthm}
\usepackage{graphicx}
\usepackage{caption}
\usepackage{subcaption}
\usepackage{bbm}
\usepackage{mathrsfs}
\usepackage{MnSymbol}
\newtheorem{theorem}{Theorem}
\newtheorem{lemma}{Lemma}

\newtheorem{definition}{Definition}
\newtheorem{assumption}{Assumption}
\usepackage{color}
\usepackage{enumerate}
\usepackage[usenames,dvipsnames,svgnames,table]{xcolor}
\usepackage[colorlinks=true,
            linkcolor=red,
            urlcolor=blue,
            citecolor=blue]{hyperref}
\usepackage{fancyhdr}
\usepackage{dlfltxbcodetips}
\usepackage{color}
\usepackage{sgame}
\usepackage[left=1.0in,right=1.0in,top=1.05in,bottom=1.0in]{geometry}
\usepackage{appendix}
\usepackage{tikz, pgfplots}
\usepackage{tikz-3dplot}

\usepackage{soul,color}
\usepackage{ushort}

\usepackage{natbib}

\usepackage{enumitem}

\usepackage{array,booktabs,multirow,rotating,tabu}
\makeatletter
\newcommand{\thickhline}{%
	\noalign {\ifnum 0=`}\fi \hrule height 1.5pt
	\futurelet \reserved@a \@xhline
}
\newcolumntype{"}{@{\hskip\tabcolsep\vrule width 1pt\hskip\tabcolsep}}
\makeatother

\usepackage{tabularx}

\newcommand\clearrow{\global\let\rowmac\relax}
\clearrow
\makeatletter
\newcommand\EightPtClose{\@setfontsize\EightPtClose\@viiipt{9}}
\newcommand\TenPtType{\@setfontsize\TenPtType\@xpt\@xiipt}
\def\notesize{\TenPtType}
\def\notesize{\EightPtClose}
\newenvironment{figurenotes}[1][\vspace{1em}Note]{\begin{minipage}[t]{\linewidth}\notesize{\itshape#1: }}{\end{minipage}}
\makeatother

\usetikzlibrary{decorations.pathreplacing}
\usetikzlibrary{arrows}
\usetikzlibrary{calc}

\usepackage{relsize}
\vspace{\bigskipamount}

\theoremstyle{definition}

\theoremstyle{plain}
\usetikzlibrary{arrows}
\usetikzlibrary{calc}

\newcommand{\q}{\mathbf{q}}

\newcommand{\mbb}{\mathbb}
\newcommand{\E}{\mbb{E}}
\newcommand{\RC}{\texttt{Rochet-Chon\'e}}

\theoremstyle{plain}

\newtheorem{cor}{Corollary}
\theoremstyle{definition}

\newtheorem{exmp}{Example}[section]

\newtheorem{proper}{Properties}

\makeatletter
\newcommand{\distas}[1]{\mathbin{\overset{#1}{\kern\z@\sim}}}%
\newsavebox{\mybox}\newsavebox{\mysim}
\newcommand{\distras}[1]{%
  \savebox{\mybox}{\hbox{\kern3pt$\scriptstyle#1$\kern3pt}}%
  \savebox{\mysim}{\hbox{$\sim$}}%
  \mathbin{\overset{#1}{\kern\z@\resizebox{\wd\mybox}{\ht\mysim}{$\sim$}}}%
}
\makeatother

\newcommand{\lowmathcal}[1]{{\operatorname{\text{\usefont{U}{BOONDOX-cal}{m}{n}#1}}}}

\newtheorem{innercustomgeneric}{\customgenericname}
\providecommand{\customgenericname}{}
\newcommand{\newcustomtheorem}[2]{%
  \newenvironment{#1}[1]
  {%
   \renewcommand\customgenericname{#2}%
   \renewcommand\theinnercustomgeneric{##1}%
   \innercustomgeneric
  }
  {\endinnercustomgeneric}
}

\newcustomtheorem{customassump}{Assumption}
\usetikzlibrary{decorations.pathreplacing,calligraphy}
\usepackage{setspace}
\begin{document}

\title{Identification and Estimation of Multidimensional Screening\thanks{
We thank participants at several seminars and conferences for helpful feedback and comments. We owe special thanks to Yao Luo for his feedback and generosity in sharing his estimation codes.}
}

\author{Gaurab Aryal\thanks{ Department of Economics, Boston University, \href{mailto:aryalg@bu.edu}{ aryalg@bu.edu}.} \qquad
 Federico Zincenko\thanks{College of Business, Department of Economics, University of Nebraska--Lincoln, \href{mailto:fzincenko2@unl.edu}{fzincenko2@unl.edu}.}
}

\date{\today}

\maketitle

\begin{abstract}
We study the identification and estimation of a multidimensional screening model in which a monopolist sells a product with multiple continuous attributes to consumers with private information about their multidimensional preferences. Optimal screening excludes ``low-type" consumers, bunches ``medium-type" consumers, and perfectly screens the ``high types." Assuming a quadratic, additively separable cost function, we identify sufficient conditions to recover the joint distribution of preferences and marginal costs from data on individual choices and payments in a single market. We propose estimators for these objects, establish their asymptotic properties, and assess their small-sample performance using Monte Carlo experiments. Finally, we illustrate our method using synthetic data on Chinese wireless service constructed using estimates from \cite{clx19}.

\vspace{.1in}


\end{abstract}

\maketitle

\doublespacing
\section{Introduction\label{section: introduction}}

We study the identification and estimation of a model of {multidimensional screening (price discrimination)}. 
In this setting, a monopolist sells a product with multiple continuous attributes to consumers with different but possibly correlated tastes for each attribute. Consumers know their preferences, but the seller knows only the joint distribution of preferences and the costs. The seller ``screens'' consumers by offering a menu of products and payments that maximize expected profit, subject to truth-telling and participation constraints.  

We thus contribute to the literature on ``empirical mechanism design," which studies markets with asymmetric information. However, this literature is based on \cite{MussaRosen1978} and \cite{MaskinRiley1984} and focuses predominantly on settings with one-dimensional preferences or costs; see, e.g., \cite{Wolak1994},  \cite{ChiapporiSalanie2003}, \cite{Leslie2004}, \cite{PerrigneVuong2011, LuoPerrigneVuong2018, AttanasioPastorino2020}, \cite{DHaultfoeuilleFevrier2020}, and \cite{AnHongZhang2023}.  

In contrast, we consider an environment in which a seller sells a product with multiple attributes to consumers with private information about their preferences of equal dimensionality, leading to multidimensional screening \citep{Armstrong1996, RochetChone1998}.
Multidimensional screening provides nuanced insights that would otherwise be missed. For instance, the quality distortion under one-dimensional screening and its welfare effects in \cite{MussaRosen1978} may be more pronounced with multidimensional preferences because now it is always profitable to exclude consumers with low willingness to pay (i.e., types) and bunch together some medium types. Under one-dimensional screening, exclusion and bunching can be suboptimal. Consequently, incorrectly using a one-dimensional screening model may bias the welfare estimate upward and the extent of asymmetric information downward. However,  systematic analysis of multidimensional screening is lacking because it is considered intractable. 
 
We begin to fill this gap by developing empirical strategies to identify and estimate the multidimensional screening model of \cite{RochetChone1998} (henceforth, \RC) using only information on the distribution of individual payments and choices in a market. Identification of this model is challenging for several reasons. First, the multidimensional screening model is hard and typically lacks a closed-form solution. Second, by construction, product qualities and the payments are endogenous functions of the seller's costs and preference distribution. Third, as we mentioned earlier, in equilibrium some consumer types are excluded, and others (medium types) are bunched, so the map between preferences and choices is not bijective but many-to-one. Furthermore, estimation is challenging because it is unclear which smoothness conditions on the parameters guarantee the smoothness of the data distribution needed to construct valid estimators. 

The {\RC} characterization provides the key theoretical structure that
allows us to address these identification challenges. In particular, it tells us
which regions of the allocation space correspond to exclusion, bunching, and
perfect screening, and it provides the no-distortion-at-the-boundary condition
that links marginal utilities to marginal costs. These implications are
especially useful in our single-market setting, where \emph{all} product attributes are
endogenously chosen by the seller, and conventional sources of exogenous variation or instrumental variables are difficult to obtain.

Based on this characterization, our identification strategy follows three steps. First, we identify sets of low-, medium-, and high-quality products so that we can classify consumers as low-, medium-, or high-type consumers. In particular, the low-quality set is straightforward to identify because it is the observed outside option. To differentiate between medium- and high-quality products, we rely on the fact that, due to bunching, any medium-quality product has a positive mass relative to any high-quality product. It is important to note that, for this classification exercise, we do not need to rely on the exact form of bunching (which would require solving the seller's optimization problem), only that there is bunching, i.e., multiple medium types choosing the same product. 

Second, having identified the perfectly screened region, we take the consumers classified in the first step as buying high-quality products and identify their joint type density.
To do so, we first identify the gradient of the payment function. Then, since gross utility is multiplicatively separable in (unobserved) consumer preferences and product choice, the demand-side optimality condition implies that marginal payments evaluated at choices equal marginal utilities, which \emph{is} the consumer preference. Thus, choices and payments for high-quality products identify the joint density of the high types.

Next, because we do not use the optimality conditions for bunching, we need additional structure to extend the preference distribution beyond the high type to the entire support. To this end, we can assume that the joint density of types is real analytic (defined shortly below) or belongs to a certain parametric family. In both cases, the distribution preference extends uniquely from high types to its entire support.  

Third, we use a robust feature of the supply-side optimality condition on the boundary to identify the cost function. In particular, in equilibrium, there is \emph{no distortion at the top}, i.e., at the upper boundary of the type space, the (unobserved) marginal cost equals the marginal utility. Moreover, because we identify marginal utility in the first step, we can identify marginal cost along the boundary of high-quality products. Therefore, we can identify the constant marginal cost parameters under the assumption of a quadratic cost function without complementarities across product attributes.   

Our identification strategy is constructive and, as such, informs our estimation approach. Assuming the pricing function is known, and that types follow a parametric distribution, we propose an MLE-based estimator of the type density based on observations of high-type consumers. We show that the resulting estimator is consistent and asymptotically normal. Further, we propose consistent estimators of the marginal cost parameters. Then, we use Monte Carlo experiments to evaluate the small-sample properties of our estimators. 

Finally, we illustrate our empirical approach using synthetic data based on the
Chinese wireless telecommunications market studied by \cite{clx19}. In this
setting, consumers choose quantities of two services, voice and 4G data, providing
a natural interpretation of the multidimensional screening. Using
the pricing functions and parameter estimates reported in \cite{clx19}, we
simulate individual-level choices and payments and apply our estimators
to recover the joint distribution of preferences for voice and data and the
corresponding marginal cost parameters.

In summary, we provide a methodology for studying a market with multidimensional preference heterogeneity and estimating its welfare effects.
Although we focus on a single product with multiple characteristics, our approach applies to other settings, e.g., a multi-product monopolist offering quantity discounts and wage contracting between a monopsony firm and workers with multidimensional skills. Thus, we contribute to the empirical literature that considers multidimensional preferences, e.g., \cite{IvaldiMartimort1994}, \cite{MiraveteRoller2004}, \cite{FinkelsteinMcGarry2006}, \cite{CohenEinav2007}, \cite{Pioner2009}, \cite{NevoTurnerWilliams2016}, \cite{clx19}, \cite{AryalGabrielli2020}, \cite{Luo2023}, \cite{AryalMurryWilliams2023}, and \cite{AryalPerrigneVuongXu2023}.
 
Several issues arise when using our approach. First, the multidimensional screening problem is complex, and its detailed solution can be hard to pin down. We therefore do not rely on the full set of supply-side optimality
conditions. Instead, from the supply side, we use only broad features of optimal screening: low types are excluded, medium types are bunched, high types are perfectly screened, and there is no distortion at the relevant boundary. We then combine these features with the consumers' optimality conditions to obtain our identification results. Second, researchers may not know the payment rule, and observed payments may
contain measurement error. We show that we can relax both assumptions. In particular, we propose a piecewise polynomial estimator of the pricing function and its derivatives that converges sufficiently fast to accommodate an unknown payment rule in our
estimation procedure.
 
Finally, we assume that preferences and products share the same dimensions and that we observe all product characteristics. In other words, we assume no unobserved product quality. 
The applicability of this assumption depends on the empirical setting. For example, this assumption may be reasonable when studying (e.g., labor) contracts in which researchers obtain complete contractual data from the seller, or in a multi-product monopoly setting in which the researcher observes all products (e.g., basic food products). However, it may not be reasonable in other settings, e.g., airline pricing, where we typically observe only some ticket restrictions. In such settings, the literature usually assumes that the missing quality is independent of the observed quality, normalizes the corresponding ``random coefficient," and uses instrumental variables \citep[e.g.,][]{Berry1994}. This approach is infeasible for multidimensional screening, where all qualities are jointly determined, and when data come from a single market, finding instruments can be challenging.    

\vspace{.1in}

\textbf{Notations.} All vectors are considered column vectors unless otherwise stated. For a set $\mathscr{S}$, let $\partial\mathscr{S}$ denote its boundary, while $\mathrm{int} (  \mathscr{S} )$ and $\mathrm{cl} (\mathscr{S})$ denote its interior and closure, respectively.
For any two vectors ${\bf a}\in\mathbb{R}^J$ and ${\bf b}\in\mathbb{R}^J$, write ${\bf a}\leq{\bf b}$  if and only if $a_j\leq b_j$ for all $j=1,\ldots, J$. Write also ${\bf a} < {\bf b}$  if and only if ${\bf a}\leq{\bf b}$ and $a_j <  b_j$ for some $j=1,\ldots, J$. Moreover, denote $[ {\bf a} , {\bf b}  ] = [ a_1, b_1 ] \times \dots \times [ a_J  , b_J  ]$. Let $\boldsymbol{\iota}_J$ be a $J\times1$ vector of ones, and let $\mathrm{diag}({\bf a})$ denote a square diagonal matrix with the elements of vector ${\bf a}$ on the main diagonal and zero elsewhere. Let ${\bf a}^\top$ and $ {\bf a} \cdot {\bf b}$ denote the transpose of ${\bf a}$ and the inner product between $ {\bf a}$ and ${\bf b}$, respectively. Let ${\bf a}\odot {\bf b}$ denote element-wise multiplication between two vectors $\bf a$ and $\bf b$, or matrices. Denote further $\mathbb{N}_0 = \mathbb{N} \cup \{ 0 \}$ and ${\bf a}^{\bf b} = (  a_1^{b_1}  , \dots , a_J^{b_J}) $ for ${\bf b} \in \mathbb{N}_0^J$, adopting the convention $0^0 = 1$. Write also $\| {\bf a} \|_q  = (  \sum_{j=1}^J | a_j  |^q)^{1/q}$ for $1\leq q < \infty$, as well as $\| {\bf a}  \|_\infty = \max_{j=1,\dots,J} | a_j | $. Let $ \mathscr{B} ( \mathbf{t} , \epsilon )$ be an open ball of radius $\epsilon > 0$ centered at $\mathbf{t} $ by $\| \cdot \|_2$ and let $d_H$ be the Hausdorff distance between sets. For a real-valued function $\psi$ of several variables, let $ \nabla \psi$ and $\mathcal{H}\psi$ denote its gradient vector and Hessian matrix, respectively. Let also $\|  \psi\|_{\mathscr{S},\infty}$ be its sup-norm over a set $\mathscr{S}$. With a slight abuse of notation, when $\psi$ is a vector-valued function, write $\|  \psi\|_{\mathscr{S},\infty}= \sup_{\mathbf{t} \in \mathscr{S}} \|  \psi (  \mathbf{t} )  \|_\infty $ and let $\nabla \psi$ denote its Jacobian matrix. Finally, given a $J_1 \times J_2$ matrix $\mathbf{A} = ( A_{j_1,j_2} )$, write $\| \mathbf{A}  \|_{\infty} = \max_{j_1, j_2} |  A_{j_1,j_2} |$ and let $ \rm{det}(A) $ be its determinant. Again, with a slight abuse of notation, when $\psi$ is a matrix-valued function, write $\|  \psi\|_{\mathscr{S},\infty}= \sup_{\mathbf{t} \in \mathscr{S}} \|  \psi (  \mathbf{t} )  \|_\infty $.

\textbf{Normal Vector.}  
For a given set $\mathscr{S} \subset \mathbb{R}^J$, consider $\mathbf{t} \in \partial\mathscr{S}$ and suppose that, in an open neighborhood of $\mathbf{t}$, the boundary $\partial\mathscr{S}$ admits a continuously differentiable parameterization
${\psi}:(0,1)^{J-1}\rightarrow\mathbb{R}^J$
such that $\mathbf{t}=\boldsymbol{\phi}(\mathbf{u}_0)$ for some $\mathbf{u}_0\in(0,1)^{J-1}$ and $\operatorname{rank} \left( \nabla {\psi}(\mathbf{u}_0) \right) =J-1$. The outward unit normal vector $\vec{\lowmathcal{n}}(\mathbf{t})\in\mathbb{R}^J$ is defined as the unique vector satisfying $\left\|\vec{\lowmathcal{n}}(\mathbf{t})\right\|_2=1$ and
$\vec{\lowmathcal{n}}(\mathbf{t})^\top\frac{\partial\boldsymbol{\phi}}{\partial u_j}(\mathbf{u}_0)
= 0, \forall \ j=1,\ldots,J-1$,
with its orientation chosen so that it points outward from $\mathscr{S}$. In words, $\vec{\lowmathcal{n}} (\mathbf{t})$ is perpendicular to the tangent plane of the surface $\partial \mathscr{S}$ at $\mathbf{t}$. See \citet[Section 10.46]{rudin76} for further details.

\textbf{Real Analytic Function.} A real-valued function $\psi(\cdot)$ defined on an open set $ \mathscr{U} \subseteq \mathbb{R}^J$ is a \emph{real analytic} on $ \mathscr{U}$ if for each $\mathbf{t} \in  \mathscr{U}$ there exist an open ball $ \mathscr{B} ( \mathbf{t} , \epsilon) \subset \mathscr{U}$ and a power series such that $\psi(\boldsymbol{\zeta}) =\sum_{\mathbf{k} \in \mathbb{N}_0^J } a_{\mathbf{k}}  (\boldsymbol{\zeta} - \mathbf{t})^{\mathbf{k}}$ and $\sum_{\mathbf{k}} | a_{\mathbf{k}} |  |\boldsymbol{\zeta} - \mathbf{t}|^{\mathbf{k}} < \infty$ for all $\boldsymbol{\zeta}\in  \mathscr{B} ( \mathbf{t} , \epsilon )$.

\section{The Model} \label{section:model}

A monopolist sells an indivisible good with multiple continuous characteristics to consumers who buy at most one unit. 
The product is characterized by a vector of continuous attributes $\mathbf{q}  = (q_1,\dots, q_J) \in \mathbb{R}^{J}$, with $2\leq J < \infty$. Consumers preferences for these attributes is described by a $J$-dimensional random vector $\boldsymbol{\theta} = (\theta_{1},\ldots, \theta_{J})\in \mathbb{R}_{+}^{J}$. We assume that when a type-$\boldsymbol{\theta}$ consumer buys a product with attributes $\mathbf{q}$ and pays $p$, her utility can be written as $U ( \mathbf{q} ;  \boldsymbol{\theta},p  )  :=  \boldsymbol{\theta} \cdot \mathbf{q}  - p  =  \sum_{j=1}^J \theta_j \times q_j -p$. 
Thus, $\boldsymbol{\theta}$ and $\mathbf{q}$ are both $J$-dimensional, and the utility is quasilinear in payment and multiplicatively separable in $\boldsymbol{\theta}$ and $\mathbf{q}$. The components of $\boldsymbol{\theta}$ are the marginal willingness to pay for the corresponding component of $\q$. 

Although we refer to $\q$ as a vector of product attributes, it can also represent several continuous
goods or services jointly supplied by the same seller. 
For example, in our application in Section \ref{sec:application}, the product is telephone service and has two dimensions, i.e., $\q=(q_1,q_2)$, where $q_1$ is voice usage and $q_2$ is data usage.
What matters for the analysis is the multidimensionality
of the endogenous allocation chosen through the screening mechanism rather
than whether the components of $\q$ are conventionally described as
``attributes'' or as separate goods.

Let the total mass of consumers in the population be normalized to one, and $\boldsymbol{\theta}$ be distributed as  $F_{\boldsymbol{\theta}}$. Furthermore, suppose the technology exhibits constant returns to scale so 
the unit cost of producing $\mathbf{q} \in \mathbb{R}^{J}$ is $C(\mathbf{q}) \geq 0$ and suppose that an outside option $\mathbf{q}_0 \in \mathbb{R}^{J}$ is available at price $p_0\geq 0$. Before proceeding, we make the following regularity assumption. 

\begin{assumption}  \label{assump:reg}

The distribution $F_{\boldsymbol{\theta}}$ and cost function $C(\cdot)$ satisfy the following conditions.

\begin{enumerate}[label=(\alph*)]

\item  The support of $F_{\boldsymbol{\theta}}$ is a convex compact subset $\mathscr{S}_{\boldsymbol{\theta}} \subset \mathbb{R}_{+}^{J}$ with nonempty interior. Moreover, $F_{\boldsymbol{\theta}}$ admits a density $f_{\boldsymbol{\theta}}>0$ that is continuously differentiable on $\mathscr{S}_{\boldsymbol{\theta}} $.

\item $C(\cdot)$ is twice continuously differentiable on $\mathbb{R}^J$, strictly convex (i.e., the eigenvalues of the Hessian matrix are uniformly bounded away from zero and infinity), and $p_0\geq C(\q_0)$. 

\end{enumerate}

\end{assumption} 

\noindent We remark that Assumption \ref{assump:reg}-(a) is a standard assumption in the literature, where we assume that the type space $\mathscr{S}_{\boldsymbol{\theta}}$ is convex and the distribution has positive mass everywhere. Assumption \ref{assump:reg}-(b) assumes that the marginal cost increases with the quality (for any dimension). Furthermore, $p_0\geq C(\mathbf{q}_0)$ simplifies the problem because the seller's optimal strategy is to offer the outside option, and the incentive compatibility constraint introduced below is well-defined for all consumers. Thus it follows that $\forall \ \mathbf{t} \in \mathscr{S}_{\boldsymbol{\theta}}$, the function $\mathbf{q} \mapsto \mathbf{t}\cdot \mathbf{q}   - C ( \mathbf{q} ) $ admits a unique maximum on $\mathbb{R}^J $. 

The seller does not observe individual $\boldsymbol\theta$, but knows its distribution $F_{\boldsymbol\theta}$, the cost function $C(\cdot)$, and the utility function $U(\cdot)$. Following the revelation principle and taxation principle, the seller's problem consists in choosing a product space $\mathscr{S} \subseteq \mathbb{R}^J$, an  allocation rule $\tilde{\mathbf{q}}: \mathscr{S}_{\boldsymbol{\theta}} \longrightarrow \mathscr{S} $, and a (nonlinear) pricing function $p: \mathscr{S} \rightarrow\mathbb{R}$ that maximizes expected profit
\begin{eqnarray}
	\int  \left\{ p [ \tilde{\mathbf{q}} ( \mathbf{t} )]  - C[ \tilde{\mathbf{q}} ( \mathbf{t })   ]   \right\} d F_{\boldsymbol\theta}(  \mathbf{t} ), 
	\label{eq:optiprob} 
\end{eqnarray}
subject to the incentive compatibility constraints, i.e., $ 
  \tilde{\mathbf{q}} ( \mathbf{t}) = \arg\max_{\mathbf{q} \in \mathscr{S} } \  U [ \mathbf{q} ; \mathbf{t}  ,  p (\mathbf{q}  )  ]$,  and the participation constraints, i.e., $U \{ \tilde{\mathbf{q}} (\mathbf{t} ) ;  \boldsymbol\theta , p [ \tilde{\mathbf{q}} ( \mathbf{t})] \} \geq  \mathbf{t}  \cdot \mathbf{q}_0  - p_0:=u_0$, for every $\mathbf{t} \in \mathscr{S}_{\boldsymbol{\theta}}$. 
 
Then, under Assumption \ref{assump:reg}, it follows from Theorem 1' in {\RC} that there exists a unique (deterministic) solution to this multidimensional screening problem defined in (\ref{eq:optiprob}), which includes the product space $\mathscr{S}_{\mathbf{Q}} \subseteq \mathbb{R}^J$, an allocation rule $\mathfrak{q}: \mathscr{S}_{\boldsymbol{\theta}}\rightarrow \mathscr{S}_{\mathbf{Q}}$, and a pricing function $ \mathfrak{p}:  \mathscr{S}_{\mathbf{Q}} \rightarrow \mathbb{R}$ such that for all $ \mathbf{t} \in \mathscr{S}_{\boldsymbol{\theta}}$, ${\mathfrak{q}} ( \mathbf{t} ) = \arg\max_{\mathbf{q} \in \mathscr{S}_{\mathbf{Q}}  } \  U [ \mathbf{q};  \mathbf{t},  \mathfrak{p} (\mathbf{q}  )  ]$.

The key insight from {\RC} is that the solution always entails exclusion, where some consumers with low willingness to pay (i.e., the low types) are excluded, \emph{and} bunching of the medium types.
In particular, under optimal screening, the seller divides the type space $ \mathscr{S}_{\boldsymbol{\theta}}$ into three (low, medium, and high) subsets such that: 
\begin{enumerate}
\item  [(i)] \emph{Lowest}-types are excluded and offered only the outside option $\{\q_{0}\}$ at price $p_0$; 
\item  [(ii)] \emph{Medium}-types are bunched and offered ``medium quality'' product. This set is further divided into equivalence classes such that all types in a class choose the same product; 
\item  [(iii)] \emph{High}-types are perfectly screened and allocated a unique product.
\end{enumerate}

Although we do not use all key properties of optimal screening for identification and estimation, we list them here for completeness. We encourage the interested reader to consult {\RC} for formal statements and proofs of these results.

\begin{proper}[from {\RC}]
\label{proper:equi}
The following conditions hold under Assumption \ref{assump:reg}. \begin{enumerate}[label=(\alph*)]

\item For each $  \mathbf{t} \in\mathscr{S}_{\boldsymbol{\theta}}  $,  $\mathfrak{q} ( \mathbf{t} )$ is well-defined. Moreover, $\mathfrak{q}$ is continuous on $\mathscr{S}_{\boldsymbol{\theta}}$ and almost everywhere differentiable. Consequently, $\mathscr{S}_{\mathbf{Q}}$ is path-connected.

\item $\mathbf{q}_0 \in \mathscr{S}_{ \mathbf{Q}}$ and, for each $  \mathbf{q} \in \mathscr{S}_{ \mathbf{Q}}  $, $\mathfrak{q}^{-1} ( \left\{ \mathbf{q}  \right\} ) $ is convex.

\item $\mathscr{S}_{\boldsymbol{\theta}}$ can be partitioned into three disjoint subsets: $\mathscr{S}_{\boldsymbol{\theta}} = \mathscr{S}_{\boldsymbol{\theta},0} \cupdot \mathscr{S}_{\boldsymbol{\theta},1} \cupdot \mathscr{S}_{\boldsymbol{\theta},2} $, where 
\begin{enumerate}
\item $ \mathscr{S}_{\boldsymbol{\theta},0}  : = \{ \mathbf{t} \in \mathscr{S}_{\boldsymbol{\theta}}  :   \mathfrak{q} ( \mathbf{t}  )  = \mathbf{q}_0  \}  =  \{ \mathbf{t} \in \mathscr{S}_{\boldsymbol{\theta}} : \mathfrak{u} ( \mathbf{t}  ) = u_0  \}$;

\item $ \mathscr{S}_{\boldsymbol{\theta},1}  : =  \cup_{\mathbf{q} \in  \mathscr{B} } \mathscr{S}_{\boldsymbol{\theta}} ( \mathbf{q}  )$ with $\mathscr{B}  = \{ \mathbf{q} \in  \mathscr{S}_{\mathbf{Q}} : \     \mathbf{q} \neq \mathbf{q}_0  ,  \  \# \mathfrak{q} ^{-1}( \{  \mathbf{q}\} ) > 1    \}$;

\item $ \mathscr{S}_{\boldsymbol{\theta},2}  : =  \cup_{\mathbf{q} \in  \mathscr{B}^c \backslash \{ \mathbf{q}_0\} } \mathscr{S}_{\boldsymbol{\theta}} ( \mathbf{q}  )$.    

\end{enumerate}

\item For every $\mathbf{t} \in \mathscr{S}_{\boldsymbol{\theta},2}$, $\mathfrak{q} (\mathbf{t} ) $ satisfies the following Euler-Lagrange condition:
\begin{equation}
  \left[ J + 1 -  \sum_{j=1}^J \sum_{j^\prime = 1}^J \frac{\partial^2 C\left[  \mathfrak{q} ( \mathbf{t} )  \right]}{\partial q_j \partial q_{j^\prime}}  \times \frac{\partial  \mathfrak{q}_{j^\prime} ( \mathbf{t} )}{\partial t_j}  \right] f_{\boldsymbol{\theta}} ( \mathbf{t}  )  +( \mathbf{t}   - \nabla C [  \mathfrak{q} ( \mathbf{t} )  ] ) \cdot   \nabla  f_{\boldsymbol{\theta}} ( \mathbf{t}  )    = 0.
\label{eq:intid}
\end{equation}Moreover, there is \emph{no distortion on top}, and hence \begin{equation}
( \mathbf{t}   - \nabla C [   \mathfrak{q} ( \mathbf{t} )  ] ) \cdot  \vec{\lowmathcal{n}} (\mathbf{t})  = 0, \quad \forall\mathbf{t} \in \mathscr{S}_{\boldsymbol{\theta},2} \cap \partial \mathscr{S}_{\boldsymbol{\theta}}.
\label{eq:boundid}
\end{equation}
\end{enumerate}
\end{proper}
\noindent Eq.\ (\ref{eq:intid}) characterizes the optimality condition, which follows from the following two observations. Let $\mathfrak{u} ( \mathbf{t}  )  =   U [ {\mathfrak{q}} ( \mathbf{t}  )   ; \mathbf{t}  ,  \mathfrak{p} ( {\mathfrak{q}} ( \mathbf{t}  )   )  ]$ denote the indirect utility of type $ \mathbf{t} \in \mathscr{S}_{\boldsymbol{\theta}}$ under ${\mathfrak{q}}(\cdot)$. First, \cite{Rochet1987} shows that the optimal screening mechanism satisfies the incentive compatibility constraint on $\tilde{\mathscr{S}}_{\boldsymbol{\theta}}\subseteq\mathscr{S}_{\boldsymbol{\theta}}$ if and only if ${\mathfrak q}( \mathbf{t}  )=\nabla {\mathfrak u}( \mathbf{t}  )$ and ${\mathfrak u}(\cdot)$ is convex and continuous on $\tilde{\mathscr{S}}_{\boldsymbol{\theta}}$. This relationship allows reformulating the seller's problem as choosing the utility $\check{u}$ that each type gets to maximize expected profit $\Pi(\check{u})=\int \{ \mathbf{t}\cdot \nabla {\check{u}}( \mathbf{t}) - C \left[ \nabla \check{u} ( \mathbf{t} )\right] - \check{u}(  \mathbf{t} ) \} d F_{\boldsymbol\theta}(  \mathbf{t} )  $ subject to the individual rationality constraint and to the condition that $\check{u}$ is convex continuous on $\mathscr{S}_{\boldsymbol{\theta}}$. Second, $\check{u}$ is optimal in the perfect screening region, if $\Pi(\check{u}+\varepsilon h)\leq \Pi(\check{u})$ for any feasible utility deviation $\check{u} +\varepsilon h$. Then $\check{u}$ is optimal if, for any convex variation $h$ in the perfect screening region, the marginal loss in the seller's profit at $\check{u}$ is zero, which is the Euler-Lagrange condition \citep[see][Page 179]{Luenberger1969} which is given by  Eq.\ (\ref{eq:intid}).

It is useful to relate Eq.\ (\ref{eq:intid}) to the familiar one-dimensional screening condition. Abstracting from the boundaries, consider an open subset of the perfectly screened region, and take $J=2$. Let
$\mathfrak u(\mathbf{t})$ denote the indirect utility of type $\mathbf{t}$.
Incentive compatibility implies $\mathfrak q(\mathbf{t})=  \nabla \mathfrak u(\mathbf{t})$. 
After substituting the incentive-compatibility condition into expected profit and letting $\mathfrak u_j=\partial\mathfrak u/\partial t_j$, the integrand of expected profit \eqref{eq:optiprob} becomes 
$$
G(\mathbf{t},\mathfrak u,\mathfrak u_1,\mathfrak u_2)=\left[t_1\mathfrak u_1+t_2\mathfrak u_2-C(\mathfrak u_1,\mathfrak u_2)-\mathfrak u\right]f_{\boldsymbol\theta}(\mathbf{t}).
$$
Then the Euler--Lagrange condition is $
 \frac{\partial G}{\partial\mathfrak u} - \frac{\partial}{\partial t_1} \left(
     \frac{\partial G}{\partial\mathfrak u_1} \right) - \frac{\partial}{\partial t_2} \left(
     \frac{\partial G}{\partial\mathfrak u_2} \right) =0$,
which, using $\mathfrak q(\mathbf{t})=\nabla\mathfrak u(\mathbf{t})$, becomes
$$
 f_{\boldsymbol\theta}(\mathbf{t}) + \sum_{j=1}^2
 \frac{\partial}{\partial t_j} \left\{ \left[ t_j- \frac{\partial C(\mathfrak q(\mathbf{t}))}{\partial q_j} \right] f_{\boldsymbol\theta}(\mathbf{t}) \right\} =0.
$$
Expanding the derivatives gives Eq.~(\ref{eq:intid}) for $J=2$.

When $J=1$, the same condition reduces to $
 f(t)+\frac{d}{dt} \left\{ \left[t-C'(q(t))\right]f(t) \right\} =0.$
Integrating from $t$ to the upper endpoint $\overline t$ and using no distortion at the top, $\overline t-C'(q(\overline t))=0$, gives
$$
C'(q(t)) = t-\frac{1-F(t)}{f(t)},
$$
which is the familiar inverse-hazard-rate condition in one-dimensional nonlinear pricing models such as \cite{MaskinRiley1984}. The key difference for $J>1$ is that Eq. (\ref{eq:intid}) is a partial differential equation defined on the perfectly screened region, whose boundary is itself shaped by exclusion and bunching. Thus, the multidimensional problem does not reduce to the scalar inverse-hazard representation available in the one-dimensional case.

Next, we present two illustrative examples, one from {\RC} and the other from \cite{Wilson1993}. Example \ref{example} illustrates the nature of screening and that, in general, there is bunching but no closed-form expression for the optimal allocation rule $\mathfrak{q}(\cdot)$, and Example \ref{example2} is a special case where there is no bunching, i.e., $\mathscr{S}_{\boldsymbol{\theta},1}=\emptyset$. Denote $\mathscr{S}_{\mathbf{Q},l} = \mathfrak{q}(  \mathscr{S}_{\boldsymbol{\theta},l})$ for $l=0,1,2$.

\begin{exmp}[From \RC]  \label{example}

\emph{Let the consumer type $\boldsymbol{\theta}$ be uniformly distributed on the unit square $\mathscr{S}_{\boldsymbol{\theta}} = [0,1]^2$. So, consumers have bi-dimensional preferences for a product with two attributes $\q=(q_1,q_2)$. Let the cost function be $C(\q)= \tilde\beta(q_{1}^{2}+ q_{2}^{2})/2$, for some $\tilde\beta >0$, and let $\q_{0}=0$ and $p_{0}=0$. 
The profit-maximizing seller divides consumers into three subsets $\mathscr{S}_{\boldsymbol{\theta}} = \mathscr{S}_{\boldsymbol{\theta},0} \cupdot \mathscr{S}_{\boldsymbol{\theta},1} \cupdot \mathscr{S}_{\boldsymbol{\theta},2}$ such that all low-type consumers in $\mathscr{S}_{\boldsymbol{\theta},0}$ are excluded, the medium-type consumers in $\mathscr{S}_{\boldsymbol{\theta},1}$ are bunched and offered medium-quality products in $\mathscr{S}_{\boldsymbol{Q},2}$ shown by the red line in Figure \ref{fig:1}. For instance, the figure shows that the subset of types indexed by $\varphi^*$ is incentivized to choose the same quantity $\q_1$. Finally, each high-type consumers in $\mathscr{S}_{\boldsymbol{\theta},2}$ is allocated a unique $\q \in \mathscr{S}_{\boldsymbol{Q},2}$, shown as the shaded region in  Figure \ref{fig:1}. For our purposes, the exact shape of $\mathscr{S}_{\boldsymbol{\theta},2}$ is not as pertinent as the fact that it is a convex subset of $\mathscr{S}_{\boldsymbol{\theta}}$ and the allocation $\mathfrak{q}(\cdot)$ restricted to $\mathscr{S}_{\boldsymbol{\theta},2}$ is bijective.}
\end{exmp}

\begin{figure}[t!]
\centering
\caption{Optimal Product Space in Example \ref*{example}}
\begin{tikzpicture}[scale=4]
\draw [blue, ultra thick, fill= lightgray]  (0.25,0.25) -- (0.25,0.75)-- (0.75,0.75)--(0.75,0.25)--(0.25,0.25);
\draw [red, ultra thick]  (-0.10,-0.10)  -- (0.25,0.25);
\filldraw[green] (-0.10,-0.10) circle (0.6pt);
\node [below] at (-0.10,-0.10){$\mathscr{S}_{\mathbf{Q},0}=\{ \q_{0}\}$};
\node [right] at (0.1,0.1){$\q_{1}=\mathfrak{q}(\varphi^*)$};
\draw [very thick,decorate, decoration = {calligraphic brace}] (-0.10,0.0) --  (0.2,0.3);
\node [left] at (0.09,0.22){$\mathscr{S}_{\mathbf{Q},1} $};
\filldraw[black] (0.1,0.1) circle (0.4pt);
\node at  (0.5,0.5)  {$\mathscr{S}_{\mathbf{Q},2} $};
\end{tikzpicture}
\begin{figurenotes}
A schematic representation of the product space for Example \ref{example}, reproduced from \RC. Here, $\mathscr{S}_{\mathbf{Q}}  = \cupdot_{l=1,2,3} \mathscr{S}_{\mathbf{Q},l}$ is the product space where for optimal allocation $\mathfrak{q}(\cdot)$, the set $\mathscr{S}_{\mathbf{Q},l} = \mathfrak{q} (  \mathscr{S}_{\boldsymbol{\theta},l} )$ for $l=0,1,2$. Here, $\varphi^*$ indexes the subset of medium types incentivized to choose the same $\q=\mathfrak{q}(\cdot)$.
\end{figurenotes}\label{fig:1}
\end{figure}

\begin{exmp}[From \citealp{Wilson1993}] \label{example2}
\emph{Suppose that $\boldsymbol{\theta}=(\theta_1,\theta_2)$ is uniformly distributed on the positive quarter disk $\mathscr{S}_{\boldsymbol{\theta}}=\{ \mathbf{t} \in\mathbb{R}_{+}^{2}:  {t} _{1}^{2}+  {t} _{2}^{2}\leq1\}$. Let the cost function be  $C(\q)=(q_{1}^{2}+q_{2}^{2})/2$, while the outside option is $\q_0=(0,0)$ and its price is $p_0 = 0$. Under optimal screening, (i) low-type consumers with type in $\mathscr{S}_{\boldsymbol{\theta},0}=\{\mathbf{t} \in \mathscr{S}_{\boldsymbol{\theta}}: {t}_{1}^{2}+{t}_{2}^{2}\leq 1/3\}$ are excluded and choose the outside option $\q_0$, (ii) there is no bunching, so $\mathscr{S}_{\boldsymbol{\theta},1}= \emptyset$, and (iii) consumers with type in $\mathscr{S}_{\boldsymbol{\theta},2}=\{\mathbf{t} \in \mathbb{R}_{+}^{2}: 1/3 < {t}_{1}^{2}+{t}_{2}^{2}\leq 1\}$ choose $\q=\mathfrak{q}(\boldsymbol{\theta})$ according to the optimal allocation function 
\begin{eqnarray}
\mathfrak{q}(\boldsymbol{\theta})=\left(\begin{array}{c}\mathfrak{q}_{1}(\boldsymbol{\theta})\\\mathfrak{q}_{2}(\boldsymbol{\theta})\end{array}\right)=\left(\begin{array}{c}0.5\times \max\left\{0, 3-\frac{1}{\theta_{1}^{2}+\theta_{2}^{2}}\right\}\times \theta_{1}\\0.5\times \max\left\{0, 3-\frac{1}{\theta_{1}^{2}+\theta_{2}^{2}}\right\}\times \theta_{2}\end{array}\right). \label{eq:q_wilson}
\end{eqnarray}
Thus, using $\mathfrak{q}(\boldsymbol{\theta})= \nabla \mathfrak{u}(\boldsymbol{\theta})$, we can determine $\mathfrak{u}(\boldsymbol{\theta})$ and therefore $\mathfrak{p}( \q )= \boldsymbol{\theta}\cdot \q-\mathfrak{u}(\boldsymbol{\theta})$. We also note that, the pricing function $\mathfrak{p}(\cdot)$ is twice continuously differentiable on $\mathrm{int} (  \mathscr{S}_{\boldsymbol{\theta},2})$ because $\mathfrak u(\cdot)$ and $\mathfrak q(\cdot)$ are twice continuously differentiable on $\mathrm{int} (  \mathscr{S}_{\boldsymbol{\theta},2})$.
} 
\end{exmp}

These examples help visualize the key features of multidimensional screening at a high level. They also show that no closed-form solution exists for the optimal allocation when there is bunching, and it can be determined only numerically. 
In our setting, without a closed-form solution, it is (a priori) unclear which constraints imposed on the primitives are sufficient for identification and estimation. So, we impose the following high-level assumptions on the primitive and the outcomes. 

\begin{assumption}  \label{assump:idcost2}

The following conditions hold:

\begin{enumerate}[label=(\alph*)]

\item The closure of $\mathrm{int} ( \mathscr{S}_{\boldsymbol{\theta},2} )$ satisfies $ \mathscr{S}_{\boldsymbol{\theta},2} \subseteq \mathrm{cl}[\mathrm{int}( \mathscr{S}_{\boldsymbol{\theta},2}  )    ] $.

 \item The pricing function $\mathfrak{p}(\cdot)$ is twice continuously differentiable on $\mathrm{int} (  \mathscr{S}_{\mathbf{Q},2})$ and $ \mathcal{H} \mathfrak{p}( \mathbf{q})$ is nonsingular for every $ \mathbf{q} \in \mathrm{int} (  \mathscr{S}_{\mathbf{Q},2})$. \label{assump:idcost22}

\item There exist $\mathbf{t}_j^\prime \in \mathscr{S}_{\boldsymbol{\theta},2} \cap \partial \mathscr{S}_{\boldsymbol{\theta}}  $, $j = 1,\dots,J$, such that the normal vectors $\{ \vec{\lowmathcal{n}} (\mathbf{t}_1^\prime  ) , \dots,   \vec{\lowmathcal{n}} (\mathbf{t}_J^\prime  ) \}$ are linearly independent.

\item  There exist ${\mathbf{t}}_j^{\prime\prime} \in \mathrm{int}( \mathscr{S}_{\boldsymbol{\theta},2}  )  $, $j = 1,\ldots,J$, such that $\mathcal{D}_{\beta}  :=  \mathcal{D}_{fq} -  \mathcal{D}_f^\top \mathcal{N}^{-1} (\mathcal{N} \odot \mathcal{Q}^\top )$ is nonsingular, where 
\begin{eqnarray*}
{\mathcal{D}}_{fq}  &  =  &    \left( \begin{array}{ccc} 
\nabla_1 [ f_{ \boldsymbol{\theta} } ( {\mathbf{t}}_1^{\prime\prime} ) \mathfrak{q}_1 ( {\mathbf{t}}_1^{\prime\prime} ) ]  &  \cdots  &  \nabla_J [ f_{ \boldsymbol{\theta} } ( {\mathbf{t}}_1^{\prime\prime} ) \mathfrak{q}_J ( {\mathbf{t}}_1^{\prime\prime} ) ]   \\
\vdots &   \ddots   & \vdots \\
\nabla_1 [ f_{ \boldsymbol{\theta} } ( {\mathbf{t}}_J^{\prime\prime} ) \mathfrak{q}_1 ( {\mathbf{t}}_J^{\prime\prime} ) ]  &  \cdots  &  \nabla_J [ f_{ \boldsymbol{\theta} } ( {\mathbf{t}}_J^{\prime\prime} ) \mathfrak{q}_J ( {\mathbf{t}}_J^{\prime\prime} ) ] 
\end{array} \right)_{[ J \times J ]} ,
 \end{eqnarray*}
\noindent ${\mathcal{D}}_f  =   \left(  \nabla f_{ \boldsymbol{\theta} } ( {\mathbf{t}}_1^{\prime\prime} )  \   \cdots \  \nabla f_{ \boldsymbol{\theta} } ( {\mathbf{t}}_J^{\prime\prime} ) \right)_{[ J \times J ]}$,
$ \mathcal{N}  =  \left( \vec{\lowmathcal{n}} (\mathbf{t}_1^\prime  ) \  \cdots \   \vec{\lowmathcal{n}} (\mathbf{t}_J^\prime  )  \right)^\top,$ and $ \mathcal{Q}   =\left(  \mathfrak{q}( \mathbf{t}_1^\prime )  \   \cdots \  \mathfrak{q}( \mathbf{t}_J^\prime )  \right)_{[ J \times J ]}.$
\end{enumerate}
\end{assumption}

Assumption \ref{assump:idcost2} is a high-level assumption because it delineates specific characteristics of the equilibrium outcomes $ \mathscr{S}_{\boldsymbol{\theta},2} $, $\mathfrak{p} (\cdot)$, and $\mathfrak{q} (\cdot)$. Specifically, part (a) refers to topological aspects of $ \mathscr{S}_{\boldsymbol{\theta},2} $ and can be regarded as a regularity condition. It holds in Examples \ref{example} and \ref{example2}, and in both cases the inclusion is strict, i.e., $ \mathscr{S}_{\boldsymbol{\theta},2} \subset \mathrm{cl}[\mathrm{int}( \mathscr{S}_{\boldsymbol{\theta},2}  ) ]$.
Providing an example where part (a) is violated is challenging because of the complexity of the model and the difficulty in deriving closed-form expressions of the equilibrium outcomes as a function of structural parameters. Building such an example requires developing a new theory of mechanism design, which is beyond the scope of our research. In this regard, this paper's main contribution is a tractable empirical framework that addresses the challenges posed by the {\RC} model.

Part (b) imposes a smoothness condition on the pricing function $\mathfrak{p} (\cdot)$ and a nonsingularity condition on its Hessian matrix $ \mathcal{H} \mathfrak{p}( \cdot )$, both of which are identified as discussed in the next section. The nonsingularity condition rules out, e.g., locally linear pricing functions on $\mathrm{int} (  \mathscr{S}_{\mathbf{Q},2})$. Roughly speaking, the nonlinearity of the pricing function captures the discount that the seller offers for those who choose high-quality products and is a hallmark of screening models \citep{Wilson1993}.

Parts (c) and (d) of Assumption \ref{assump:idcost2} are regularity conditions that help identify the cost function. More specifically, part (c) refers to the shape of $ \mathscr{S}_{\boldsymbol{\theta}}$ and its subset $ \mathscr{S}_{\boldsymbol{\theta},2}$. This assumption clearly holds for Examples \ref{example} and \ref{example2}.
As suggested by {\RC}, $ \mathscr{S}_{\boldsymbol{\theta},2} \cap \partial \mathscr{S}_{\boldsymbol{\theta}}  $ should be nonempty because, under optimal screening, the seller always prefers to perfectly screen high-type consumers whose types are located in the north-east region of $\partial \mathscr{S}_{\boldsymbol{\theta}}$. Part (d) ensures that the derivative of the type density and the allocation rule meet specific conditions; for instance, these conditions can be compared to those in \citet[Section 3]{matzkin2015estimation}. Part (d) holds in Example \ref{example2} as shown in Section \ref{section:mc} below, where we also provide concrete values for the vectors ${\mathbf{t}}_j^{\prime}$ and ${\mathbf{t}}_j^{\prime\prime} $  (see Figure \ref{figmc}).

The next lemma follows as a consequence of Assumptions \ref{assump:reg} and \ref{assump:idcost2}-(b).

\begin{lemma}	\label{lem:foc}
Suppose that Assumptions \ref{assump:reg} and \ref{assump:idcost2}-(b) hold. Then, 
\begin{eqnarray}
\nabla  \mathfrak{p} \left[   \mathfrak{q} (  \mathbf{t}  )   \right]   =   \mathbf{t}, \quad \forall \ \mathbf{t} \in \mathrm{int}( \mathscr{S}_{\boldsymbol{\theta},2}  ) 
\label{eq:lemfoc}
\end{eqnarray}
and we can write $\mathrm{int}( \mathscr{S}_{\boldsymbol{\theta},2}  )  = \nabla  \mathfrak{p} [   \mathrm{int}( \mathscr{S}_{\mathbf{Q},2}  )  ]$. Consequently, $\mathfrak{q} (  \cdot  )$ is continuously differentiable on $ \mathrm{int}( \mathscr{S}_{\boldsymbol{\theta},2}  ) $ and $\nabla \mathfrak{q} (   \mathbf{t}  )$ is nonsingular for every $ \mathbf{t} \in \mathrm{int}( \mathscr{S}_{\boldsymbol{\theta},2}  ) $.
\end{lemma}
\noindent This result formalizes the idea that the incentive compatibility constraints for the high types imply that, at the chosen $\q$, the marginal price equals the marginal utility, which, given the multiplicatively separable preferences, is the consumer's type. The continuity of the allocation rule (Property \ref{proper:equi}-(a)), Assumption \ref{assump:idcost2}-(b) about the pricing function, and the Invariance of Domain Theorem ensure that the marginal price is well defined for all high types. 
We complete this section with an assumption about the marginal cost function.

\begin{assumption}  \label{assump:idcost1}

The marginal cost function $\nabla C(\cdot)$ is linear and separable across attributes, i.e., there are vectors $\boldsymbol{\alpha}  \in \mathbb{R}^J_+ $ and $\boldsymbol{\beta} \in \mathbb{R}^J_{++}$ such that $\nabla C ( \mathbf{q} ) =  \boldsymbol{\alpha}  + \boldsymbol{\beta} \odot  \mathbf{q}$ for all $\mathbf{q} \in \mathscr{S}_{\mathbf{Q}}$.

\end{assumption}

This assumption, which is similar to \cite{LuoPerrigneVuong2018} for $J=1$, implies that the cost function is quadratic and additively separable in each dimension, so the marginal cost of producing $q_j$, $\nabla_j C ( \mathbf{q})= \alpha_j + \beta_j q_j$, depends only on $q_j$. 
It also implies no cost complementarity across any two quality dimensions, and the vector $(\alpha_j, \beta_j)$ characterizes the marginal cost, $j=1,\ldots, J$. In keeping with the theoretical literature on screening, Assumption \ref{assump:idcost1} rules out stochastic costs. If the cost is stochastic due to an idiosyncratic component observed by the seller but not by the researcher, the seller would incorporate those cost shocks into the screening mechanism.

\section{Identification}  \label{sec:id}

In this section, we determine sufficient conditions under which we can identify the $J$-variate preference distribution $F_{\boldsymbol\theta}$ and the marginal cost function $ \nabla C:\mathbb{R}^J\rightarrow \mathbb{R}^J$, from the joint distribution,  $F_{\mathbf{Q}, P}$, of choices and payments $(\mathbf{Q}, P ) \in\mathbb{R}^J\times\mathbb{R}$, that satisfy the optimality conditions $\mathbf{Q} =  \mathfrak{q}( \boldsymbol\theta ), P = \mathfrak{p} (\mathbf{Q}  )$, and ${\boldsymbol\theta}\sim F_{\boldsymbol\theta}$. We assume that the researcher knows $F_{\mathbf{Q}, P}$ and the outside option, $\q_0$.

Our identification strategy consists of four steps. First, we identify the pricing function. For that purpose, note that the conditional distribution $F_{P\mid \mathbf{Q}} (\cdot\mid\mathbf{q} )$ is degenerate at $\mathfrak{p} ( \mathbf{q}  )$ because the pricing function is unique and deterministic, i.e., $P=\mathfrak{p} ( \mathbf{q}  )$ with probability one (w.p.1) for any $\mathbf{Q}  = \mathbf{q}$. This  identifies $\mathfrak{p}(\cdot)$ on $\mathscr{S}_{\mathbf{Q}}$, which is the support of $F_{\mathbf{Q}} $.

Second, as mentioned earlier, the solution to the consumer's problem $\max_{\mathbf{q}\in\mathscr{S}_{\mathbf{Q}}}U [ \mathbf{q}; {\boldsymbol\theta},  \mathfrak{p} ( \mathbf{q}  )]$ induces a mapping, $\mathscr{S}_{{\boldsymbol\theta}}\mapsto \mathscr{S}_{\mathbf{Q}}$, which is bijective only for the subset $\mathscr{S}_{{\boldsymbol\theta},2}$. For the other types, exclusion and bunching mean that more than one type chooses the same product. So, for the next step, we separately identify the subsets $\{ \mathscr{S}_{\mathbf{Q},l}:  l=1,2 \}$ from the joint distribution of choices $F_{\mathbf{Q}}$, noting that $\mathscr{S}_{\mathbf{Q},0}=\{\q_0\}$ is the outside option and hence identified. In some cases, determining the set of medium-quality products, $\mathscr{S}_{\mathbf{Q},1}$, is relatively straightforward. Consider the next stylized example for additional intuition.

\begin{exmp}\emph{ 
Suppose that $J=2$, the product line is as shown in Figure \ref{fig:price}, and the outside option is $\q_0=(0,0)$. For instance, in the telecommunications industry, a product could be a plan with two dimensions: voice and internet data. Some consumers use only voice data up to a certain point, i.e., $0< q_{1} \leq \bar{q}_1$ and $q_2=0$ for some $\bar{q}_1 > 0$. This identifies $\mathscr{S}_{\mathbf{Q},1}=\{\q \in \mathbb{R}_+^2 :  0< q_{1} \leq \bar{q}_1 , q_{2}=0\}$ as the red line in Figure \ref{fig:price}. Moreover, types are multidimensional, so points over the red line are expected to be ``heavier'' than the ones in the gray rectangle.}    
 \end{exmp}

\begin{figure}[t!]
\caption{Example of Product Space}
  \tdplotsetmaincoords{75}{150}
\pgfmathsetmacro{\rvec}{.5}
\pgfmathsetmacro{\thetavec}{90}
\pgfmathsetmacro{\phivec}{45}
\centering
\parbox{5cm}{
\begin{tikzpicture}[scale=4.5]
\draw[thick,->] (0,0) -- (1,0); 
\draw[thick,->] (0,0) -- (0,1) node[anchor=north east]{$q_2$}; 
\draw [ ultra thick, red] (0,0)  -- (0.45,0) ;
\draw [blue,  ultra thick, fill= lightgray]  (0.45,0) -- (0.45,.85)-- (.85,.85)--(.85,0)--(0.45,0);
\filldraw[green] (0.45,0) circle (0.4pt);
\node at  (1.02,-0.07){${q}_1$};
\node at  (0.85,-0.07){$\overline{q}_1$};
\node at  (0.65,0.4)  {$\mathscr{S}_{\mathbf{Q},2}$};
\node at  (0.65,0.2){\texttt{perfect}};
\node at  (0.65,0.1){\texttt{screening}};

\node at  (0.2,0.07) {$\mathscr{S}_{\mathbf{Q},1}$};
\node at  (0.22,-0.07){\texttt{bunching}};
\filldraw[green] (0,0) circle (0.4pt);
\node at  (-0.25,0){$\mathscr{S}_{\mathbf{Q},0}=\{\q_{0}\}$};
\end{tikzpicture}
}
\begin{figurenotes}
An stylized example where $\mathscr{S}_{\mathbf{Q},1}=\{\q \in \mathbb{R}_+^2 :  0< q_{1} \leq \bar{q}_1 , q_{2}=0\}$.
\end{figurenotes}
\label{fig:price}
\end{figure}  

More generally, the next lemma shows that we can separate medium-quality products from high-quality ones by noting that a medium-quality product is concentrated among more consumers (because of bunching) and has a higher mass than a high-quality product. In other words, to cream-skim ``one" high type, the seller degrades the attributes of products meant for ``several" medium types and offers them the degraded option. 

\begin{lemma}	\label{le:idsubsets}
Under Assumptions \ref{assump:reg} and \ref{assump:idcost2}, the sets $\{ \mathscr{S}_{\mathbf{Q},l}:  l=0,1,2 \}$ are identified.
\end{lemma}

Lemma \ref{le:idsubsets} is interesting in and of itself, and will be important for identifying the preference density  $f_{\boldsymbol\theta}$ later. The proof of this lemma is in Appendix \ref{section:proofs} and requires knowing only the joint distribution of choices $F_{\mathbf{Q}}$ and the outside option $\q_0$. Essentially, we exploit the idea that the distribution of $\mathbf{Q}$ may change sharply between the bunching and perfectly screened regions, assigning positive probability to lower-dimensional spaces in the former. To see this, e.g., consider Figure \ref{fig:1} and note that a line segment contained in the red bunching region can have strictly positive probability even though it has zero two-dimensional Lebesgue measure. By contrast, a line segment of the same length contained in the perfectly screened region has probability zero.

In the third step of the identification process, we identify the type density $f_{\boldsymbol\theta}$ on the interior of $\mathscr{S}_{\boldsymbol{\theta},2}$ using the supply-side pricing function. For that purpose, observe that by combining Lemma \ref{lem:foc} with prior identification findings, we can identify $\mathrm{int} ( \mathscr{S}_{\boldsymbol{\theta},2} )$ as $\mathrm{int}(\mathscr{S}_{\boldsymbol{\theta},2}) = \{  \nabla \mathfrak{p}(\q):  \q \in \mathrm{int} ( \mathscr{S}_{\mathbf{Q},2} )  \}$. Then, $f_{\boldsymbol\theta}$ can be recovered from the equality 
\begin{eqnarray}
\boldsymbol{\theta} = \nabla \mathfrak{p}( \mathbf{Q}),  \  \  \text{for}  \  \boldsymbol{\theta}  \in \mathrm{int} ( \mathscr{S}_{\boldsymbol{\theta},2} ) ,
\label{eq:gpvtype}
\end{eqnarray}which follows from Lemma \ref{lem:foc} and the fact that $\mathbf{Q}$ are optimal choices. In words, under perfect screening, a consumer of type $\boldsymbol{\theta} $ chooses $\q$ such that the marginal utility from the choice equals the marginal price; hence, $\nabla_{\q}U [ \mathbf{Q};  \boldsymbol{\theta}, \mathfrak{p}(\mathbf{Q})   ] = 0$ and therefore choices provide precise information about how $\boldsymbol{\theta} $ is distributed. This idea is formalized in the next lemma.

\begin{lemma} \label{le:fid}
Under Assumptions \ref{assump:reg} and \ref{assump:idcost2},  $f_{\boldsymbol\theta}$ is identified on $\mathrm{int} ( \mathscr{S}_{\boldsymbol{\theta},2} )$.
\end{lemma}

Two observations are noteworthy. First, this identification result is independent of the cost function. Second, we can extend the identified region in Lemma \ref{le:fid} to include the bunching region, $\mathscr{S}_{\boldsymbol{\theta},1}$,  if we allow an additional structure on $f_{\boldsymbol\theta}$. For instance, it can be shown that $f_{\boldsymbol\theta}$ is nonparametrically identified on $\mathscr{S}_{\boldsymbol{\theta}}$ if $f_{\boldsymbol\theta}$ is assumed to be real analytic on $\mathrm{int}( \mathscr{S}_{\boldsymbol{\theta}} )$.\footnote{Similar assumption has been used to identify nonparametric IV \citep{NeweyPowell2003}, random coefficient Logit \citep{FoxKimRyanBajari2012}, and random utility \citep{FoxGandhi2013} models.} Analyticity implies that when two real analytic functions coincide on an open subset of $\mathrm{int}( \mathscr{S}_{\boldsymbol{\theta}} )$, they must also coincide on $\mathrm{int}( \mathscr{S}_{\boldsymbol{\theta}} )$. Thus, $f_{\boldsymbol\theta}$ must have a unique extension over $\mathrm{int}( \mathscr{S}_{\boldsymbol{\theta}} )$ from $\mathrm{int} ( \mathscr{S}_{\boldsymbol{\theta},2} )$. Since $f_{\boldsymbol\theta}$ is continuous and $\mathscr{S}_{\boldsymbol{\theta}}$ is convex and compact, which implies $\mathscr{S}_{\boldsymbol{\theta}} = \mathrm{cl}[\mathrm{int} ( \mathscr{S}_{\boldsymbol{\theta}}  )   ]$, it is also straightforward to extend this identification result to $\mathscr{S}_{\boldsymbol{\theta}}$.

In the fourth and last step, we identify the marginal cost function $\nabla C$ on $\mathscr{S}_{\mathbf{Q}}$. We focus on marginal costs because, in equilibrium, observed prices (and therefore choices) depend on marginal costs, not on fixed costs. In other words, we cannot identify the fixed cost from choices and payments. We use the supply-side optimality conditions, Eqs.\ (\ref{eq:intid}) and (\ref{eq:boundid}), to identify the marginal cost function. 

In particular, Eq.\ (\ref{eq:intid}) implies that for every $\q \in  \mathscr{S}_{\boldsymbol{\bf Q},2} $, marginal revenue equals marginal cost, and marginal revenue is a functional of $F_{\boldsymbol{\theta}}$. Eq.\ (\ref{eq:boundid}) implies that the marginal cost of producing goods for high-type consumers, whose types are at the boundary region $\mathscr{S}_{\boldsymbol{\boldsymbol{\theta},2}} \cap \partial \mathscr{S}_{\boldsymbol\theta}$, is equal to their marginal utilities. 
The next lemma formalizes this idea.

\begin{lemma}	\label{le:costfun}
Suppose that Assumptions \ref{assump:reg}-\ref{assump:idcost1} hold and that the values of $(\mathbf{t}_j^\prime , \vec{\lowmathcal{n}} (\mathbf{t}_j^\prime  ) )$, $j=1,\dots,J$, from Assumption \ref{assump:idcost2}-(c) are known. Then, $\nabla C$ is identified on $\mathscr{S}_{\mathbf{Q}}$; specifically, \begin{equation*}
\boldsymbol{\beta}  =     \mathcal{D}_{\beta}^{-1}  \left[ (J  + 1)  \left(\begin{array}{c}
 f_{ \boldsymbol{\theta} } ( {\mathbf{t}}_1^{\prime\prime} ) \\
 \vdots \\
  f_{ \boldsymbol{\theta} } ( {\mathbf{t}}_J^{\prime\prime} )
\end{array}  \right)    +    \left(\begin{array}{c}
({\mathbf{t}}_1^{\prime\prime} -  \mathcal{N}^{-1} \mathbf{T}  ) \cdot \nabla f_{ \boldsymbol{\theta} } ( {\mathbf{t}}_1^{\prime\prime} ) \\
 \vdots \\
( {\mathbf{t}}_J^{\prime\prime}  -  \mathcal{N}^{-1} \mathbf{T}  ) \cdot \nabla  f_{ \boldsymbol{\theta} } ( {\mathbf{t}}_J^{\prime\prime} )
\end{array}  \right)     \right]  \     \text{and} \     \boldsymbol{\alpha}     =    \mathcal{N}^{-1} \left[ \mathbf{T}    - \left( \mathcal{N}  \odot \mathcal{Q}^\top \right) \boldsymbol{\beta} \right]       ,
\end{equation*}
where $\mathbf{T} = \left( \mathbf{t}_1^\prime \cdot \vec{\lowmathcal{n}} (\mathbf{t}_1^\prime  )  ,  \dots ,  \mathbf{t}_J^\prime \cdot\vec{\lowmathcal{n}} (\mathbf{t}_J^\prime  ) \right)_{[ J \times 1]}$.

\end{lemma}

The proof of this lemma, which is provided in the Appendix \ref{section:proofs}, consists of plugging in the functional form of $\nabla C$ given by Assumption \ref{assump:idcost1} into Eqs.\ (\ref{eq:intid}) and (\ref{eq:boundid}), and then solving the resulting system of linear equations by using the conditions provided in Assumptions \ref{assump:idcost2} and \ref{assump:idcost1}. We remark that, in practice, the condition that $(\mathbf{t}_j^\prime, \vec{\lowmathcal{n}} (\mathbf{t}_j^\prime  ) )$ is known is reasonable, as high types that lie at the north-east of the boundary $\partial \mathscr{S}_{\boldsymbol{\boldsymbol{\theta}}}$ are usually perfectly screened and $\mathrm{cl}[\mathrm{int}( \mathscr{S}_{\boldsymbol{\boldsymbol{\theta},2}}  )]$ has been identified above. We also note that $\boldsymbol{\beta} $ is overidentified as there exist infinitely many ${\mathbf{t}}_j^{\prime\prime} \in \mathrm{int}( \mathscr{S}_{\boldsymbol{\theta},2}  ) $ satisfying Assumption \ref{assump:idcost2}-(d).

\subsection{Discussion}
Previous identification strategies rely mainly on {\RC} results on a model of multidimensional screening without solving the full multidimensional screening problem. 
This structure nonetheless imposes restrictions on the dimensions of
consumer heterogeneity and product characteristics and on the form of utility.
Other restrictions, most notably the parametric structure
on marginal costs, are not required by the model but are sufficient for the point identification from the equilibrium conditions
available in a single market.
In this section, we clarify the role of these assumptions.

\subsubsection{Dimensionality}
A central feature of the model is that the dimension of the consumer's private
type coincides with the dimension of the allocation. 
This equal-dimensional structure is motivated by the fact that otherwise perfect screening can be ruled out for purely dimensional reasons.
Moreover, the same restriction has a particularly transparent interpretation for our
identification argument, for which we recall that the consumer's first-order condition \eqref{eq:gpvtype} shows that the $J$ derivatives of the observed pricing schedule provide $J$ equations that recover the $J$ components of the consumer's private type. 

Suppose first that $\dim({\boldsymbol\theta})>\dim(\q)$, so the seller has fewer instruments than types, and some aggregation of types is unavoidable, leading to bunching everywhere in the type space. In other words, there will be two types of consumers ${\boldsymbol\theta}\neq {\boldsymbol\theta}'$ that choose the same product $\q$, so price derivatives cannot pin down the consumer type. Identification of the full
distribution of types would consequently require additional structure, such as
an index restriction specifying which combinations of the components of
$\theta$ affect demand, or additional observable choices that provide
independent information about the missing dimensions. An example such setting is the identification of risk and risk preference parameters in \cite{CohenEinav2007} and \cite{AryalPerrigneVuongXu2023}, when $\dim({\boldsymbol\theta})=2$ and $\dim(\q)=1$. 

Conversely, if $\dim(\q)>\dim({\boldsymbol\theta})$, the seller has more screening
instruments than there are primitive dimensions of private heterogeneity. 
While this case appears less natural for most empirical applications, $\dim(\q)>\dim({\boldsymbol\theta})$ does not by itself
invalidate our identification argument. Some components of $\q$ may be redundant, or
the higher-dimensional allocation may enter preferences only through a
lower-dimensional, known index. For example, several observable product
attributes may jointly determine a single utility-relevant quality index. After
an appropriate reparameterization, the effective allocation space can then have
the same dimension as the type space, where our approach is applicable. More generally, when the additional allocation
dimensions are economically distinct, the appropriate identification argument
depends on the particular screening model and on the restrictions linking those
instruments to preferences. Thus, the relevant issue is not the raw number of
observed allocation variables, but the dimension of the utility-relevant
allocation through which private information is screened. See, for example,
\cite{Basov2005} for a broader treatment of multidimensional screening models
and the role of dimensionality.

These arguments also explain why observing all economically relevant product attributes is important for our identification strategy. To see this, suppose that the true allocation is $\q=(\q^{o},q^{u})$, where $\q^{o}$ is observed by the econometrician and the one-dimensional $q^{u}$ is not. The relevant pricing schedule is then a function of the full allocation, $\mathfrak{p}(\q^{o},q^{u})$. So, while applying \eqref{eq:gpvtype} gives $ {\boldsymbol\theta}=\nabla_{\q}\mathfrak{p}(\q^{o},q^{u})$, without observing $q^{u}$ we cannot evaluate the gradient and therefore our identification approach does not apply. 

Another reason an unobserved product characteristic is particularly consequential in this setting is that, in standard demand models, product characteristics are often decomposed into observed attributes and an unobserved quality component. The latter is typically assumed to be independent of the observed product characteristics, with its coefficient implicitly normalized to one.
Under appropriate exclusion restrictions, such as in \cite{BajariBenkard2005}, the resulting price
endogeneity can then be addressed using instrumental variables. In principle,
a similar strategy could be adopted here if one were willing to
impose analogous restrictions and had access to suitable time-varying instruments that shift the pricing schedule without affecting preferences.

We do not pursue this approach for two reasons. First, finding such instruments
in our setting of one market would be difficult because the seller jointly chooses the
pricing schedule and all product characteristics as part of the screening
mechanism. More importantly, any restrictions on 
unobserved quality would be difficult to reconcile
with the economics. The observed characteristics are
chosen by  a profit-maximizing seller and are likely correlated and enter utility with
heterogeneous coefficients, whereas the single unobserved characteristic
would have to be treated differently. 
We therefore consider a restrictive but internally consistent setting in which all characteristics are observed.

\subsubsection{Consumer Heterogeneity}

We model consumer preferences by a finite-dimensional parameter ${\boldsymbol\theta}\sim F_{\boldsymbol\theta}(\theta_1,\ldots,\theta_J)$, allowing for arbitrary dependence among willingness to pay for the different allocation dimensions. For example, in a two-dimensional application, consumers with a high willingness to pay for $q_1$ may also systematically have a high willingness to pay for $q_2$, or vice versa.

Thus far, $ {\boldsymbol\theta} $ summarizes all privately observed heterogeneity relevant to the screening problem. A natural extension is to introduce consumer characteristics $X$ that are observable to both the econometrician and the seller, allowing the distribution of private types to depend on these characteristics, i.e., ${\boldsymbol\theta}\mid X=x \sim F_{\boldsymbol\theta\mid X}(\cdot\mid x)$.
We remark that this formulation does not require the conditional support of ${\boldsymbol\theta}$ to be invariant with respect to $x$. Instead, we can allow the support $\boldsymbol\theta$ to be $\mathscr{S}_{\boldsymbol\theta}(x)$, so that both the distribution and the set of possible private types may vary with observed consumer characteristics. For instance, if $X$ represents income, it may be reasonable for the conditional type distribution of higher-income consumers to place less weight on, or even exclude, types associated with very low willingness to pay. Such dependence can be accommodated by allowing $\mathscr{S}_{\boldsymbol\theta}(x)$ to vary with $x$, so our framework does not require high-income consumers to exhibit the same range of private types as low-income consumers.

For a given $x$ in the support of $X$, $x \in \mathscr{S}_X$, consider the conditional distribution $F_{\boldsymbol\theta\mid X}(\cdot\mid x)$ as our object of interest. Assume that the econometrician knows the conditional c.d.f.\ of $(\mathbf{Q}, P )$ given $X=x$, $F_{\mathbf{Q}, P | X} ( \cdot, \cdot \mid x)$, and suppose that $F_{\boldsymbol\theta\mid X}(\cdot\mid x)$ satisfies Assumption \ref{assump:reg}, noting that $\mathscr{S}_{\boldsymbol\theta}(x)$ may not coincide with the unconditional support $\mathscr{S}_{\boldsymbol\theta}$, or with the conditional support associated with any other value of $x \in  \mathscr{S}_X$.
Whether our identification results extend directly to this setting depends on how the seller may use the observed characteristic $x$ when constructing the pricing schedule $\mathfrak{p}$.

On the one hand, suppose that the seller can offer a separate menu to consumers with each value of $x$, with consumers restricted to the menu corresponding to their observed characteristics. That is, a consumer with characteristics $x^\prime\neq x$ cannot select from the menu designed for consumers with characteristics $x$.
Assume further that the relevant pricing and quantity schedules, $\mathfrak{p}(\cdot;x)$ and $\mathfrak{q}(\cdot;x)$, are the result of the seller separately maximizing \eqref{eq:optiprob} where the expectation is with respect to $F_{\boldsymbol\theta | X}(  \mathbf{t} \mid  x)$, i.e, $\int  \left\{ p [ \tilde{\mathbf{q}} ( \mathbf{t} )]  - C[ \tilde{\mathbf{q}} ( \mathbf{t })   ]   \right\} d F_{\boldsymbol\theta | X}(  \mathbf{t} \mid  x)$, for each $x \in  \mathscr{S}_X$, with respect to $(\tilde{\mathbf{q}}, p)$ and subject to the corresponding IC and participation constraints. 

In this case, the seller combines second- and third-degree price discrimination, and, conditional on $X=x$, the model reduces to the screening problem studied in previous sections. Consequently, for a given $x \in \mathscr{S}_X$, our identification results for $F_{\boldsymbol\theta\mid X}(\cdot\mid x)$ and the cost parameters extend immediately, provided that Assumption \ref{assump:idcost2} holds with $\mathscr{S}_{\boldsymbol\theta,j}(x)$, $\mathscr{S}_{\mathbf{Q},j}(x)$, $\mathfrak{p}(\cdot;x)$, and $f_{\boldsymbol\theta\mid X}(\cdot\mid x)$ replacing $\mathscr{S}_{\boldsymbol\theta,j}$, $\mathscr{S}_{\mathbf{Q},j}$, $\mathfrak{p}$, and $f_{\boldsymbol\theta}$, respectively.

On the other hand, suppose the seller cannot offer different menus to consumers with different values of $x$ or, more generally, does not maximize its expected profit separately for each $x$.\footnote{Such restrictions on price discrimination can arise in practice even when consumer characteristics are observable. For example, \citet{DellaVignaGentzkow2019} document that large U.S. retail chains typically charge nearly uniform prices across stores despite differences in local consumer demographics, while \citet{Weiss2000} discusses Amazon's decision not to base customer-specific prices on demographic information.} In this case, the structure of the optimal mechanism--including the occurrence and location of perfect screening and bunching--must be characterized for this richer screening problem and cannot, in general, be inferred by conditioning the analysis above on $X=x$. Because these properties of the optimal mechanism play a central role in our identification strategy, our existing results do not immediately extend to this case. Characterizing optimal mechanisms in related environments remains an active area of research in economic theory \cite[]{BergmannHeumannWang2026}.
However, to the best of our knowledge, the corresponding problem with multidimensional preference types has not yet been fully characterized. 

Thus, observable consumer characteristics do not, by themselves, pose a problem for our approach. If $X$ affects the distribution of private types and the seller can condition the menu on $X$, we can apply our analysis separately conditional on each $X=x$, subject to the corresponding conditional-support and regularity conditions. The difficulty arises when the seller does not solve the problem separately for each $x$. In that case, for example, consumers with different conditional type distributions must be screened through a common menu, and the resulting mechanism falls outside the class of screening problems we consider.

\subsubsection{Utility Specification}
We choose our baseline utility specification for its model tractability. 
The linear-index structure in private types, i.e., $\boldsymbol{\theta} \cdot \mathbf{q}$, underlies the convex-analytic characterization of incentive compatibility constraints that are integral in \RC. However, the linearity of utility in $q_j$ is less restrictive than the utility may 
suggest. We can consider utility function of the more general form  $\mathfrak{u} ( \mathbf{q} ;  \boldsymbol{\theta},p) =  \boldsymbol{\theta} \cdot v(\mathbf{q})  - p $, where $ \boldsymbol{v}(q)=\left(v_1(q),\ldots,v_J(q)\right)$ is a $J-$dimensional vector of one-to-one functions that are common knowledge by the players and also known to the econometrician. Defining $\widetilde \q=\boldsymbol{v}(\q)$, we can re-write the indirect utility function as  $\mathfrak{u} ( \mathbf{q} ;  \boldsymbol{\theta},p)= \boldsymbol{\theta} \cdot\widetilde\q-p$, and apply {\RC} to determine the solution.  

Our identification argument can also suggest how the assumption that
$\boldsymbol{v}(\q)$ is known by the econometrician might be relaxed in the presence of additional exogenous
variation. To see this, in a perfectly screened region,
the consumer's first-order conditions \eqref{eq:gpvtype} give $\partial_j\mathfrak p_z(\q)=
\theta_j v_j'(q_j), j=1,\ldots, J$. 
With data from a single market, the type $\theta_j$ and marginal utility
$v_j'(q_j)$ cannot in general be separately identified from the
observed marginal price.

An additional exogenous cost ``shifter" can potentially break this
substitutability between $\theta_j$ and $v_j'(\cdot)$. Suppose that $Z=\{z_1, z_2\}$ changes the seller's costs and hence the
pricing schedule, but is excluded from consumer preferences and does not
change the distribution of ${\boldsymbol\theta}$. If the same type ${\boldsymbol\theta}$ chooses $\q_1({\boldsymbol\theta})$ and $\q_2({\boldsymbol\theta})$ under $Z=z_1$ and $Z=z_2$, respectively, then we have $ \frac{\partial_j\mathfrak p_1(\q_1({\boldsymbol\theta}))}{
 \partial_j\mathfrak p_2(\q_2({\boldsymbol\theta}))}=\frac{v_j'(q_{1j}({\boldsymbol\theta}))
 }{v_j'(q_{2j}({\boldsymbol\theta}))}.$
Thus, if allocations corresponding to the same latent type can be matched
across the two regimes, observed changes in marginal prices provide functional
equations for the unknown marginal utilities. In the one-dimensional case,
such matching follows directly from rank invariance and the two observed
choice distributions, yielding an identification argument analogous in spirit
to \cite{GPV2009} which can be estimated using  \cite{Zincenko2017}. With multidimensional types and allocations, however, there
is no canonical scalar rank, and identifying the same-type matching across
markets requires additional structure. Developing such a multivariate
matching argument is beyond the scope of the present paper.

There is a deeper theoretical reason for maintaining the linear-index structure in types. Let $\mathfrak{u}({\boldsymbol\theta})$ denote the indirect utility generated by an incentive-compatible mechanism. Under the bilinear specification, incentive compatibility implies, almost everywhere, $\q({\boldsymbol\theta})=\nabla_{{\boldsymbol\theta}}\mathfrak{u}({\boldsymbol\theta})$ and implementability is characterized by the convexity of $U$. This representation is fundamental to the Rochet--Chon\'e analysis. In particular, it converts multidimensional incentive compatibility into a convexity restriction and allows us to characterize exclusion, bunching, and perfect screening geometrically. Once we depart from the linearity in private types, this characterization is generally lost. More general multidimensional single-crossing environments generate nontrivial changes in the mechanism-design problem underlying our identification argument.\footnote{For example, \cite{Ghili2023} studies nonadditive valuations
in a discrete bundling problem. He notes that characterizing
optimal bundling with both nonadditive valuations and multidimensional
heterogeneity remains an open problem.}

\subsubsection{Cost Specification}

Assumption \ref{assump:idcost1} rules out cross-dimensional dependence of marginal costs across attributes. We impose this restriction mainly for tractability. In particular, it assumes the cross-partials are zero, i.e., ${\partial^2 C\left[\mathfrak{q}(\mathbf{t})\right]} /{\partial q_j \partial q_{j^\prime}}=0$, $j\neq j^\prime$, from Eq.\ (\ref{eq:intid}). This yields the relatively simple, testable identification conditions in Assumption \ref{assump:idcost2}, along with the convenient representation of the cost parameters in Lemma \ref{le:costfun}. These results follow by substituting the linear and separable cost specification into Eqs.\ (\ref{eq:intid})-(\ref{eq:boundid}). Notice that, on the supply side, this identification argument does not exploit the bunching region; instead, it combines the optimality conditions in the perfect-screening region with the no-distortion-at-the-top condition in Eq.\ (\ref{eq:boundid}).

The restriction in Assumption \ref{assump:idcost1}, however, is not essential for identification. The general characterization of the seller's problem allows for a nonseparable cost function, and the identification argument can therefore be extended to accommodate cross-dimensional dependence in marginal costs. The price of this generality is that we lose the simple closed-form identification conditions of Assumption \ref{assump:idcost2} and the convenient representation in Lemma \ref{le:costfun}. Instead, we can establish identification through a higher-level uniqueness condition.

Specifically, consider the following general specification for the marginal cost function
\begin{eqnarray*} 
\nabla C(\mathbf{q}) = \boldsymbol{\alpha} + \sum_{\mathbf{h}} \boldsymbol{\beta}_{\mathbf{h}}\, \Pi\left(\mathbf{q}^{\mathbf{h}}\right),
\end{eqnarray*}
where $\boldsymbol{\beta}_{\mathbf{h}}\in\mathbb{R}^J$, $\mathbf{q}^{\mathbf{h}}=(q_1^{l_1},\dots,q_J^{l_J})$, and $\Pi\left(\mathbf{q}^{\mathbf{h}}\right) = \prod_{j=1}^J q_j^{l_j}$, while the sum $\sum_{\mathbf{h}}$ is taken over all $\mathbf{h}=(h_1,\dots,h_J)\in\mathbb{N}_0^J$ satisfying $ 1\leq\sum_{j=1}^J h_j\leq D_c $ for some fixed $D_c\in\mathbb{N}$, with the coefficients restricted so that the resulting cost function satisfies Assumption \ref{assump:reg}. (b).

Thus, the marginal cost of attribute $j^\prime$ may depend on the quantities of attributes $j\neq j^\prime$. Hence, the specification accommodates, for example, a setting of a smartphone in which the marginal cost of processing power depends on display quality.

To establish identification conditions under this more general specification, note first that $\frac{\partial^2 C(\mathbf{q})} {\partial q_j \partial q_{j^\prime}} = \sum_{\mathbf{h}} \beta_{\mathbf{h},j^\prime}\, h_j\, q_j^{h_j-1} \prod_{k\neq j}q_k^{h_k},$
where $\beta_{\mathbf{h},j^\prime}$ denotes the $j^\prime$-th component of $\boldsymbol{\beta}_{\mathbf{h}}$. Thus, substituting the general marginal-cost specification into Eq.\ (\ref{eq:intid}) yields
\begin{multline*}
\left[ J+1 - \sum_{j=1}^J \sum_{j^\prime=1}^J \frac{\partial\mathfrak{q}_{j^\prime}(\mathbf{t})} {\partial t_j} \sum_{\mathbf{h}} \beta_{\mathbf{h},j^\prime}\, h_j\, \mathfrak{q}_j(\mathbf{t})^{h_j-1} \prod_{k\neq j} \mathfrak{q}_k(\mathbf{t})^{h_k} \right] f_{\boldsymbol{\theta}}(\mathbf{t}) \\ 
+ \ \left( \mathbf{t} - \boldsymbol{\alpha} - \sum_{\mathbf{h}} \boldsymbol{\beta}_{\mathbf{h}}\, \Pi\left( \mathfrak{q}(\mathbf{t})^{\mathbf{h}} \right) \right) \cdot \nabla f_{\boldsymbol{\theta}}(\mathbf{t}) = 0 \, ,
\qquad \forall\, \mathbf{t}\in\mathscr{S}_{\boldsymbol{\theta},2} \, .
\end{multline*}
Similarly, substituting the general marginal-cost specification into Eq.\ (\ref{eq:boundid}) yields
\begin{equation*}
\left( \mathbf{t} - \boldsymbol{\alpha} - \sum_{\mathbf{h}} \boldsymbol{\beta}_{\mathbf{h}}\, \Pi\left( \mathfrak{q}(\mathbf{t})^{\mathbf{h}} \right) \right) \cdot \vec{\lowmathcal{n}}(\mathbf{t}) = 0, \qquad \forall\, \mathbf{t} \in \mathscr{S}_{\boldsymbol{\theta},2} \cap \partial\mathscr{S}_{\boldsymbol{\theta}}.
\end{equation*}

These two equations provide a system of restrictions that is linear in the cost parameters, $\boldsymbol{\alpha}$ and $\{\boldsymbol{\beta}_{\mathbf{h}}\}_{\mathbf{h}}$. Moreover, all the remaining objects in these equations are identified. In particular, the consumer's first-order conditions in Lemma \ref{lem:foc} identify the allocation rule $\mathfrak{q}$. So, the derivatives $ {\partial\mathfrak{q}_{j^\prime}(\mathbf{t})}/{\partial t_j}$ can then be recovered by applying the implicit function theorem to the consumer's first-order conditions; note that Assumption \ref{assump:reg} is not required for the consumer's first-order conditions in Lemma \ref{lem:foc}.

Hence, identifying the cost parameters under this richer specification reduces to establishing the uniqueness of $ \boldsymbol{\alpha}$ and $(\boldsymbol{\beta}_{\mathbf{h}} )_{\mathbf{h}} $ in the two systems of equations above. Equivalently, because these restrictions are linear in the cost parameters, identification can be stated in terms of an appropriate full-rank condition on the functions multiplying $\boldsymbol{\alpha}$ and $(\boldsymbol{\beta}_{\mathbf{h}} )_{\mathbf{h}} $. This provides a high-level identification condition for a substantially richer class of cost functions. Relative to Assumption \ref{assump:idcost1}, the cost of this additional flexibility is that the conditions in Assumption \ref{assump:idcost2}. (d), and the representation in Lemma \ref{le:costfun}, are replaced by a higher-level uniqueness condition.
Therefore, our baseline specification should be viewed as a tractability restriction rather than as an essential requirement for identification.

\section{Estimation\label{section:estimation}}

In this section, we propose estimators for the joint density $f_{\boldsymbol{\theta}}$ and the marginal cost parameters $(\boldsymbol{\alpha},\boldsymbol{\beta})$ based on a random sample of equilibrium choices and payments $\{( \mathbf{Q}_i, P_i ): i=1,\dots, n \}$, where $\mathbf Q_i = \mathfrak{q} (  \boldsymbol{\theta}_i) $, $P_i  = \mathfrak{p} ( \mathbf Q_i  )$, and $\{ \boldsymbol{\theta}_1, \dots,  \boldsymbol{\theta}_n \}$ is an i.i.d.\ sample from $F_{\boldsymbol{\theta}}$. We assume that the pricing function $\mathfrak{p}$, the support $\mathscr{S}_{\boldsymbol\theta}$, and both regions $\mathscr{S}_{\boldsymbol\theta,2}$ and $\mathscr{S}_{\mathbf{Q},2}$ are known, so here we use them to build our estimators; in Appendix \ref{section:p_hat}, we present a version of our estimator in which these objects are estimated. We provide asymptotic properties of the proposed estimators as $n \rightarrow \infty$.

Lemma \ref{lem:foc} and Eq.\ (\ref{eq:gpvtype}) suggest that the types $\{\boldsymbol{\theta}_i:  \mathbf{Q}_i \in \mathrm{int} ( \mathscr{S}_{\mathbf{Q},2} ) \}$ can be recovered, without estimation errors, using the gradient of the pricing function $ \nabla \mathfrak{p}$:
\begin{eqnarray}
\label{eq:gpvesti}
\boldsymbol{\theta}_i =  \nabla \mathfrak{p} (  \mathbf{Q}_i  ) \  \  \text{for}  \  \mathbf{Q}_i \in \mathrm{int} ( \mathscr{S}_{\mathbf{Q} ,2} ).
\end{eqnarray}
These values represent consumer preferences, and they are of interest. 
While we can use the Kernel-based method to estimate the density of high-type consumers, we cannot automatically extend such a nonparametric estimate to the entire support $\mathscr{S}_{\boldsymbol\theta}$, so we propose a parametric approach.

Let $\Gamma \subset \mathbb{R}^D$, $D \in \mathbb{N}$, be our parameter space, assumed to be convex, compact, and with a nonempty interior. Let $\tilde{\mathcal{S}} \subseteq \mathscr{S}_{\boldsymbol\theta,2}$ be a nonempty and open subset. Now we introduce a family of density functions known up to a parameter $\boldsymbol{\gamma} \in \Gamma$ that satisfy certain regularity conditions for Maximum Likelihood Estimation (MLE).

\begin{definition}		\label{defF}
Let $\mathscr{F}$ be a collection of continuous density functions $f (\cdot ; \boldsymbol{\gamma}) : \mathscr{S}_{\boldsymbol{\theta}}  \rightarrow \mathbb{R}_+$ indexed by $\boldsymbol{\gamma} \in \Gamma$, that satisfy the next conditions: \begin{enumerate}

\item[(i)] $\int f (\cdot ; \boldsymbol{\gamma}) = 1$ for every $\boldsymbol{\gamma} \in \Gamma$.

\item[(ii)]  If $( \boldsymbol{\gamma}  ,  \boldsymbol{\gamma}^\prime ) \in \Gamma^2$ and $
\frac{ f (\mathbf{t} ;  \boldsymbol{\gamma} ) }{  \int_{\tilde{\mathcal{S}}}   f \left( \mathbf{t}^\prime ; \boldsymbol{\gamma}  \right) d \mathbf{t}^\prime    } = \frac{  f (\mathbf{t} ;  \boldsymbol{\gamma}^\prime ) }{ \int_{\tilde{\mathcal{S}}}   f \left( \mathbf{t}^\prime ; \boldsymbol{\gamma}^\prime  \right) d \mathbf{t}^\prime   }	\quad \forall \  \mathbf{t} \in  \tilde{\mathcal{S}}$ then $\boldsymbol{\gamma}  =  \boldsymbol{\gamma}^\prime$.

\item[(iii)] $f (\cdot;  \cdot )$ admits continuous partial derivatives of second order on $\mathscr{S}_{\boldsymbol\theta}   \times  \Gamma $.

\item[(iv)] For each $ \boldsymbol{\gamma} \in \Gamma$,  $  \E \left\{  \left[ \nabla_{\boldsymbol{\gamma}}  \mathfrak{L} ( \boldsymbol{\theta} ; \boldsymbol{\gamma} ) \right]  \left[ \nabla_{\boldsymbol{\gamma}}  \mathfrak{L} ( \boldsymbol{\theta} ; \boldsymbol{\gamma} ) \right] ^\top  \right\}$ is positive definite, where 
\begin{equation*}
\mathfrak{L} ( \mathbf{t} ; \boldsymbol{\gamma} ) =   \left\{  \begin{array}{l l}    \log \left[ f \left( \mathbf{t} ; \boldsymbol{\gamma}  \right)   \right]   - \log  \left[  \int_{\tilde{\mathcal{S}}}   f \left( \mathbf{t}^\prime ; \boldsymbol{\gamma}  \right) d \mathbf{t}^\prime    \right]   & \text{if} \  \mathbf{t} \in  \tilde{\mathcal{S}}  , \\
0 & \text{otherwise}.
\end{array}\right.
\end{equation*}
\end{enumerate}
\end{definition}

The researcher must specify the parametric family $\mathscr{F}$ and the subset $\tilde{\mathcal{S}}$, with the latter independent of $\boldsymbol{\gamma}$. So $\mathscr{F}$ and $\tilde{\mathcal{S}}$ can be regarded as input parameters. A multivariate exponential family on $\mathscr{S}_{\boldsymbol\theta}$ can serve as an example of $\mathscr{F}$. Specifically, given $D$ different vectors $\mathbf{k}_1 , \mathbf{k}_2 , \dots , \mathbf{k}_D \in \mathbb{N}_0^J \backslash \{ (0,\dots,0)\}$ and a constant $c>0$, we can consider the following multivariate exponential family over $ \mathscr{S}_{\boldsymbol\theta} $:
\begin{eqnarray}    \label{eq:expofam}
\mathscr{F}_{\mathrm{exp}} =  \left\{  f(\cdot ; \boldsymbol{\gamma} ) :   \    \boldsymbol{\gamma} \in   [-c , c]^D ,   \  f(\mathbf{t}  ; \boldsymbol{\gamma} )  = \exp\left( \sum_{d=1}^D \gamma_d \mathbf{t}^{ \mathbf{k}_d} \right)  \Bigg{/}  \int_{\mathscr{S}_{\boldsymbol\theta}} \exp\left( \sum_{d=1}^D \gamma_d \mathbf{v}^{ \mathbf{k}_d}\right) d \mathbf{v}   \  \forall \mathbf{t} \in   \mathscr{S}_{\boldsymbol\theta} \right\} .
\end{eqnarray}
Note that, within this family, we have that $ \boldsymbol{\gamma}_0   = (0,\dots , 0) \in \mathbb{R}^D$ in both Examples \ref{example} and \ref{example2}. We can construct other examples of $\mathscr{F}$ from copula families such as Gaussian or Archimedean copulas. In Section \ref{section:mc} below, we provide an additional example of $\mathscr{F}$ that we build upon the family of Beta distributions, as well as an example of $\tilde{\mathcal{S}} $ (see Figure \ref{figmc} below).

To construct valid estimators, $\mathscr{F}$ has to include the true p.d.f.\ $f_{\boldsymbol{\theta}}$, which we assume next.

\begin{assumption}
We have that $ f_{\boldsymbol{\theta}} (\cdot)  = f (\cdot ;  \boldsymbol{\gamma}_0 )  \in \mathscr{F}$ for some $ \boldsymbol{\gamma}_0 \in \mathrm{int} (\Gamma )$.
\label{assump:inte}
\end{assumption}

Then we propose estimating  $\boldsymbol{\gamma}_0 $ by $
\hat{\boldsymbol{\gamma}}   = \arg\max_{\boldsymbol{\gamma} \in \Gamma  }  \   
\sum_{i \in \mathscr{I}_2}  \mathfrak{L} [ \nabla\mathfrak{p}(\mathbf{Q}_i)   ,  \boldsymbol{\gamma} ]$, 
where $\mathscr{I}_2 = \{ i = 1, \dots n : \, \mathbf{Q}_i  \in \mathscr{S}_{\mathbf{Q},2}  \}$. The empirical criterion function, $ \sum_{i \in \mathscr{I}_2} \mathfrak{L}  [ \nabla\mathfrak{p}(\mathbf{Q}_i)   ,  \boldsymbol{\gamma} ]$, can be interpreted as a conditional-on-$(\boldsymbol{\theta} \in \tilde{\mathcal{S}})$ log-likelihood function that uses only the recovered types $\boldsymbol{\theta}_i =  \nabla \mathfrak{p} (  \mathbf{Q}_i  )$ in $\tilde{\mathcal{S}}$ by construction of $\mathfrak{L}$ in Definition \ref{defF}.(iv). This conditioning is important for the regularity of the MLE problem. In particular, although the product distribution's support may depend on $\boldsymbol{\gamma}$, the specified log-likelihood above targets the density of $\boldsymbol{\theta}$ over the fixed set $\tilde{\mathcal{S}}$. The set  $\tilde{\mathcal{S}}$ does not vary with $\boldsymbol{\gamma}$, so the parameter-dependent-support issue that can lead to irregular MLE behavior does not arise.
Thus, following standard arguments on MLE (see the proof of Theorem \ref{thm:asymp} below), we can show that the proposed estimator satisfies the first-order conditions 
\begin{eqnarray}
\sum_{i \in \mathscr{I}_2}   \nabla_{ \boldsymbol{\gamma}}  \mathfrak{L} \left( \boldsymbol{\theta}_i , \hat{\boldsymbol{\gamma}} \right)  = 0  \quad  \text{w.p.a.1}.
\label{eq:mlefoc}
\end{eqnarray}

The asymptotic properties of $\hat{\boldsymbol{\gamma}}$ are provided in the next theorem, for which we denote $
{\Sigma}  =  \E \left\{  \left[ \nabla_{\boldsymbol{\gamma}}  \mathfrak{L} ( \boldsymbol{\theta} ; \boldsymbol{\gamma}_0 ) \right]  \left[ \nabla_{\boldsymbol{\gamma}}  \mathfrak{L} ( \boldsymbol{\theta} ; \boldsymbol{\gamma}_0 ) \right] ^\top  \right\}  \ \text{and}$  $\hat{\Sigma} = \frac{1}{n}  \sum_{i \in \mathscr{I}_2} \left[ \nabla_{ \boldsymbol{\gamma}}  \mathfrak{L}\left( \boldsymbol{\theta}_i , \hat{\boldsymbol{\gamma}} \right)  \right]  \left[ \nabla_{ \boldsymbol{\gamma}}  \mathfrak{L} \left( \boldsymbol{\theta}_i , \hat{\boldsymbol{\gamma}} \right)  \right] ^\top.$

\begin{theorem}	\label{thm:asymp}
Suppose that Assumptions \ref{assump:reg}, \ref{assump:idcost2}, and \ref{assump:inte} hold. Then, $\hat{\boldsymbol{\gamma}}$ is consistent and asymptotically normal, i.e., $\hat{\boldsymbol{\gamma}} \overset{p}{\rightarrow} \boldsymbol{\gamma}_0 $ and $
\sqrt{n} \left( \hat{\boldsymbol{\gamma}} -  \boldsymbol{\gamma}_0 \right) \overset{d}{\rightarrow}   N({\boldsymbol 0}, {\Sigma}^{-1})$. Moreover, $ \hat{\Sigma} \overset{p}{\rightarrow}  {\Sigma} $ and therefore $\hat{\Sigma}^{-1}$ is a consistent estimator of the asymptotic variance.
\end{theorem}

The proof is provided in Appendix \ref{section:proofs}. From this theorem, we can consistently estimate the density of types over the whole support by setting $\hat{f} (\mathbf{t}):= \tilde{f} (\mathbf{t}; \hat{\boldsymbol{\gamma}} )$, $\mathbf{t} \in \mathscr{S}_{\boldsymbol\theta}$.  In contrast to Lemma \ref{le:costfun}, this theorem does not require a parametric assumption on the cost function. Note also that, even though this theorem establishes the validity of the plug-in method for estimating $\Sigma$, the smooth nature of $\hat{\boldsymbol{\gamma}}$, given by Eq.\ (\ref{eq:mlefoc}), suggests that the variance-covariance matrix of $\hat{\boldsymbol{\gamma}}$ and the variance of $\hat{f} (\mathbf{t})$ can be estimated by the Jackknife method, which we use in Section \ref{section:mc} below.

Next, we construct an estimator of the marginal cost parameters $(\boldsymbol{\alpha}, \boldsymbol{\beta})$ or, equivalently, the coefficients $(\alpha_j, \beta_j)$, $j=1,\ldots, J$, based on the identification result of Lemma \ref{le:costfun} that suggests the use of the plug-in method. 

The proposed estimators $(\hat{\boldsymbol{\alpha}} ,\hat{\boldsymbol{\beta}})$ of the parameters $(\boldsymbol{\alpha}, \boldsymbol{\beta})$ can be computed in four steps as follows. First, choose $J$ vectors $\mathbf{t}_j^\prime$, $j = 1,\dots, J$, satisfying Assumption \ref{assump:idcost2}-(c) and compute the outward normal vectors $\vec{\lowmathcal{n}} (\mathbf{t}_j^\prime  )$ and $\mathcal{N} = ( \vec{\lowmathcal{n}} (\mathbf{t}_1^\prime  ) \  \cdots \   \vec{\lowmathcal{n}} (\mathbf{t}_J^\prime  )  )^\top$: note that this computation is feasible because in this section we assume that $\mathscr{S}_{\boldsymbol\theta}$ and $\mathscr{S}_{\boldsymbol\theta,2}$ are known. Second, compute $\mathcal{Q} =   (  \mathfrak{q}( \mathbf{t}_1^\prime )  \   \cdots \  \mathfrak{q}( \mathbf{t}_J^\prime )  )$ with $ \mathfrak{q}( \mathbf{t}_j^\prime ) =  ( \nabla {\mathfrak{p}}  )^{-1} ( \mathbf{t}_j^\prime  )$. Third, pick a sufficiently small constant $c >0$ and choose the vectors $\mathbf{t}^{\prime\prime}_j$ from Assumption \ref{assump:idcost2}-(d), $j=1,\dots, J$, so that $| \mathrm{det} (  \hat{\mathcal{D}}_{\beta}  )  |   > c$, where  
\begin{eqnarray*}
\hat{\mathcal{D}}_{\beta} &=&  \hat{\mathcal{D}}_{fq}  -  \hat{\mathcal{D}}_f^\top \mathcal{N}^{-1} (\mathcal{N} \odot \mathcal{Q}^\top ); \qquad \hat{\mathcal{D}}_f       =    \left(  \nabla \hat{f} ( {\mathbf{t}}_1^{\prime\prime} )  \   \cdots \  \nabla \hat{f} ( {\mathbf{t}}_J^{\prime\prime} ) \right); \quad \text{and}\\
\hat{\mathcal{D}}_{fq}    & = &      \left( \begin{array}{ccc} 
\nabla_1 [ \hat{f} ( {\mathbf{t}}_1^{\prime\prime} ) {\mathfrak{q}}_1 ( {\mathbf{t}}_1^{\prime\prime} ) ]  &  \cdots  &  \nabla_J [ \hat{f} ( {\mathbf{t}}_1^{\prime\prime} ) {\mathfrak{q}}_J ( {\mathbf{t}}_1^{\prime\prime} ) ]   \\
\vdots &   \ddots   & \vdots \\
\nabla_1 [ \hat{f} ( {\mathbf{t}}_J^{\prime\prime} ) {\mathfrak{q}}_1 ( {\mathbf{t}}_J^{\prime\prime} ) ]  &  \cdots  &  \nabla_J [ \hat{f} ( {\mathbf{t}}_J^{\prime\prime} ) {\mathfrak{q}}_J ( {\mathbf{t}}_J^{\prime\prime} ) ] 
\end{array} \right).
\end{eqnarray*}

Fourth, estimate $\boldsymbol{\beta}$ and $\boldsymbol{\alpha}$ by the plug-in method:
\begin{equation*}
\hat{\boldsymbol{\beta}}  =     \hat{\mathcal{D}}_{\beta}^{-1}  \left[ (J  + 1)  \left(\begin{array}{c}
\hat{f} ( {\mathbf{t}}_1^{\prime\prime} ) \\
 \vdots \\
 \hat{f}( {\mathbf{t}}_J^{\prime\prime} )
\end{array}  \right)    +    \left(\begin{array}{c}
({\mathbf{t}}_1^{\prime\prime} -  {\mathcal{N}}^{-1} {\mathbf{T} } ) \cdot \nabla \hat{f} ( {\mathbf{t}}_1^{\prime\prime} ) \\
 \vdots \\
( {\mathbf{t}}_J^{\prime\prime}  -  {\mathcal{N}}^{-1} {\mathbf{T}}  ) \cdot \nabla  \hat{f} ( {\mathbf{t}}_J^{\prime\prime} )
\end{array}  \right)     \right]    \ \text{and} \    \hat{\boldsymbol{\alpha}}     =         \mathcal{N}^{-1} \left[ \mathbf{T}    - \left( \mathcal{N}  \odot \mathcal{Q}^\top \right) \hat{\boldsymbol{\beta}} \right] ,
\end{equation*}
respectively. The next corollary establishes the consistency of these estimators, following Lemma \ref{le:costfun} and Theorem \ref{thm:asymp}.

\begin{cor}	\label{corol:cost}
If Assumptions \ref{assump:reg}-\ref{assump:inte} hold, then $\hat{\boldsymbol{\beta}} \overset{p}{\rightarrow}  \boldsymbol{\beta}$ and $\hat{\boldsymbol{\alpha}}  \overset{p}{\rightarrow} \boldsymbol{\alpha}$.
\end{cor}

To complete this section, we highlight that since the estimators $( \hat{\boldsymbol{\alpha}}, \hat{\boldsymbol{\beta}} )$ can be regarded as a continuously differentiable function of $\hat{\boldsymbol{\gamma}}$, the variance-covariance matrix of $( \boldsymbol{\alpha}, \hat{\boldsymbol{\beta}} )$ can be estimated by the Jackknife method. Previous results also suggest that critical values from a standard normal distribution can be used to build valid confidence intervals, as $\sqrt{n}\Sigma^{-1}(\hat{\boldsymbol{\gamma}} - \boldsymbol{\gamma}_0)$ converges in distribution to a standard normal distribution.

\subsection{Monte Carlo Experiments} \label{section:mc}

This section presents Monte Carlo experiments to evaluate the proposed estimators' finite-sample performance. We use the proposed estimator's bias and standard deviation, as well as the coverage probabilities of the confidence intervals, as our evaluation criteria.

The complexity of optimal multidimensional screening limits the set of
tractable data-generating processes. 
We use Example \ref{example2},
which provides a closed-form solution for the optimal allocation
$\mathfrak q(\cdot)$ in Eq.~(\ref{eq:q_wilson}), and return at the end of this
section to discuss this choice.

\begin{figure}[tt]

\caption{Design Points and Input Parameters for Monte Carlo Experiment} 

{\centering

\includegraphics[scale=0.35]{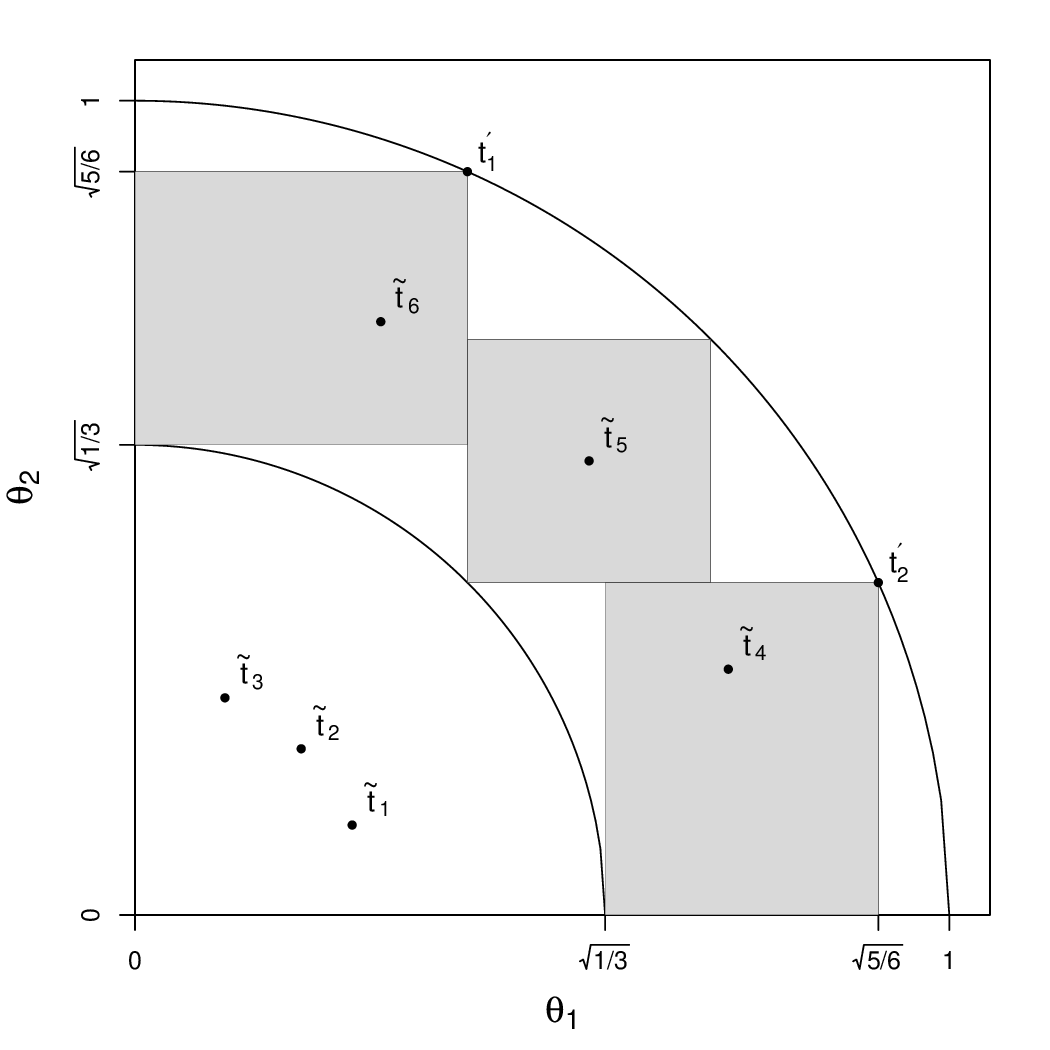}

}

\begin{figurenotes}
Subset $\tilde{\mathcal{S}}$ marked in gray. We set $\mathbf{t}_1^{\prime\prime} = \tilde{\mathbf{t}}_4$, and $\mathbf{t}_2^{\prime\prime} = \tilde{\mathbf{t}}_6$.

\end{figurenotes}
\label{figmc}

\end{figure}

So, the values of the true parameters are given by the uniform distribution on the positive quarter disk, i.e., $f_{\boldsymbol\theta} (  \mathbf{t}) = 4/\pi$ for all $ \|  \mathbf{t} \|_2 \leq 1$, $\boldsymbol{\alpha} = (0, 0)$, and $\boldsymbol{\beta} = (1, 1)$. We treat the pricing function $\mathfrak{p}(\cdot)$, $\mathscr{S}_{\boldsymbol\theta}   =  \{ \mathbf{t} \in \mathbb{R}^2_{+}:  \|  \mathbf{t} \|_2 \leq 1 \}$, and $\mathscr{S}_{\boldsymbol\theta,2} = \{  \mathbf{t} \in  \mathscr{S}_{\boldsymbol\theta}:  \|  \mathbf{t} \|_2^2 > 1/3 \}$ as known so that their estimation errors do not affect the finite-sample performance of our proposed estimators. We set $\tilde{\mathcal{S}} \subset \mathscr{S}_{\boldsymbol\theta,2} $ to be a union of three disjoint rectangles; specifically, 
\begin{eqnarray*}
\tilde{\mathcal{S}} =  \left( 0 , \sqrt{\frac{1}{6}}  \right) \times  \left( \sqrt{\frac{1}{3}}  , \sqrt{\frac{5}{6}} \right)  \bigcupdot     \left(\sqrt{\frac{1}{6}} , \sqrt{\frac{1}{2}}   \right) \times \left(\sqrt{\frac{1}{6}} , \sqrt{\frac{1}{2}}   \right)  \bigcupdot   \left( \sqrt{\frac{1}{3}}  , \sqrt{\frac{5}{6}} \right)  \times \left( 0 , \sqrt{\frac{1}{6}}  \right) . 
\end{eqnarray*}
The set $\tilde{\mathcal{S}}$ is given by the gray area in Figure \ref{figmc}. We generate choices from $n=500$ consumers. 

We employ 1,000 replications and follow the next steps in each replication:
\begin{enumerate}
\item Generate $n$ values of $\boldsymbol{\theta}_i$ from a uniform distribution on the positive quarter disk and compute their corresponding equilibrium quantities $\mathbf{Q}_i$ from the allocation functions. 

\item Compute $\hat{\boldsymbol{\gamma}}$ considering the following two families of distributions in which $D=2$: the exponential family $\mathscr{F}_{\mathrm{exp}}$ from Eq.\ (\ref{eq:expofam}) with $\mathbf{k}_1 = (1,0)$ and $\mathbf{k}_2 = (0,1)$; the truncated Beta family 
$\mathscr{F}_{\mathrm{Beta}}  =  \left\{  f(\cdot ; \boldsymbol{\gamma} ) :   \    \boldsymbol{\gamma} \in  \Gamma_{\mathrm{Beta}} ,   \  f(\mathbf{t}  ; \boldsymbol{\gamma} )  = \frac{ f_{\mathrm{Beta}} (t_1  ; \boldsymbol{\gamma}) f_{\mathrm{Beta}} (t_2  ; \boldsymbol{\gamma}) }{  \int_{\mathscr{S}_{\boldsymbol\theta}}  f_{\mathrm{Beta}} (v_1  ; \boldsymbol{\gamma}) f_{\mathrm{Beta}} (v_2  ; \boldsymbol{\gamma})  d \mathbf{v}}   \  \forall \mathbf{t} \in   \mathscr{S}_{\boldsymbol\theta} \right\},$ 
where $\Gamma_{\mathrm{Beta}} \subset \mathbb{R}_{++}^2$ is a compact subset that includes $ (1,1)$, the value of the true parameter within the Beta family, $f_{\mathrm{Beta}} (t; \boldsymbol{\gamma}) = t^{\gamma_1-1}(1-t)^{\gamma_2-1}/\mathrm{B}( \gamma_1, \gamma_1) \mathds{1} ( t \in [0,1])$ is the Beta density with parameters $\gamma_1>0$ and $\gamma_2 > 0$, and $\mathrm{B}( \gamma_1, \gamma_2)$ is the Beta function. 

\item Compute the standard errors of $\hat\gamma_1$ and $\hat\gamma_2$ using the plug-in and Jackknife methods.

Then, compute 95\% Confidence Intervals (CIs) for each coefficient of $\boldsymbol{\gamma}_0 = ( \gamma_{0,1}, \gamma_{0,2})$, using the quantile from the standard normal as the critical value ($\approx 1.96$) together with the previously computed standard errors.

\item Compute $\hat{f} ( \tilde{\mathbf{t}}_m  )$ for $m=1,\dots,6$, where $\tilde{\mathbf{t}}_1 , \dots, \tilde{\mathbf{t}}_6$ are design points that can be visualized in Figure \ref{figmc} and, more specifically, are defined as follows: $\tilde{\mathbf{t}}_1 = \tilde\varphi  ( 1/( 2\sqrt{3} ) ,  1/4 )$, $\tilde{\mathbf{t}}_2 = \tilde\varphi  ( 1/( 2\sqrt{3} ) ,  1/2 )$, $\tilde{\mathbf{t}}_3 = \tilde\varphi  ( 1/( 2\sqrt{3} ) ,  3/4 )$, $\tilde{\mathbf{t}}_4 = \tilde\varphi  ( 1/2 + 1/( 2\sqrt{3} )  ,  1/4 )$, $\tilde{\mathbf{t}}_5 = \tilde\varphi  ( 1/2 + 1/( 2\sqrt{3} )  ,  1/2 )$, and $\tilde{\mathbf{t}}_6 = \tilde\varphi  ( 1/2 + 1/( 2\sqrt{3} )  ,  3/4 )$, where $\tilde\varphi ( r, a) = ( r \cos(a \pi/2) ,  r \sin(a \pi/2) )$ is the transformation from polar coordinates in the positive quarter disk to Cartesian coordinates.
Then compute 95\% CIs for each $f_{\boldsymbol\theta} (\tilde{\mathbf{t}}_m)$, $m=1,\dots,6$, using the standard normal quantile as the critical value and Jackknife standard errors.

\item Compute $( \hat{\boldsymbol{\alpha}}, \hat{\boldsymbol{\beta}})$ setting $\mathbf{t}_1^\prime = (\sqrt{1/6} , \sqrt{5/6}   )$, $\mathbf{t}_2^\prime = ( \sqrt{5/6}    ,  \sqrt{1/6} )$, $\mathbf{t}_1^{\prime\prime} = \tilde{\mathbf{t}}_4$, and $\mathbf{t}_2^{\prime\prime} = \tilde{\mathbf{t}}_6$. Figure \ref{figmc} shows these points.
Then, compute 95\% CIs for each coefficient of $({\boldsymbol{\alpha}},{\boldsymbol{\beta}}) = (\alpha_1, \alpha_2, \beta_1, \beta_2)$ using the quantile from the standard normal as the critical value and Jackknife standard errors.

\end{enumerate}

\begin{table}[t!] 	
\caption{Performance of $\hat{\boldsymbol\gamma}$ and $\hat{f} ( \cdot)$ when $n=500$: bias, standard deviation, and coverage probability. \label{table:MC_density}}
\begin{tabular}{lllllllll}
\toprule
 & \multicolumn{2}{c !{}}{$\hat{\boldsymbol\gamma}$}        & \multicolumn{6}{c !{}}{$\hat{f} (\cdot) $}                                 \\ 
  {\bf Family}      &        $\hat{\gamma}_1$ & $\hat{\gamma}_2$       & $\hat{f} (  \tilde{\mathbf{t}}_1 )$   &  $\hat{f} (  \tilde{\mathbf{t}}_2 )$ & $\hat{f} (  \tilde{ \mathbf{t} }_3 )$   &  $\hat{f} (\tilde{ \mathbf{t} }_4 )$    &  $\hat{f} ( \tilde{\mathbf{t}}_5 )  $ &  $\hat{f} (  \tilde{ \mathbf{t} }_6 )$  \\
       \midrule
\multirow{3}{*}{\bf Exponential}  &	0.028	&	0.036	&	-0.009	&	-0.010	&	-0.007	&	-0.001	&	0.006	&	0.003	\\
&  (0.422)	&	(0.424)	&	(0.239)	&	(0.222)	&	(0.239)	&	(0.115)	&	(0.134)	&	(0.117)	\\
&	[0.947]	&	[0.946]	&	[0.934]	&	[0.932]	&	[0.934]	&	[0.939]	&	[0.933]	&	[0.953]	\\

\hline

\multirow{3}{*}{\bf Beta} &	0.010	&	0.005	&	-0.005	&	-0.002	&	-0.005	&	-0.003	&	0.004	&	-0.003	\\
&	(0.072)	&	(0.168)	&	(0.204)	&	(0.197)	&	(0.204)	&	(0.076)	&	(0.104)	&	(0.076)	\\
&	[0.947]	&	[0.953]	&	[0.942]	&	[0.947]	&	[0.942]	&	[0.948]	&	[0.950]	&	[0.948]	\\
\bottomrule
\end{tabular}
\begin{figurenotes}
This table presents the estimated bias and standard deviation (in parentheses) of $\hat{\boldsymbol\gamma}$ and $\hat{f}(\tilde{\mathbf{t}}_m)$, $m=1,\dots,6$, as well as coverage probabilities of 95\% CIs (in square brackets), obtained from a Monte Carlo experiment with 1,000 replications. True value of the parameters: $\boldsymbol{\gamma}_0 = (0,0)$ under exponential family, $\boldsymbol{\gamma}_0 = (1,1)$ under Beta family, and $\hat{f}(\tilde{\mathbf{t}}_m) = 4/\pi$ for all $m$.
\end{figurenotes}
\end{table}

Tables \ref{table:MC_density} and \ref{table:MC_cost} report the simulation results. Specifically, Table \ref{table:MC_density} reports the estimated bias and standard deviation (in parentheses) of the estimators $\hat{\boldsymbol{\gamma}}$ and $\hat{f}(\tilde{\mathbf{t}}_m)$, $m=1,\dots,6$. As shown, the bias and standard deviation are small. We also note that $\hat{f}(\cdot )$ performs better at the design points, $ \tilde{\mathbf{t}}_4$, $\tilde{\mathbf{t}}_5$, and $\tilde{\mathbf{t}}_6$, that belong to $\mathscr{S}_{\boldsymbol{\theta},2}$. Table \ref{table:MC_density} also provides estimated coverage probabilities (in square brackets) of the 95\% CIs using Jackknife standard errors.\footnote{When asymptotic-based standard errors are used instead, CIs for $\gamma_{0,1}$ and $\gamma_{0,2}$ exhibit coverage probabilities of 0.947 and 0.943 (0.958 and 0.948), respectively, under the exponential (Beta) family.} As noted, all these estimates are very close to the targeted 0.95.

Table \ref{table:MC_cost} presents the estimated bias and standard deviation (in parentheses) of the cost-parameter estimators, along with estimates of the confidence-interval coverage probabilities. The bias and standard deviation are also small across all cases, and the coverage probabilities closely approximate 0.95.

\begin{table}[ht!] 	
\begin{center}
\caption{Performance of $\hat{\boldsymbol{\alpha}}$ and $\hat{\boldsymbol{\beta}}$ when $n=500$: bias, standard deviation, and coverage probability. \label{table:MC_cost}}
\begin{tabular}{llllll}
\toprule
                &       \multicolumn{2}{l}{$\qquad\hat{\boldsymbol{\alpha}}$} & \multicolumn{2}{l}{$\qquad\hat{\boldsymbol{\beta}}$} \\
{\bf Family}                & $\hat{\alpha}_1$ & $\hat{\alpha}_2$ & $\hat{\beta}_1$ & $\hat{\beta}_2$ \\
\midrule
\multirow{3}{*}{\bf Exponential} &	0.011	&	-0.029	&	-0.003	&	0.028	\\
&	(0.128)	&	(0.195)	&	(0.080)	&	(0.168)	\\
&	[0.952]	&	[0.961]	&	[0.945]	&	[0.963]	\\
\hline
\multirow{3}{*}{\bf Beta} &	0.014	&	-0.014	&	-0.011	&	0.011	\\
&	(0.058)	&	(0.110)	&	(0.035)	&	(0.100)	\\
&	[0.953]	&	[0.926]	&	[0.932]	&	[0.930]	\\
\bottomrule
\end{tabular}
\end{center}
\begin{figurenotes}
This table presents the estimated bias, standard deviation (parentheses) of $\hat{\boldsymbol\alpha}$ and $\hat{\boldsymbol\beta}$, and coverage probabilities of 95\% CI (square brackets) from a Monte Carlo experiment with 1,000 replications and true parameter values $\boldsymbol{\alpha} = (0, 0)$ and $\boldsymbol{\beta} = (1, 1)$.
\end{figurenotes}

\end{table}

\paragraph{Note.} It is worth emphasizing that our preference and cost estimators use observations from the perfectly screened region. Bunching matters for distinguishing regions of the type space where the allocation is many-to-one from those where it is one-to-one. Once we identify the perfectly screened region, our estimation procedure does not use the structure of the bunching region. We therefore use Example \ref{example2}, which has a closed-form solution that makes it easy to simulate the data. We did not use the canonical {\RC} Example \ref{example} with bunching because
\cite{McCannZhang2023b} show that its proposed bunching solution is not self-consistent, and using a numerical solution in  \citep{EkelandMoreno2010} as the DGP would therefore confound the finite-sample performance of our estimators with approximation error from solving the multidimensional screening problem.

\section{Empirical Application\label{sec:application}}
In this section, we illustrate our method using synthetic data generated from the estimates in \citet{clx19}. They study the demand for wireless services under nonlinear pricing in a Chinese city, where consumers choose continuous quantities of voice ($q_1$, measured in minutes) and 4G data ($q_2$, measured in megabytes). We use their estimated model to generate two-dimensional usage choices and the corresponding total payments, and then apply our identification and estimation procedure to these observations.

In \citet{clx19}, consumers first choose a service plan and then, conditional on that plan, choose their voice and data usage. Their empirical application considers two plans. Plan 1 has a monthly access fee of \rmb58 and includes 500 MB of data and 50 voice minutes, while Plan 2 has an access fee of \rmb88 and includes 700 MB of data and 200 voice minutes. Both plans charge
\rmb0.29 per MB for additional data and \rmb0.19 per minute for additional voice usage. Formally, plan $k\in\{1,2\}$ is characterized by $(T_k,A_k,\tau)$, where $T_k >0$ is the access fee, $A_k=(A_{k1}, A_{k2})$ is the vector of usage allowances, and $\tau=(\tau_1,\tau_2)$ is the vector of marginal prices for usage above the allowances, with the first component of each vector corresponding to voice and the second to data.

Our analysis does not concern the discrete plan choice itself. Instead, for each consumer $i$, we use the plan $k_i$ chosen by the consumer together with the consumer's final usage $\mathbf{q}_i=(q_{i1},q_{i2})$ to determine the total payment $P_i=T_{k_i}+\sum_{j=1}^2\max\{0,q_{ij}-A_{k_i j}\}\tau_j$, where $q_{i1}$ denotes consumer $i$'s voice usage and $q_{i2}$ denotes data usage.
After generating the choices and payments $(\mathbf{q}_i, P_i)$, we apply our method. 

The two plans imply two piecewise-linear pricing functions,
$\mathfrak{p}_1(q_1,q_2)$ and $\mathfrak{p}_2(q_1,q_2)$. 
To map it to our model of a single pricing function assigning a payment to each allocation $(q_1,q_2)$, we
combine the two plan-specific pricing functions by taking their lower envelope,
$\mathfrak{p}^{\,L}(q_1,q_2)=\min\{\mathfrak{p}_1(q_1,q_2),\mathfrak{p}_2(q_1,q_2)\}$, which assigns a single payment to each pair $(q_1,q_2)$.
To generate sufficiently smooth pricing function, we approximate $\mathfrak{p}^{\,L}$ by a degree-5 bivariate Bernstein polynomial on $[0,q_1^M]\times[0,q_2^M]$, where $q_1^M$ and $q_2^M$ are the corresponding choices. We denote this smooth approximation by $\mathfrak{p}(q_1,q_2)$ and use it as the pricing function in our analysis.

Our analysis concerns the continuous usage choices and the associated preference heterogeneity. In \citet{clx19}, the preference for service $l$ is $\tilde{\theta}_{ilt}=\tilde{\theta}_{il}+\tilde{\sigma}_{ilt}$, where $\tilde{\theta}_{il}$ is time invariant and $\tilde{\sigma}_{ilt}$ is a time-varying preference shock. We interpret the resulting final preference coefficients $(\tilde{\theta}_{i1t},\tilde{\theta}_{i2t})$ as the two-dimensional type $\boldsymbol{\theta}_i=(\theta_{i1},\theta_{i2})$ in our model and allow them to be correlated.

\begin{figure}[t!!]
\caption{Pooled Quantities and Types in Synthetic Datasets.}
{\centering
\includegraphics[scale=.65]{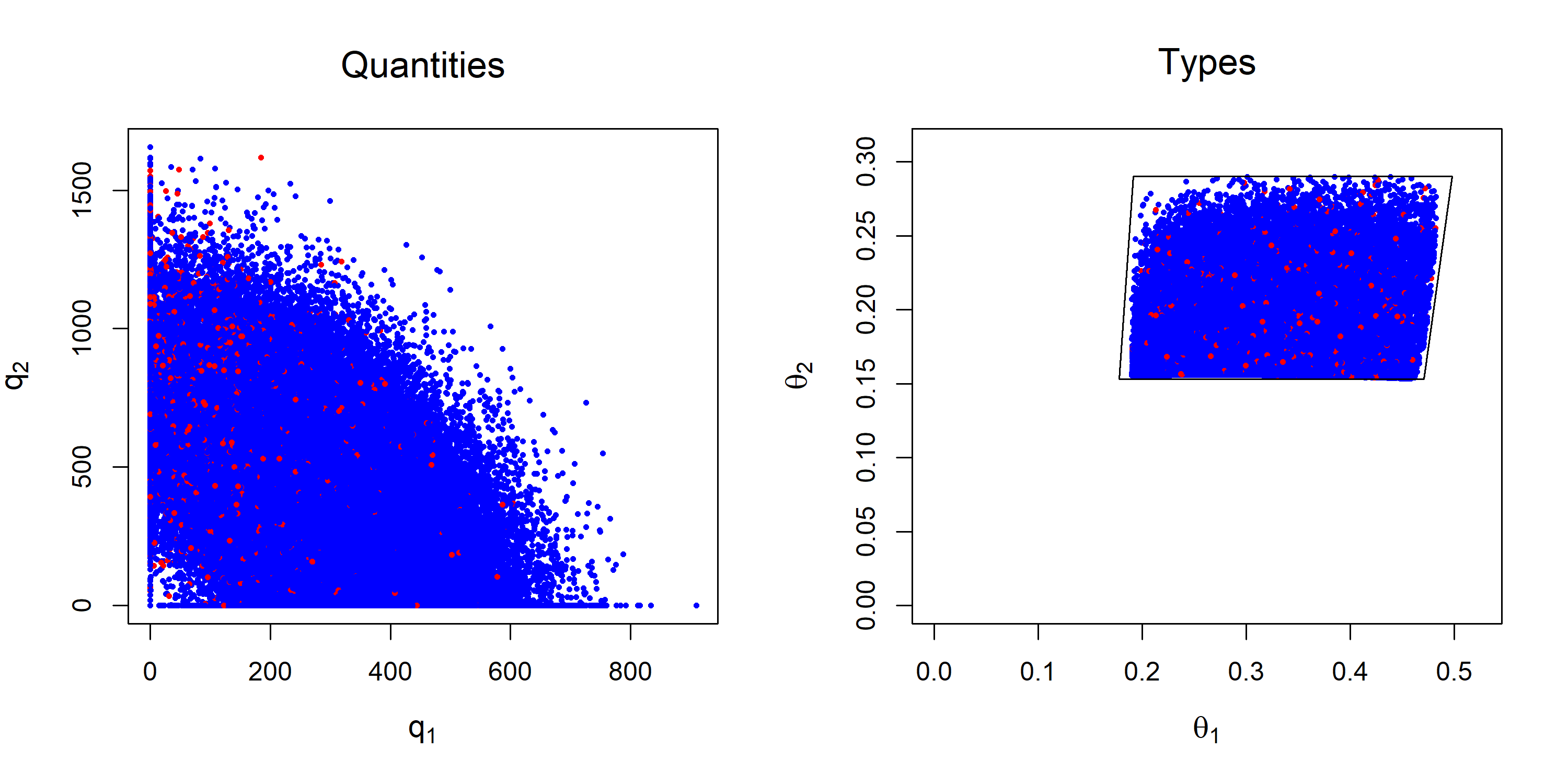}
}
\begin{figurenotes}
Quantity $q_1$ corresponds to voice and $q_2$ to data. Choices for Plan 1 are shown in 
red and for Plan 2 in blue.
\end{figurenotes}
\label{figexasyn}
\end{figure}

Using the pricing functions and parameter estimates reported in
\citet[Tables III and VI]{clx19}, we simulate 200 datasets, each containing
$n=500$ choices of voice and data usage. The left panel of Figure
\ref{figexasyn} shows the pooled quantity choices. For completeness, we use
different colors to indicate whether an observation originated from Plan 1 or
Plan 2. This distinction is only descriptive and in our subsequent analysis, the
two plan-specific pricing functions are combined through their lower envelope,
whose smooth approximation provides the single pricing function
$\mathfrak{p}(\cdot)$ used in our model.

For each dataset, we recover the types associated with voice and data preferences
from Eq.\ (\ref{eq:lemfoc}), using the smooth pricing function
$\mathfrak{p}(\cdot)$ constructed above and restricting attention to choices with
$q_{i1},q_{i2}>0$. The right panel of Figure \ref{figexasyn} shows the resulting
types.

Based on the results shown in the right panel of Figure \ref{figexasyn}, the support 
$\mathscr{S}_{\boldsymbol\theta}$ of $\boldsymbol{\theta} = (\theta_1, \theta_2)$ is 
specified as a trapezoid, which satisfies Assumption 
\ref{assump:reg}. The parallel lower and upper edges lie at $\theta_2 \approx 0.153$ and 
$\theta_2 \approx 0.290$, respectively, while the left and right edges are defined by the 
linear functions $\theta_2 = -1.625 + 10\theta_1$ and $\theta_2 = -2.2 + 5\theta_1$. 

To estimate the joint density of types over this support, we set $D=4$ and consider the 
next two families of distributions $\mathscr{F}$, which are similar to those employed in 
Section \ref{section:mc}. Let $\psi_{T}$ be the map that transforms the unit square 
$[0,1]^2$ into the aforementioned trapezoid, i.e.\ the support $\mathscr{S}_{\boldsymbol
\theta}$ of $\boldsymbol\theta$, with the first unit-square coordinate determining the vertical position in the trapezoid and the second determining the horizontal position conditional on the first. Then, we define
\begin{eqnarray}
\mathscr{F}_{\mathrm{Exp2}} & = & \left\{ f(\cdot ; \boldsymbol{\gamma} ) : \ 
    \boldsymbol{\gamma} \in \Gamma_{\mathrm{Exp2}} , \ 
    f( \mathbf{t} ; \boldsymbol{\gamma} ) = f_{\mathrm{Exp2}} \left( \psi_{T}^{-1} ( 
    \mathbf{t} ) ; \boldsymbol{\gamma} \right) \left| \mathrm{det} \left( \nabla 
    \psi_{T}^{-1} ( \mathbf{t} ) \right) \right| \ \ \forall \mathbf{t} \in 
    \mathscr{S}_{\boldsymbol\theta} \right\},\label{eq:exp2_synthetic}\qquad \\
\mathscr{F}_{\mathrm{Be2}} & = & \left\{ f(\cdot ; \boldsymbol{\gamma} ) : \ 
    \boldsymbol{\gamma} \in \Gamma_{\mathrm{Be2}} , \ f(\mathbf{t} ; \boldsymbol{\gamma} 
    ) = f_{\mathrm{Be2}} \left( \psi_{T}^{-1} ( \mathbf{t} ) ; \boldsymbol{\gamma} 
    \right) \left| \mathrm{det} \left( \nabla \psi_{T}^{-1} ( \mathbf{t} ) \right) 
    \right| \ \ \forall \mathbf{t} \in \mathscr{S}_{\boldsymbol\theta} \right\},\label{eq:be2_synthetic}\qquad
\end{eqnarray}
where
\begin{equation*}
f_{\mathrm{Exp2}} ( \mathbf{x} ; \boldsymbol{\gamma} ) =
\left\{\begin{array}{ll}
\dfrac{\exp\left[ \gamma_1 x_1 + \gamma_2 x_1^2 + \gamma_3 x_2 + \gamma_4 x_2^2 
    \right]}{ \int_0^1 \int_0^1 \exp \left( \gamma_1 v_1 + \gamma_2 v_1^2 + \gamma_3 
    v_2 + \gamma_4 v_2^2 \right) dv_1 dv_2} & \text{if} \ \mathbf{x} = (x_1 , x_2 ) 
    \in [0,1]^2 , \\[6pt]
0 & \text{otherwise},
\end{array} \right.
\end{equation*}
and $f_{\mathrm{Be2}} ( \mathbf{x} ; \boldsymbol{\gamma} ) = f_{\mathrm{Beta}} \left( 
x_1 ; ( \gamma_1 , \gamma_2 ) \right) f_{\mathrm{Beta}} \left( x_2 ; ( \gamma_3 , 
\gamma_4 ) \right)$, while $\Gamma_{\mathrm{Exp2}}$ and $\Gamma_{\mathrm{Be2}}$ are 
compact subsets of $\mathbb{R}^4$ and $\mathbb{R}_{++}^4$, respectively. Under either family, the trapezoid transformation $\psi_T$ induces dependence between
$\theta_1$ and $\theta_2$, consistent with Figure \ref{figexasyn}.

For each synthetic dataset, we estimate the density parameters $\boldsymbol{\gamma}_0$ 
via MLE and the cost parameters $(\boldsymbol{\alpha}, \boldsymbol{\beta})$ following the 
procedure described in the previous section. Specifically, we set $\mathbf{t}_1^\prime = 
\psi_{T}(1/4, 1)$ and $\mathbf{t}_2^\prime = \psi_{T}(1, 1/4)$, with their corresponding 
outward normal vectors given by $\vec{\lowmathcal{n}}(\mathbf{t}_1^\prime) = (0, 1)$ and 
$\vec{\lowmathcal{n}}(\mathbf{t}_2^\prime) = (5, -1)/\sqrt{1 + 5^2}$. The interior 
points used for estimation are chosen as $\mathbf{t}_1^{\prime\prime} = \psi_{T}(1/4, 
3/4)$ and $\mathbf{t}_2^{\prime\prime} = \psi_{T}(3/4, 1/4)$.

Table \ref{table:eapp} reports the mean estimates across the 200 datasets,
together with the 0.025 and 0.975 quantiles of their empirical distributions, shown in
square brackets. The width of these intervals provides a measure of the sampling
variation of the estimates across datasets.
Narrow confidence intervals for the $\mathrm{Be2}$ family, reflect high precision and consistency across replications, 
whereas wider intervals, as seen for the $\mathrm{Exp2}$ parameters, suggest greater 
sampling variability.

\begin{table}[tt]
\begin{center}
\caption{Estimates and Confidence Intervals. \label{table:eapp}}
\begin{tabular}{lcccc}
\toprule
{\bf Density} & \multicolumn{4}{c}{$\qquad\hat{\boldsymbol{\gamma}}$} \\
{\bf parameters} & $\hat{\gamma}_1$ & $\hat{\gamma}_2$ & $\hat{\gamma}_3$ & 
    $\hat{\gamma}_4$ \\
\midrule
\multirow{2}{*}{\bf Exp2}
    & 2.84       & $-6.85$    & $-6.01$    & 1.49       \\
    & $[1.56,\ 4.20]$ & $[-8.87,\ -5.04]$ & $[-7.46,\ -4.61]$ & $[-0.20,\ 3.08]$ \\[8pt]
\multirow{2}{*}{\bf Be2}
    & 1.55       & 3.29       & 0.83       & 3.29       \\
    & $[1.39,\ 1.73]$ & $[2.74,\ 3.94]$ & $[0.76,\ 0.91]$ & $[2.81,\ 3.87]$ \\[8pt]
\toprule
{\bf Cost} & \multicolumn{2}{c}{$\qquad\hat{\boldsymbol{\alpha}}$} & 
    \multicolumn{2}{c}{$\qquad\hat{\boldsymbol{\beta}}$} \\
{\bf parameters} & $\hat{\alpha}_1$ & $\hat{\alpha}_2$ & $\hat{\beta}_1$ & 
    $\hat{\beta}_2$ \\
\midrule
\multirow{2}{*}{\bf Exp2}
    & 0.427      & $-0.006$   & 0.0005     & 0.0001     \\
    & $[0.321,\ 0.651]$ & $[-0.195,\ 0.083]$ & $[0.0003,\ 0.0006]$ & 
    $[0.0001,\ 0.0002]$ \\[8pt]

\multirow{2}{*}{\bf Be2}
    & 0.262      & 0.159      & 0.0006     & 0.0000     \\
    & $[0.214,\ 0.298]$ & $[0.137,\ 0.177]$ & $[0.0006,\ 0.0007]$ & 
    $[0.0000,\ 0.0001]$ \\
\bottomrule
\end{tabular}
\end{center}
\begin{figurenotes}
The table reports mean estimates across 200 datasets. The intervals shown below the means are the 0.025 and 0.975 quantiles of the estimates across these datasets.
\end{figurenotes}
\end{table}

\begin{figure}[t!!]
\centering
\caption{Estimated Joint Densities of Consumer Preference\label{fig:joint_pdfs}}
\includegraphics[scale=0.1]{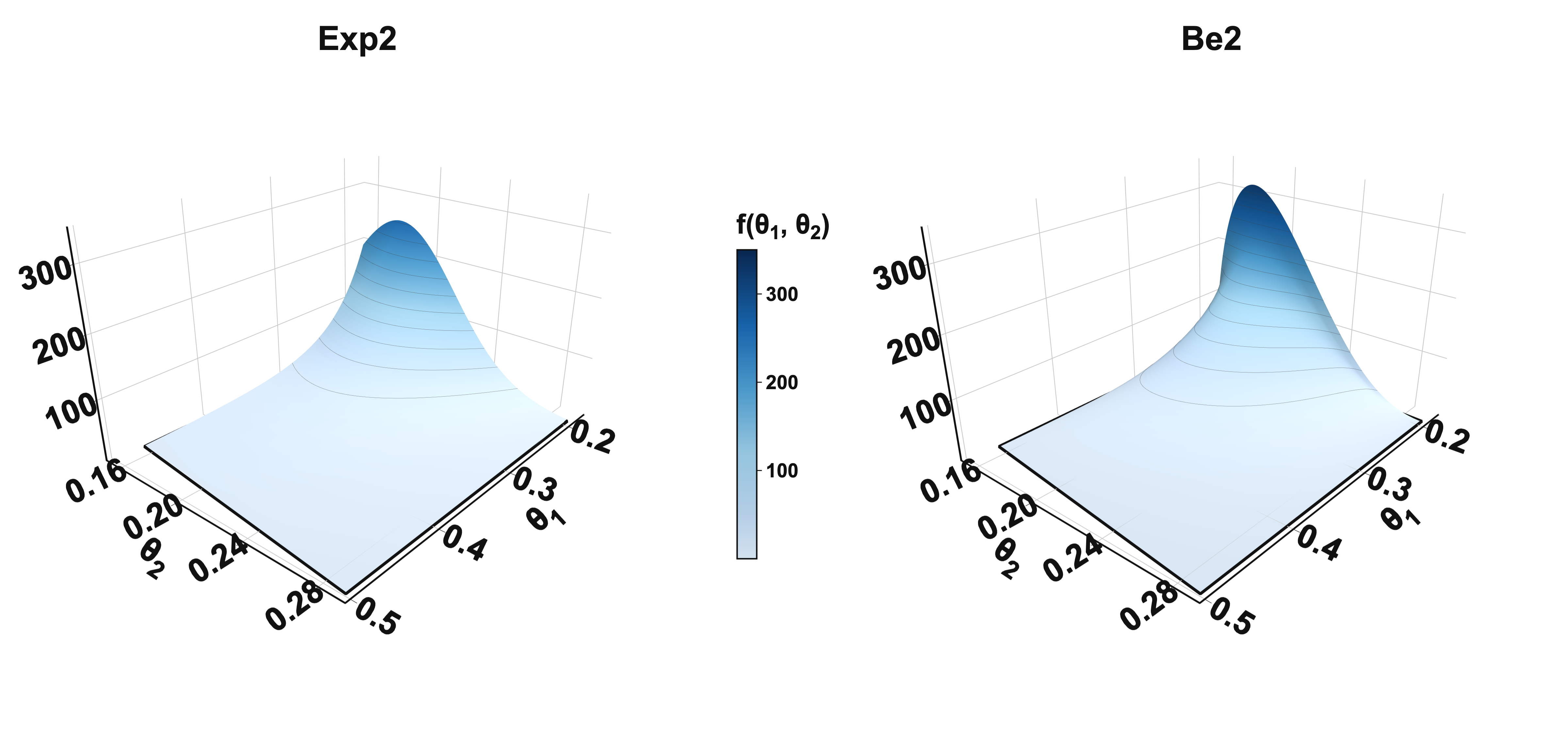}
\begin{figurenotes}
Estimated joint densities of consumer preferences $(\theta_1,\theta_2)$ under the Exp2 and Be2 families defined in Eqs.  (\ref{eq:exp2_synthetic}) and (\ref{eq:be2_synthetic}), using the parameter estimates reported in Table \ref{table:eapp}.
\end{figurenotes}
\end{figure}

\paragraph{Density Parameter Estimates.}
For the $\mathrm{Exp2}$ family, $\boldsymbol{\gamma}=\mathbf{0}_{4\times1}$ corresponds to a uniform distribution on the trapezoid, while for the $\mathrm{Be2}$ family the uniform benchmark is $\boldsymbol{\gamma}=\mathbf{1}_{4\times1}$. Table \ref{table:eapp} indicates substantial departures from these benchmarks under both specifications. For the $\mathrm{Exp2}$ family, $\hat{\gamma}_1>0$ and $\hat{\gamma}_2<0$ imply a nonmonotone profile in the first transformed dimension, whereas $\hat{\gamma}_3<0$ implies a declining profile over most of the second dimension. For the $\mathrm{Be2}$ family, the estimates also imply substantial nonuniformity along both dimensions. Figure \ref{fig:joint_pdfs} displays the corresponding estimated joint densities, which broadly reproduce the concentration of simulated types visible in the right panel of Figure \ref{figexasyn}.

{The implied mean type} is $\mathrm{E}(\boldsymbol{\theta}) \approx (0.2394,0.1953)$ under the $\mathrm{Exp2}$ specification and $\mathrm{E}(\boldsymbol{\theta}) \approx (0.2421,0.1969)$ under the $\mathrm{Be2}$ specification. Both specifications imply a small positive correlation between the two type dimensions, with $\mathrm{corr}(\theta_1,\theta_2)\approx 0.057$ for $\mathrm{Exp2}$ and $0.060$ for $\mathrm{Be2}$, suggesting that those who use more voice also weakly use more data.

\paragraph{Cost Parameter Estimates.}
The estimates of $\boldsymbol{\beta}=(\beta_1,\beta_2)$ are close to zero under
both distributional families, implying that marginal costs vary little with voice
and data usage over the relevant range of the product space. The estimates are also
similar across the two specifications.

For the intercepts $\boldsymbol{\alpha}=(\alpha_1,\alpha_2)$, both specifications
produce a larger estimate for voice than for data. Under the $\mathrm{Exp2}$ family,
$\hat{\alpha}_1=0.427$, while $\hat{\alpha}_2=-0.006$ with an empirical quantile
interval of $[-0.195,0.083]$, so the latter is effectively close to zero. Under the
$\mathrm{Be2}$ family, the corresponding estimates are
$\hat{\alpha}_1=0.262$ and $\hat{\alpha}_2=0.159$, with substantially narrower
intervals. Thus, while the levels depend somewhat on the distributional
specification, the broad pattern of a larger voice intercept and nearly flat marginal
cost schedules is similar across the two families.

\section{Conclusion\label{section:discussion}}

In this paper, we consider a screening model with multidimensional unobserved consumer preferences and determine when individual-level data on choices and payments are sufficient to identify the joint distribution of consumer preferences and the cost function. Based on these identification arguments, we propose estimators and establish their asymptotic properties. A Monte Carlo experiment and estimation using synthetic data show that the estimators have desirable small-sample properties. 

We note that we have not included consumer characteristics in our analysis. Insofar as the characteristics affect the preference distribution, we can view our analysis as conditional on them. On the other hand, the estimation would require adjustments, such as applying a local MLE approach (see, e.g., \cite{fan1998local}). This interpretation assumes that the seller can offer (possibly) a distinct menu based on those characteristics. Otherwise, we must be careful about how ``third-degree" screening interacts with ``second-degree" screening, which was the only focus of our study, and adapt our identification framework accordingly.

For future extensions, we note that our method excludes multidimensional screening with more than one seller \citep{Stole2007} and cannot capture how incomplete information and imperfect competition interact to affect welfare. While some work has addressed this topic \citep[e.g.,][]{IvaldiMartimort1994, MiraveteRoller2004, McManus2007, BusseRysman2005, AryalGabrielli2020}, developing a systematic empirical framework to study such markets \citep[e.g.,][]{MahoneyWeyl2017} remains promising area of future research.

\vspace{1in}

\renewcommand{\thesection}{\Alph{section}} \setcounter{section}{0}
\renewcommand{\thesubsection}{\Alph{section}.\arabic{subsection}} \setcounter{subsection}{0}
\renewcommand{\theequation}{\Alph{section}.\arabic{equation}} \setcounter{equation}{0}
\renewcommand{\theassumption}{\Alph{section}.\arabic{assumption}} \setcounter{assumption}{0}
\renewcommand{\thelemma}{\Alph{section}.\arabic{lemma}} \setcounter{lemma}{0}
\renewcommand{\thetheorem}{\Alph{section}.\arabic{theorem}} \setcounter{theorem}{0}
\renewcommand{\theHsection}{\Alph{section}.\arabic{section}}

\section{Appendix: Extensions \label{section:p_hat}}

\subsection{Estimation under unknown $\mathfrak{p}(\cdot)$ and $\mathscr{S}_{\boldsymbol{\theta},2}$}

Here, we provide estimators of $f_{\boldsymbol{\theta}}(\cdot)$ and $(\boldsymbol{\alpha},\boldsymbol{\beta})$ for an scenario in which the pricing function $\mathfrak{p}(\cdot)$ and the no-bunching region $\mathscr{S}_{\boldsymbol{\theta},2}$ are unknown. However, we continue to treat $\mathscr{S}_{\mathbf{Q},2}$ and $\mathscr{S}_{\boldsymbol{\theta}}$ as known, as the former can be identified from Lemma \ref{le:idsubsets}.

We start by constructing an estimator of $\mathfrak{p}(\cdot)$, for which we recall that $P = \mathfrak{p} ( \mathbf{Q}  )$, therefore $\Pr ( P = \mathfrak{p} ( \mathbf{q} )  \mid  \mathbf{Q} = \mathbf{q} ) =1$. Thus, estimating $\mathfrak{p}(\cdot)$ is a nonparametric irregular estimation problem. Our estimator of $\mathfrak{p}(\cdot)$ is based on the boundary estimator of \cite{KorostelevTsybakov1993}. Next, we describe its implementation procedure and then provide its asymptotic properties. 

Let $S \in \mathbb{N}$ and $(M_n)_{n \in \mathbb{N}}$ be a sequence of positive integers such that $
M_n \asymp  [  \frac{n}{\log(n)} ]^{\frac{1}{S+1+J}}.$
For each $n \in \mathbb{N}$, consider the set $\mathcal{R}_n: = \bigcupdot_{m=1}^{M_n^J} \mathcal{R}_{n,m} \subseteq \mathscr{S}_{\mathbf{Q},2}$, where $\mathcal{R}_{n,1}, \dots, \mathcal{R}_{n, M_n^J} $ are disjoint rectangles with nonempty interior and each side proportional to $1/M_n$. Thus, the volume of each rectangle $\mathcal{R}_{n,m}$ is proportional to $1/M_n^{J}$, while the volume of $\mathcal{R}_n $ is proportional to a constant. Denote the lower vertex of each $\mathcal{R}_{n,m}$ by $\ushort{\mathbf{q}}_{m}$ and define a mapping $ \mathbf{q}  \mapsto m ( \mathbf{q} )$ so that $m (\cdot)$ is the index function, i.e., $m ( \mathbf{q} ) \in \{1,\dots, M_n^J\}$ and $\mathbf{q} \in \mathcal{R}_{n, m ( \mathbf{q} )} $. For notational ease, we have suppressed the dependence of $\ushort{\mathbf{q}}_{m}$ and $m ( \mathbf{q} )$ on $n$. 

Consider the set $\mathscr{L} =  \{  \boldsymbol{\ell} \in \mathbb{N}_0^J:  \ 0 \leq \boldsymbol{\iota}_J \cdot \boldsymbol{\ell} \leq S  \}$ and let $  \boldsymbol{\ell}_1 \prec \cdots \prec \boldsymbol{\ell}_L$ be its elements lexicographically ordered, where $L: =\# \mathscr{L}  $ is the cardinality of the set that does not depend on $n$ as $S$ is fixed. Let $\lowmathcal{s} (  \mathbf{q} ) = \left( \mathbf{q}^{\boldsymbol{\ell}_1} ,  \dots ,   \mathbf{q}^{\boldsymbol{\ell}_L}  \right)$ be the monomial basis and consider also the discrete subset $
\Pi_n = \left\{  \mathcal{M}_n \mathbf{z}  \  \in    \left[ - {c}_2 , c_2 \right]^L  :  \   \mathbf{z} \in \mathbb{Z}^L   \right\},$ where $\mathcal{M}_n = \frac{1}{c_1} \mathrm{diag}\left(  M_n^{(\boldsymbol{\iota}_J \cdot \boldsymbol{\ell}_1 ) - S - 1} , \dots ,  M_n^{(\boldsymbol{\iota}_J \cdot \boldsymbol{\ell}_L ) - S - 1} \right)$ and $c_1 , c_2 > 0$ are sufficiently large constants chosen by the researcher.

In this setting, for $ \mathbf{q} \in \mathcal{R}_n$, we propose estimating $\mathfrak{p} (  \mathbf{q} )$ by the piecewise polynomial \begin{equation*}
\hat{\mathfrak{p}} (  \mathbf{q}  )  =  \lowmathcal{s}\left(  \mathbf{q} - \ushort{\mathbf{q}}_{m( \mathbf{q}  )} \right)\cdot     \frac{1}{2}(  \overline{\boldsymbol{\pi}}_{m(  \mathbf{q} )}  + \underline{\boldsymbol{\pi}}_{m( \mathbf{q}  )}   )  ,
\end{equation*}where $\overline{\boldsymbol{\pi}}_m$ and $\underline{\boldsymbol{\pi}}_{m} \in \Pi_n$ are vectors of coefficients defined as follows for $m = 1,\dots, M_n^J$:
\begin{eqnarray*}
\overline{\boldsymbol{\pi}}_m &  = & \underset{\boldsymbol{\pi} \in \Pi_n}{\arg\min} \    \int_{\mathcal{R}_{n,m}}   \lowmathcal{s} \left(  \mathbf{q} - \ushort{\mathbf{q}}_m \right) \cdot  \boldsymbol{\pi}  \  d \mathbf{q}  \ \   \text{subject to}     \  \   P_i \ \leq  \  \lowmathcal{s} \left(  \mathbf{Q}_i  - \ushort{\mathbf{q}}_m \right) \cdot  \boldsymbol{\pi}  \   \  \forall \ \mathbf{Q}_i  \in  \mathcal{R}_{n,m} , \\
\underline{\boldsymbol{\pi}}_m  &  = &  \underset{\boldsymbol{\pi} \in \Pi_n}{\arg\max} \int_{\mathcal{R}_{n,m}}   \lowmathcal{s} \left(  \mathbf{q}  - \ushort{\mathbf{q}}_m \right) \cdot  \boldsymbol{\pi}  \  d \mathbf{q}  \  \   \text{subject to}     \  \   P_i \ \geq  \   \lowmathcal{s}   \left(  \mathbf{Q}_i  - \ushort{\mathbf{q}}_m \right)  \cdot  \boldsymbol{\pi}  \   \  \forall \ \mathbf{Q}_i  \in  \mathcal{R}_{n,m} .
\end{eqnarray*}
We also consider $\nabla \hat{\mathfrak{p}} (  \mathbf{q}  )  $ and $\mathcal{H} \hat{\mathfrak{p}} (  \mathbf{q} )$ as the estimators of $\nabla \mathfrak{p} (  \mathbf{q} )$ and $\mathcal{H} \mathfrak{p} (  \mathbf{q} )$, respectively, where the partial derivatives of the estimators are defined from the interior when $ \mathbf{q} \in \partial \mathcal{R}_{n, m ( \mathbf{q} ) }$.

Several remarks are noteworthy. First, when $\mathscr{S}_{\mathbf{Q},2}$ is a rectangle (as in Example \ref{example}), we can set $\mathcal{R}_n  = \mathscr{S}_{\mathbf{Q},2}$ for all $n$ by choosing the disjoint rectangles $\mathcal{R}_{n,m}$, $m = 1,\dots, M_n^J$, accordingly. In general, one can choose $\mathcal{R}_n$ so that $\lim_{n\rightarrow\infty}d_H ( \mathcal{R}_n , \mathscr{S}_{\mathbf{Q},2} ) = 0$, as we suggest below in condition (c1). Second, $\overline{\boldsymbol{\pi}}_m$ and $\underline{\boldsymbol{\pi}}_{m} \in \Pi_n$ are solutions to linear-programming problems with linear inequality constraints, where the optimization is performed over the discrete set $\Pi_n$. We adopt this discretization without loss of generality to simplify the proof. In practice, the researcher can choose $c_1, c_2 >0$ arbitrarily large, so the discretization has negligible impact on computational tasks.

To establish the uniform rate of convergence of $\hat{\mathfrak{p}} (  \cdot  )  $, $\nabla \hat{\mathfrak{p}} (  \cdot )  $, and $\mathcal{H} \hat{\mathfrak{p}} ( \cdot )$ on $\mathcal{R}_n $, we make the next high-level assumption.

\begin{assumption}  \label{assump:smooth}

The pricing function $\mathfrak{p}(\cdot)$ admits continuous and uniformly bounded partial derivatives of order $S+1$ on $\mathcal{R}_n$ for all $n \in \mathbb{N}$. Moreover, there exists a constant $\ushort{f}_{\mathbf{Q}} >0$ such that $f_{\mathbf{Q}} (  \mathbf{q} ) \geq \ushort{f}_{\mathbf{Q}}$ for all $ \mathbf{q}  \in \mathcal{R}_n $ and $n \in \mathbb{N}$.

\end{assumption} 

The next lemma provides the uniform convergence rates of the proposed estimators.

\begin{lemma}	\label{lem:boundest} 			 

Suppose that Assumptions \ref{assump:reg} and \ref{assump:smooth} hold. Then, $
 \left\|  \hat{\mathfrak{p}}  -   \mathfrak{p}   \right\|_{ \mathcal{R}_n  , \infty} = O_p \left( {M}_n^{-S-1} \right)$ and   $ \left\|  \nabla \hat{\mathfrak{p}} -   \nabla \mathfrak{p}   \right\|_{ \mathcal{R}_n  , \infty} = O_p \left( {M}_n^{-S} \right).$
Moreover, $\|  \mathcal{H} \hat{\mathfrak{p}}  -  \mathcal{H} \mathfrak{p}   \|_{ \mathcal{R}_n  , \infty}= O_p ( {M}_n^{-S+1} )$ if $S\geq2$.

\end{lemma}

From this lemma, and as suggested by Eq.\ (\ref{eq:gpvtype}), types $\{\boldsymbol{\theta}_i:  \mathbf{Q}_i \in \mathcal{R}_n \}$ corresponding to $\mathbf{Q}_i \in \mathcal{R}_n $ can be consistently estimated by
\begin{equation}
\label{eq:gpvestimates}
\hat{\boldsymbol{\theta}}_i = \left\{
\begin{array}{ll}
\nabla \hat{\mathfrak{p}} (  \mathbf{Q}_i  ) & \text{if} \  \nabla \hat{\mathfrak{p}} (  \mathbf{Q}_i  )  \in \mathscr{S}_{\boldsymbol\theta}  , \\
\arg\min_{\mathbf{t} \in \mathscr{S}_{\boldsymbol\theta}} \left\| \mathbf{t}   -  \nabla \hat{\mathfrak{p}} (  \mathbf{Q}_i  )   \right\|_2  & \text{otherwise},
\end{array}
\right.
\end{equation}
selecting the smallest element in lexicographic order if multiple minimizers exist when $\nabla \hat{\mathfrak{p}} (  \mathbf{Q}_i  )  \notin \mathscr{S}_{\boldsymbol\theta}$.

As in Section \ref{section:estimation}, we can now construct both a nonparametric and a parametric estimator for $f_{\boldsymbol{\theta}}(\cdot)$. The former consists of estimating $f_{\boldsymbol{\theta}}( \mathbf{t})$ by a kernel density estimator that uses the estimated types:
\begin{equation*}
 \check{f}_{\mathrm{np}} ( \mathbf{t})  = \frac{1}{h_f n} \sum_{i \in \check{\mathscr{I}}_2} \kappa \left( \frac{\mathbf{t} - \hat{\boldsymbol{\theta}}_i }{h_f}  \right) ,
\end{equation*}where $\check{\mathscr{I}}_2 = \{ i  = 1,\dots, n: \ \mathbf{Q}_i \in \mathcal{R}_n \}$. We construct the latter as follows. Choose a set $\tilde{\mathcal{R}} \subseteq \mathscr{S}_{\mathbf{Q},2}$ such that $\tilde{\mathcal{R}}  \subseteq \mathcal{R}_n$ for all $n \in \mathbb{N}$ and here set $\tilde{\mathcal{S}}  = \nabla \mathfrak{p} ( \tilde{\mathcal{R}}  )$. From Assumption \ref{assump:inte}, we know that $ f_{\boldsymbol{\theta}} (\cdot)  = f (\cdot ;  \boldsymbol{\gamma}_0 )  \in \mathscr{F}$ for some $ \boldsymbol{\gamma}_0 \in \mathrm{int} (\Gamma )$, so we set $\check{f} ( \mathbf{t})  = f (\mathbf{t};  \check{ \boldsymbol{\gamma} } )$, where $
\check{ \boldsymbol{\gamma} }  = \arg\max_{\boldsymbol{\gamma} \in \Gamma  }  \sum_{i \in \check{\mathscr{I}}_2}  \hat{\mathfrak{L}}_i ( \boldsymbol{\gamma} )$ and 
\begin{equation*}
 \hat{\mathfrak{L}}_i ( \boldsymbol{\gamma} )  =  \left\{ \begin{array}{l l}    \log \left[ f \left( \hat{\boldsymbol{\theta}}_i   ; \boldsymbol{\gamma}  \right)   \right]   - \log  \left[  \int_{\mathcal{R}_n}  f \left[   \nabla\hat{\mathfrak{p}} ( \mathbf{q} ) ; \boldsymbol{\gamma}    \right]  \left|  \mathcal{H} \hat{\mathfrak{p}} ( \mathbf{q} )  \right| d\mathbf{q}    \right]    &  \text{if} \ \mathbf{Q}_i \in \tilde{\mathcal{R}} ,  \\ 
 0  & \text{otherwise} .
 \end{array}\right.
\end{equation*}
Letting $\check{\Sigma} = (1/n)  \sum_{i=1}^n [ \nabla \hat{\mathfrak{L}}_i ( \check{\boldsymbol{\gamma}} ) ]  [ \nabla \hat{\mathfrak{L}}_i ( \check{\boldsymbol{\gamma}} ) ]^\top$, the next theorem establishes the asymptotic properties of $\check{\boldsymbol{\gamma}}$. 

\begin{theorem}	\label{thm:asympapp}
Suppose that Assumptions \ref{assump:reg}, \ref{assump:idcost2}, \ref{assump:inte}, and \ref{assump:smooth} hold and also $S > J+3$. Then, $\check{\boldsymbol{\gamma}} \overset{p}{\rightarrow} \boldsymbol{\gamma}_0 $ and $
\sqrt{n} \left( \check{\boldsymbol{\gamma}} -  \boldsymbol{\gamma}_0 \right) \overset{d}{\rightarrow}   N({\boldsymbol 0}, {\Sigma}^{-1})$. Moreover, $ \check{\Sigma} \overset{p}{\rightarrow}  {\Sigma} $ and therefore $\check{\Sigma}^{-1} \overset{p}{\rightarrow}  {\Sigma}^{-1}$.
\end{theorem}

The proof of this theorem is provided in Appendix \ref{section:proofs}. The proof relies on standard arguments about MLE \citep{NeweyMcFadden94} combined with the fact that $\hat{\boldsymbol{\theta}}_i$, $\hat{\mathfrak{p}} ( \cdot ) $, and $\mathcal{H} \hat{\mathfrak{p}} ( \cdot )$ converge faster than the parametric $\sqrt{n}$-rate when $S > J+3$ (Lemma \ref{lem:boundest}). We remark that the asymptotic variance of Theorem \ref{thm:asympapp} is the same as the one in Theorem \ref{thm:asymp}, as the estimation errors of the types $\boldsymbol{\theta}_i$ do not contribute to the asymptotic distribution, again, due to the fast convergence rate of $ \nabla \hat{\mathfrak{p}} ( \cdot)$.

Next, we propose estimators of the marginal cost parameters $(\boldsymbol{\alpha}, \boldsymbol{\beta})$. To establish the consistency of the proposed estimators, we first assume that the values of ${\mathbf{t}}^\prime_j$, $ \vec{\lowmathcal{n}} ({\mathbf{t}}_j^\prime  )$, $\mathbf{t}_j^{\prime\prime}$, $j = 1,\dots, J$, from Assumption \ref{assump:idcost2} are known: data-driven methods for choosing these values are presented at the end of this section. Moreover, we assume that the sequence of subsets $(\mathcal{R}_n)_{n \in \mathbb{N}} $ satisfies the next conditions: \begin{enumerate}

\item[(c1)] $d_H ( \mathcal{R}_n , \mathscr{S}_{\mathbf{Q},2} ) \rightarrow 0$ as $n \rightarrow \infty$.

\item[(c2)]  For some fixed $\epsilon >0$, $ \bigcup_{j=1}^J \mathscr{B} (\mathbf{t}_j^{\prime\prime} , \epsilon) \subset  \nabla \hat{\mathfrak{p}}  ( \mathcal{R}_n  ) $ w.p.a.1.

\end{enumerate}These conditions are technical requirements that, heuristically speaking, mean that $\mathcal{R}_n  $ should be chosen as large as possible.

The proposed estimators of $\boldsymbol{\alpha}$ and $ \boldsymbol{\beta}$ can now be computed in three steps as follows. First, estimate the vector of choices $\mathcal{Q}$ by $\check{\mathcal{Q}}  = ( \check{\mathbf{Q}}_1 \  \cdots  \ \check{\mathbf{Q}}_J )_{[ J \times J ]}$, where
\begin{equation*}
 \check{\mathbf{Q}}_j  = \underset{\mathbf{q} \in \mathcal{R}_n }{\arg\min}  \  \left\|  {\mathbf{t}}^\prime_j  -  \nabla \hat{\mathfrak{p}}  (  \mathbf{q} ) \right\|_2 \quad \text{for} \ j = 1,\dots, J .
\end{equation*}
Second, letting $\hat{\mathfrak{q}} (  \mathbf{t} ) =  ( \nabla \hat{\mathfrak{p}}  )^{-1} ( \mathbf{t}  )$ for $ \mathbf{t} \in  \nabla \hat{\mathfrak{p}}  ( \mathcal{R}  ) $, estimate $\mathcal{D}_{\beta}$ by $\check{\mathcal{D}}_{\beta}  :=  \check{\mathcal{D}}_{fq} -   \check{\mathcal{D}}_f^\top \mathcal{N}^{-1} (\mathcal{N} \odot \check{\mathcal{Q}}^\top )$, where \begin{equation*}
\check{\mathcal{D}}_{fq}    =      \left( \begin{array}{ccc} 
\nabla_1 [ \check{f} ( {\mathbf{t}}_1^{\prime\prime} ) \hat{\mathfrak{q}}_1 ( {\mathbf{t}}_1^{\prime\prime} ) ]  &  \cdots  &  \nabla_J [ \check{f} ( {\mathbf{t}}_1^{\prime\prime} ) \hat{\mathfrak{q}}_J ( {\mathbf{t}}_1^{\prime\prime} ) ]   \\
\vdots &   \ddots   & \vdots \\
\nabla_1 [ \check{f} ( {\mathbf{t}}_J^{\prime\prime} ) \hat{\mathfrak{q}}_1 ( {\mathbf{t}}_J^{\prime\prime} ) ]  &  \cdots  &  \nabla_J [ \check{f} ( {\mathbf{t}}_J^{\prime\prime} ) \hat{\mathfrak{q}}_J ( {\mathbf{t}}_J^{\prime\prime} ) ] 
\end{array} \right)_{[J \times J]} , 
\end{equation*}and $\check{\mathcal{D}}_f      =    \left(  \nabla \check{f} ( {\mathbf{t}}_1^{\prime\prime} )  \   \cdots \  \nabla \check{f} ( {\mathbf{t}}_J^{\prime\prime} ) \right)_{[ J \times J ]} $. Third, the proposed estimators of $\boldsymbol{\beta}$ and $\boldsymbol{\alpha}$ can now be computed as follows:
 \begin{equation*}
\check{\boldsymbol{\beta}}  =     \check{\mathcal{D}}_{\beta}^{-1}  \left[ (J  + 1)  \left(\begin{array}{c}
\check{f} ( {\mathbf{t}}_1^{\prime\prime} ) \\
 \vdots \\
 \check{f}( {\mathbf{t}}_J^{\prime\prime} )
\end{array}  \right)    +    \left(\begin{array}{c}
({\mathbf{t}}_1^{\prime\prime} -  {\mathcal{N}}^{-1} {\mathbf{T} } ) \cdot \nabla \check{f} ( {\mathbf{t}}_1^{\prime\prime} ) \\
 \vdots \\
( {\mathbf{t}}_J^{\prime\prime}  -  {\mathcal{N}}^{-1} {\mathbf{T}}  ) \cdot \nabla  \check{f} ( {\mathbf{t}}_J^{\prime\prime} )
\end{array}  \right)     \right]    \ \text{and} \    \check{\boldsymbol{\alpha}}     =         \mathcal{N}^{-1} \left[ \mathbf{T}    - \left( \mathcal{N}  \odot \check{\mathcal{Q}}^\top \right) \check{\boldsymbol{\beta}} \right] ,
\end{equation*}
respectively.

The next theorem establishes the consistency of these estimators. Let $\mathrm{conv}(\cdot)$ denote the convex hull of a set.

\begin{theorem}	\label{thm:costapp}
Suppose that Assumptions \ref{assump:reg}-\ref{assump:inte} and \ref{assump:smooth} hold, $S > J + 3$, and also that conditions (c1)-(c2) are satisfied. Assume further that there exists constants $0 < \underline{c}_{\ref*{thm:costapp}} \leq \bar{c}_{\ref*{thm:costapp}} <\infty$ such that, for any $n \in \mathbb{N}$ and $\mathbf{q} \in \mathrm{conv}( \mathcal{R}_n )$, the eigenvalues of $\mathcal{H} \mathfrak{p} ( \mathbf{q} )$ belong to the interval $[\ushort{c}_{\ref*{thm:costapp}} , \bar{c}_{\ref*{thm:costapp}}]$. Then, $\hat{\boldsymbol{\beta}} \overset{p}{\rightarrow}  \boldsymbol{\beta}$ and $\hat{\boldsymbol{\alpha}}  \overset{p}{\rightarrow} \boldsymbol{\alpha}$.
\end{theorem}

To conclude this section, without considering formal asymptotic aspects, we suggest a data-driven procedure for selecting the values of ${\mathbf{t}}^\prime_j$ and $\mathbf{t}^{\prime\prime}_j$, $j=1,\dots, J$, referred to in Assumption \ref{assump:idcost2}. Specifically, we recommend following the next two steps for selecting ${\mathbf{t}}^\prime_j$, $j=1,\dots, J$. \begin{enumerate}

\item Compute the set of estimated types $\hat{\mathscr{S}}_{\boldsymbol{\theta},2} = \{\hat{\boldsymbol\theta}_i  :  \mathbf{Q}_i \in \mathcal{R}_n \}$ using Eq.\ ($\ref{eq:gpvestimates}$) and choosing $\mathcal{R}_n \subset  \mathscr{S}_{\mathbf{Q},2}$ in a manner that encompasses a substantial portion of $\mathscr{S}_{\mathbf{Q},2}$.

\item Choose ${\mathbf{t}}^\prime_j \in  \partial \mathscr{S}_{\boldsymbol{\theta}}$, $j=1,\dots,J$, such that $\{ \vec{\lowmathcal{n}} ({\mathbf{t}}_1^\prime  ) , \dots,   \vec{\lowmathcal{n}} ({\mathbf{t}}_J^\prime  ) \}$ are linearly independent and $\| \hat{\boldsymbol\theta}_i^\prime   -   {\mathbf{t}}^\prime_j   \|_2 < \epsilon$ for all $j = 1,\dots,J$ and for some $\hat{\boldsymbol\theta}_i^\prime  \in \hat{\mathscr{S}}_{\boldsymbol{\theta},2}$ and $\epsilon > 0$ sufficiently small.

\end{enumerate}Finally, to choose the values of $\mathbf{t}^{\prime\prime}_j$, $j=1,\dots, J$, one can opt for values that result in a relatively large $| \mathrm{det} ( \check{\mathcal{D}}_{\beta}  )  |$.

\subsection{Discussion on Measurement Errors in Payments}

In this section, we consider the scenario where observed payments may differ
from the model-implied payments and discuss how to adapt our identification and
estimation strategies to this framework. So far, we have assumed that the
observed payment $P$ is the optimal payment implied by the model. Although
the equilibrium pricing function is a unique and deterministic function of the
product, in practice, observed payments may deviate from the model-implied
payment scheme because of measurement errors, such as discounts (coupons).
Suppose that the model-implied payment is $P^*=\mathfrak{p}(Q)$, while the
researcher observes $P=\phi(P^*,\varepsilon)$, where $\varepsilon$ is an
unobservable payment ``error" independent of $\boldsymbol\theta$, and
$\phi:\mathbb R_+\times\mathbb R\rightarrow\mathbb R_+$ is known by the
researcher. Since $Q=\q(\boldsymbol\theta)$, independence between
$\boldsymbol\theta$ and $\varepsilon$ also implies independence between $Q$
and $\varepsilon$.

The identification strategy of Section \ref{sec:id} can be adapted to this
framework if we identify ${\mathfrak p}(\cdot)$ nonparametrically, as this will
allow us to know the joint distribution $F_{\mathbf{Q}, P^\ast}$ and therefore
Lemmas \ref{le:idsubsets}-\ref{le:costfun} can be extended. We can consider the
additive structure
\begin{eqnarray}
\phi ( P^\ast ,  \varepsilon )  = P^\ast +  \varepsilon
\ \  \text{with}  \  \E (\varepsilon  ) = 0.
\label{eq:exteerrorp}
\end{eqnarray}
Alternatively, the multiplicative one
$\phi ( P^\ast ,  \varepsilon )  = P^\ast \varepsilon$ with
$\varepsilon >0$ and $\E [ \log(\varepsilon) ] = 0$. In both cases,
${\mathfrak p}(\cdot)$ can be identified by standard arguments.
Specifically, as
$P = \phi \left[ {\mathfrak p}(\mathbf{Q}) ,  \varepsilon \right]$, we have
that $\mathfrak{p} ( \mathbf{q}  ) =
\E  ( P |  \mathbf{Q}=  \mathbf{q} )$ in Eq.\ (\ref{eq:exteerrorp}), while
$\mathfrak{p} ( \mathbf{q}  ) =
\exp\{ \E [ \log(P) | \mathbf{Q}=  \mathbf{q} ] \}$ when
$\phi ( P^\ast ,  \varepsilon )  = P^\ast \varepsilon$.

The same specification can accommodate random lump-sum discounts without
requiring the error to have mean zero. In particular, let
$\varepsilon=-\mathfrak{D}$, where $\mathfrak{D}\geq0$ is a random discount independent of
$\boldsymbol\theta$. Then $P=\mathfrak{p}(Q)-\mathfrak{D}$ and
$\E(P\mid Q=\q)=\mathfrak{p}(\q)-\E(\mathfrak{D})$.
Thus, if $\E(\mathfrak{D})$ is known, we identify the pricing function. More
importantly for our identification strategy, even when $\E(\mathfrak{D})$ is unknown,
$\nabla_{\q}\E(P\mid Q=\q)=\nabla_{\q}\mathfrak{p}(\q)$.
Consequently, the marginal pricing function is identified despite the unknown
mean discount. In particular, for consumers in the perfectly screened region,
Eq. \eqref{eq:gpvtype} becomes
$\boldsymbol\theta=\nabla_{\q}\E(P\mid Q=\q)$,
so the preference-identification argument is unchanged.

A similar argument applies to random percentage discounts. Let
$\mathfrak{D}\in[0,1)$ now denote the percentage discount, so that
$P=(1-\mathfrak{D})\mathfrak{p}(Q)$.
If $\mathfrak{D}$ is independent of $\boldsymbol\theta$, and hence of $Q$, then
$\E(P\mid Q=\q)=\mathfrak{p}(\q)\E(1-\mathfrak{D}\mid Q=\q)
=\mathfrak{p}(\q)\E(1-\mathfrak{D})$.
Letting $\mu_\mathfrak{D}=\E(1-\mathfrak{D})>0$, we get
$\nabla_{\q}\E(P\mid Q=\q)
=\mu_\mathfrak{D}\nabla_{\q}\mathfrak{p}(\q)$.
Thus, unlike a lump-sum discount, an unknown percentage discount introduces a
scale normalization. If $\mu_\mathfrak{D}$ is known, both
$\mathfrak{p}(\cdot)$ and its gradient are identified:
$\mathfrak{p}(\q)=\E(P\mid Q=\q)/\mu_\mathfrak{D}$ and
$\nabla_{\q}\mathfrak{p}(\q)
=\nabla_{\q}\E(P\mid Q=\q)/\mu_\mathfrak{D}$.
Alternatively, if the support of $\mathfrak{D}$ has lower endpoint zero, the pricing
function is identified directly from
$\mathfrak{p}(\q)=\operatorname*{ess\,sup}(P\mid Q=\q)$.

More generally, these examples are special cases of the nonseparable
measurement-error model
$P=\phi[\mathfrak{p}(Q),\varepsilon]$. We refer to
\cite{Matzkin2003} for identification and estimation in nonadditive random
function models. The relevance of that framework here follows from the
independence restriction: since $Q=\q(\boldsymbol\theta)$ and
$\varepsilon$ is independent of $\boldsymbol\theta$, the payment disturbance
is also independent of $Q$. The particular additive and multiplicative
specifications above permit the simpler direct identification arguments we
have displayed; more general choices of $\phi$ require the corresponding
regularity and normalization conditions discussed in \cite{Matzkin2003}.
Economically, the independence assumption means that measurement errors or
random discounts do not vary systematically across consumer types and,
therefore, across their chosen products.

The estimation method proposed in Section \ref{section:p_hat} must be modified
because the presence of measurement errors or unobserved discounts does not
allow us to construct a super-consistent estimator of
$\mathfrak{p}(\cdot)$. Instead, $\mathfrak{p}(\cdot)$ and its partial
derivatives must be estimated nonparametrically from the relevant conditional
payment distribution; see \cite{Matzkin2003} for related estimation methods in
the general nonadditive framework. The resulting estimators converge at a
nonparametric rate, which is slower than the parametric one, and consequently
the estimators of $\boldsymbol{\gamma}_0$ and the marginal cost parameters
will generally converge at a rate slower than the parametric rate.

\renewcommand{\thesubsection}{\Alph{section}.\arabic{subsection}} \setcounter{subsection}{0}
\renewcommand{\theequation}{\Alph{section}.\arabic{equation}} \setcounter{equation}{0}
\renewcommand{\theassumption}{\Alph{section}.\arabic{assumption}} \setcounter{assumption}{0}
\renewcommand{\thelemma}{\Alph{section}.\arabic{lemma}} \setcounter{lemma}{0}
\renewcommand{\thetheorem}{\Alph{section}.\arabic{theorem}} \setcounter{theorem}{0}
\renewcommand{\theHsection}{\Alph{section}.\arabic{section}}

\section{Appendix: Proofs \label{section:proofs}}

This appendix provides the proofs of the lemmas, theorems, and corollaries stated in the main text and Appendix \ref{section:p_hat}.

\paragraph{Proof of Lemma \ref{lem:foc}.} We start by presenting two results:\begin{eqnarray}
& &  \nabla  \mathfrak{p} ( \mathbf{q}  )   =  \mathfrak{q}^{-1} (  \mathbf{q}  ) \quad \forall \   \mathbf{q} \in \mathrm{int}( \mathscr{S}_{\mathbf{Q},2}  )   \quad \text{and}  \label{eqa:foc} \\
& &  \mathfrak{q} \left[  \mathrm{int}( \mathscr{S}_{\boldsymbol{\theta},2}  )  \right]  \subseteq \mathrm{int}( \mathscr{S}_{\mathbf{Q},2}  )  .  \label{eqa:invdom} 
\end{eqnarray}Eq.\ (\ref{eqa:foc}) is an immediate consequence of the consumer's first order conditions on $\mathrm{int}( \mathscr{S}_{\mathbf{Q},2}  ) $, while Eq.\ (\ref{eqa:invdom}) arises from the fact that $\mathfrak{q} (\cdot)$ is continuous  (Properties 1.1) and by applying the Invariance of Domain theorem, which implies that $\mathfrak{q} \left[  \mathrm{int}( \mathscr{S}_{\boldsymbol{\theta},2}  )  \right] $ must be open.

Then, Eq.\ (\ref{eq:lemfoc}) arises from the following implications: \begin{equation*}
\mathbf{t} \in \mathrm{int}( \mathscr{S}_{\boldsymbol{\theta},2}  )  \   \ \underset{\text{Eq.\ (\ref{eqa:invdom})}}{\Longrightarrow}  \ \ \mathfrak{q} (\mathbf{t})   \in \mathrm{int}( \mathscr{S}_{\mathbf{Q},2}  )  \   \ \underset{\text{Eq.\ (\ref{eqa:foc})}}{\Longrightarrow}  \  \   \nabla  \mathfrak{p} [  \mathfrak{q} (\mathbf{t})  ]   =  \mathfrak{q}^{-1}  [  \mathfrak{q} (\mathbf{t})  ]  =  \mathbf{t}.
\end{equation*}The inclusion $\mathrm{int}( \mathscr{S}_{\boldsymbol{\theta},2}  )  \subseteq \nabla  \mathfrak{p} [   \mathrm{int}( \mathscr{S}_{\mathbf{Q},2}  )  ]$ is an immediate consequence of Eqs.\ (\ref{eq:lemfoc}) and (\ref{eqa:invdom}), while the other inclusion, $ \nabla  \mathfrak{p} [   \mathrm{int}( \mathscr{S}_{\mathbf{Q},2}  )  ] \subseteq \mathrm{int}( \mathscr{S}_{\boldsymbol{\theta},2}  ) $, arises from Eq.\ (\ref{eqa:foc}) and the Invariance of Domain theorem. Finally, continuity and non-singularity of $\nabla \mathfrak{q} (\cdot)$ follow from Eq.\ (\ref{eq:lemfoc}) and the Implicit Function Theorem. \qed

\paragraph{Proof of Lemma \ref{le:idsubsets}.} We start by defining a collection of curves on $\mathscr{S}_{\mathbf{Q}}$. Let $\mathscr{C}$ be the set of curves $\mathcal{C}$ that satisfy the next conditions: $\mathcal{C}$ can be parameterized by a continuously differentiable bijective function on $(0,1)$, and there exists an open set $\mathcal{N} \subset  \mathrm{int}( \mathscr{S}_{\mathbf{Q}}  )$ such that $\mathcal{C} \subset \mathcal{N}$ and $F_{\mathbf{Q}}$ admits a continuous p.d.f.\ on $\mathcal{N}$ with respect to the Lebesgue measure. Note that $\mathscr{C}$ is nonempty because so is $\mathrm{int}(  \mathscr{S}_{\boldsymbol{\theta},2} )$ and $\mathfrak{q} (\cdot)$ is continuosly differentiable on this set (Assumption \ref{assump:idcost2}-(d) and Lemma \ref{lem:foc}) and that $\mathscr{C}$ depends exclusively on the functional form of $F_{\mathbf{Q}} (\cdot)$.

Now note that $\mathscr{S}_{\mathbf{Q},0} = \{ \mathbf{q}_0 \}$ is trivially identified because the outside option is assumed known. So, pick any arbitrary $\mathbf{q} \in \mathscr{S}_{\mathbf{Q}} \backslash \mathscr{S}_{\mathbf{Q},0} $ and note that $\mathbf{q}$ must satisfy one and only one of the next conditions, which can be verified from our knowledge of $F_{\mathbf{Q}} (\cdot)$: \begin{enumerate}

\item[(c1)] $\exists \ \mathcal{C} \in \mathscr{C}$ such that $\Pr \left(   \mathbf{Q} = \mathbf{q} \ \middle| \  \mathbf{Q} \in \{ \mathbf{q}  \}   \cup  \mathcal{C} \right)   > 0$.

\item[(c2)] $\forall \ \mathcal{C} \in \mathscr{C}, \quad \Pr \left(   \mathbf{Q} = \mathbf{q} \ \middle| \  \mathbf{Q} \in \{ \mathbf{q}  \}   \cup  \mathcal{C} \right)   = 0$.

\end{enumerate}On the one hand, note that $\mathbf{q} \in \mathscr{S}_{\mathbf{Q},1}$ implies that (c1) holds. To see this, note that the curve given by $\mathcal{C}^\prime 	= \mathfrak{q} ( \mathcal{L}^\prime ) $, for some line segment $\mathcal{L}^\prime  \subset  \mathrm{int}(  \mathscr{S}_{\boldsymbol{\theta},2} )$, clearly satisfies $\mathcal{C}^\prime  \in \mathscr{C}$. Moreover, by construction of $\mathscr{S}_{\mathbf{Q},1}$ and by convexity of $\mathfrak{q}^{-1} ( \mathbf{q}  )$ (Properties \ref{proper:equi}-(b)), we must have  \begin{equation*}
\Pr \left(   \mathbf{Q} = \mathbf{q} \ \middle| \  \mathbf{Q} \in \{ \mathbf{q}  \}   \cup  \mathcal{C}^\prime \right)  = \Pr \left[  \boldsymbol{\theta} \in \mathfrak{q}^{-1} ( \mathbf{q}  )  \middle| \  \boldsymbol{\theta}  \in \mathfrak{q}^{-1} ( \mathbf{q}  )    \cup\mathcal{L}^\prime \right] >0  
\end{equation*}as $\mathfrak{q}^{-1} ( \mathbf{q}  )$ is either a line segment or a set with a nonempty interior. On the other hand, note that $\mathbf{q} \in \mathscr{S}_{\mathbf{Q},2}$ implies that (c2) holds. To see this, note that there is a unique $\tilde{\mathbf{t}} \in  \mathscr{S}_{\boldsymbol{\theta},2} $ such that $ \mathbf{q}  =  \mathfrak{q} (\tilde{\mathbf{t}} )$; hence, for any $\mathcal{C} \in \mathscr{C}$, we must have \begin{equation*}
\Pr \left(   \mathbf{Q} = \mathbf{q} \ \middle| \  \mathbf{Q} \in \{ \mathbf{q}  \}   \cup  \mathcal{C}  \right)  = \Pr \left[  \boldsymbol{\theta} = \tilde{\mathbf{t}}  \ \middle| \  \boldsymbol{\theta}  \in  \{ \tilde{\mathbf{t}} \}   \cup \mathfrak{q} ^{-1} ( \mathcal{C} ) \right] =  0  
\end{equation*}because $F_{\boldsymbol\theta}(\cdot)$ does not have mass points.


Finally, since (c1) are (c2) mutually exclusive conditions and also $\mathscr{S}_{\mathbf{Q},1}$ and $\mathscr{S}_{\mathbf{Q},2}$ are disjoint, for each $\mathbf{q} \in \mathscr{S}_{\mathbf{Q}} \backslash \mathscr{S}_{\mathbf{Q},0}$, we can check whether it satisfies either (c1) or (c2) and then classify every $\mathbf{q}\in \mathscr{S}_{\mathbf{Q}} \backslash \mathscr{S}_{\mathbf{Q},0}$ into either $\mathscr{S}_{\mathbf{Q},1}$ or $\mathscr{S}_{\mathbf{Q},2}$ as follows: $\mathscr{S}_{\mathbf{Q},1}   =   \{  \mathbf{q} \in \mathscr{S}_{\mathbf{Q}} \backslash \mathscr{S}_{\mathbf{Q},0}   :  \text{$\mathbf{q}$ satisfies (c1)} \}$ and $\mathscr{S}_{\mathbf{Q},2}   =    \{  \mathbf{q} \in \mathscr{S}_{\mathbf{Q}} \backslash \mathscr{S}_{\mathbf{Q},0}   : \  \text{$\mathbf{q}$ satisfies (c1)} \}$.  \qed

\paragraph{Proof of Lemma \ref{le:fid}.} Pick any $\mathbf{t} = ( t_1,\dots,t_J  )\in \mathrm{int} (\mathscr{S}_{\boldsymbol{\theta} ,2} ) $ together with $\mathbf{t}^\prime  = ( t_1^\prime ,\dots , t_J^\prime  )  \in \mathrm{int} (\mathscr{S}_{\boldsymbol{\theta} ,2} ) $ such that ${t}_j^\prime < {t}_j$ for all $j=1,\dots,J$ and $[\mathbf{t}^\prime  , \mathbf{t}  ]  \subset \mathrm{int} (\mathscr{S}_{\boldsymbol{\theta} ,2} ) $. Observe that
\begin{equation*}
 \Pr \left\{   \nabla \mathfrak{p}(  \mathbf{Q} )  \in    [\mathbf{t}^\prime  , \mathbf{t}  ]  \right\} =   \Pr \left( \boldsymbol{\theta} \in    [\mathbf{t}^\prime  , \mathbf{t}  ]  \right)  =  \int_{t_J^\prime}^{t_J} \dots  \int_{t_1^\prime}^{t_1}  f_{\boldsymbol{\theta}} (  w_1,\dots,w_J  ) dw_1 \dots d w_J
\end{equation*}and note that the utmost left-hand side is known because $\mathfrak{p} (\cdot)$ has already been identified and $F_{\mathbf{Q}}$ is assumed to be known. As a result, $f_{\boldsymbol{\theta}} (\mathbf{t} )   =   \partial  \Pr \left\{   \nabla \mathfrak{p}(  \mathbf{Q} )  \in    [\mathbf{t}^\prime  , \mathbf{t}  ]  \right\} / \partial t_1 \dots \partial t_J$. \qed

\paragraph{Proof of Lemma \ref{le:costfun}.} After evaluating Eq.\ (\ref{eq:boundid}) at each $\mathbf{t}_j^\prime$, and since $\mathcal{N}$ is nonsingular, it follows that $\boldsymbol{\alpha}     =      \mathcal{N}^{-1} [ \mathbf{T}    - ( \mathcal{N}  \odot \mathcal{Q}^\top) \boldsymbol{\beta}]$. Plugging in this equality into Eq.\ (\ref{eq:intid}) and noting that $\partial^2 C / (\partial q_j \partial q_{j^\prime}) = 0$ whenever $j \neq j^\prime $ (Assumption \ref{assump:idcost1}) yield
\begin{multline}
\left[ J + 1  \  - \  \left(  \nabla_1  \mathfrak{q}_1 ( {\mathbf{t}} )   , \dots ,  \nabla_J  \mathfrak{q}_J ( {\mathbf{t}} )  \right)   \cdot \boldsymbol{\beta}  \right] f_{ \boldsymbol{\theta} } ( {\mathbf{t}} )  \ +  \  {\mathbf{t}} \cdot  \nabla f_{ \boldsymbol{\theta} }  ( {\mathbf{t}} ) \\
- \ \left[  \mathcal{N}^{-1} \mathbf{T}    -  \mathcal{N}^{-1} ( \mathcal{N}  \odot \mathcal{Q}^\top) \boldsymbol{\beta} \ +   \  \mathfrak{q} ( {\mathbf{t}} ) \odot  \boldsymbol{\beta}   \right] \cdot  \nabla f_{ \boldsymbol{\theta}}  ( {\mathbf{t}} )   = 0 
\label{eq:proofid}
\end{multline}
for every ${\mathbf{t}} \in \mathrm{int} (   \mathscr{S}_{\boldsymbol{\theta},2}  )$. Finally, the desired result follows after evaluating Eq.\ (\ref{eq:proofid}) at each ${\mathbf{t}}_j^{\prime\prime}$, $j =1,\dots,J$, and noting that $\mathcal{D}_{\beta}$ is nonsingular by Assumption \ref{assump:idcost2}-(d). \qed

\paragraph{Proof of Theorem \ref{thm:asymp}.}  To establish consistency of $\hat{\boldsymbol{\gamma}}$, first, note that the  (population criterion) function $\mathcal{T}_{T\ref*{thm:asymp}}$ defined on $\Gamma$ by
\begin{equation}	\label{eqapp:crite}
\mathcal{T}_{T\ref*{thm:asymp}} ( \boldsymbol\gamma  )	: =    \E \left[ \mathfrak{L} ( \boldsymbol{\theta} ; \boldsymbol{\gamma} )     \right]  =  \E \left[ \mathfrak{L} ( \boldsymbol{\theta} ; \boldsymbol{\gamma} )   | \boldsymbol{\theta} \in  \tilde{\mathcal{S}}   \right] \Pr \left( \boldsymbol{\theta} \in  \tilde{\mathcal{S}}   \right)  
\end{equation}has a unique maximum at $\boldsymbol{\gamma}_0 \in \mathrm{int}( \Gamma)$. This result follows from the second condition of Definition \ref{defF} and by applying the strict version of Jensen's inequality to the first term on the utmost right-hand side of Eq.\ (\ref{eqapp:crite}); see, e.g., the arguments in the proof of Lemma 2.2 in \cite{NeweyMcFadden94}. Second, observe that Lemma 2.4 in \cite{NeweyMcFadden94} implies that \begin{equation}
\sup_{\boldsymbol{\gamma} \in \Gamma} \left|  \frac{1}{n} \sum_{i \in \mathscr{I}_2}  \mathfrak{L} [ \nabla\mathfrak{p}(\mathbf{Q}_i)   ,  \boldsymbol{\gamma} ]  - \mathcal{T}_{T\ref*{thm:asymp}} ( \boldsymbol\gamma  )  \right|  \overset{p}{\rightarrow} 0 ,
\end{equation}
noting that $\sum_{i \in \mathscr{I}_2}  \mathfrak{L} [ \nabla\mathfrak{p}(\mathbf{Q}_i)   ,  \boldsymbol{\gamma} ] = \sum_{i=1}^n \mathfrak{L} \left( \boldsymbol{\theta}_i   ,  \boldsymbol{\gamma} \right) $ because $  \mathfrak{L} \left( \boldsymbol{\theta}_i   ,  \boldsymbol{\gamma} \right) = 0$ whenever $i \notin \mathscr{I}_2$. Then, $\hat{\boldsymbol{\gamma}} \overset{p}{\rightarrow} \boldsymbol{\gamma}_0 $ follows immediately from Theorem 2.1 in \cite{NeweyMcFadden94}.

To prove asymptotic normality, first, observe that consistency of $\hat{\boldsymbol{\gamma}}$ and $\boldsymbol{\gamma}_0 \in \mathrm{int}( \Gamma)$ (Assumption \ref{assump:inte}) imply that $\hat{\boldsymbol{\gamma}} \in \mathrm{int}( \Gamma)$ w.p.a.1 and therefore Eq.\ (\ref{eq:mlefoc}) holds. Second, note that
\begin{eqnarray*}
\E \left[  \mathcal{H}_{\boldsymbol{\gamma}\boldsymbol{\gamma}} \mathfrak{L} \left(  \boldsymbol{\theta}  ; \boldsymbol{\gamma}_0  \right)   \right]  & = &   \E \left[  \mathcal{H}_{\boldsymbol{\gamma}\boldsymbol{\gamma}} \mathfrak{L} \left(  \boldsymbol{\theta}  ; \boldsymbol{\gamma}_0  \right)  \middle|   \boldsymbol{\theta} \in  \tilde{\mathcal{S}}   \right]    \Pr \left( \boldsymbol{\theta} \in  \tilde{\mathcal{S}}   \right)    ,   \\
  \E \left[  \mathcal{H}_{\boldsymbol{\gamma}\boldsymbol{\gamma}} \mathfrak{L} \left(  \boldsymbol{\theta}  ,   \boldsymbol{\gamma}_0  \right)  \middle|   \boldsymbol{\theta} \in  \tilde{\mathcal{S}}   \right]   &  = &  -   \E \left\{  \left[ \nabla_{\boldsymbol{\gamma}}  \mathfrak{L} ( \boldsymbol{\theta} ; \boldsymbol{\gamma}_0 ) \right]  \left[ \nabla_{\boldsymbol{\gamma}}  \mathfrak{L} ( \boldsymbol{\theta} ; \boldsymbol{\gamma}_0 ) \right] ^\top  \middle|  \boldsymbol{\theta} \in  \tilde{\mathcal{S}}    \right\} , \ \text{and therefore} \\
\Sigma    & = &    \E \left\{  \left[ \nabla_{\boldsymbol{\gamma}}  \mathfrak{L} ( \boldsymbol{\theta} ; \boldsymbol{\gamma}_0 ) \right]  \left[ \nabla_{\boldsymbol{\gamma}}  \mathfrak{L} ( \boldsymbol{\theta} ; \boldsymbol{\gamma}_0 ) \right] ^\top  \middle|  \boldsymbol{\theta} \in  \tilde{\mathcal{S}}    \right\}  \Pr \left( \boldsymbol{\theta} \in  \tilde{\mathcal{S}}   \right)      \\ 
& = & -    \E \left[  \mathcal{H}_{\boldsymbol{\gamma}\boldsymbol{\gamma}} \mathfrak{L} \left(  \boldsymbol{\theta}  ,   \boldsymbol{\gamma}_0  \right)  \middle|   \boldsymbol{\theta} \in  \tilde{\mathcal{S}}   \right]   \Pr \left( \boldsymbol{\theta} \in  \tilde{\mathcal{S}}   \right)    = -  \E \left[  \mathcal{H}_{\boldsymbol{\gamma}\boldsymbol{\gamma}} \mathfrak{L} \left(  \boldsymbol{\theta}  ; \boldsymbol{\gamma}_0  \right)   \right] ;
\end{eqnarray*}
where the second equality follows from the Information matrix equality. Then, applying the generalized mean value theorem to the right-hand side of Eq.\ (\ref{eq:mlefoc}) leads to \begin{equation*}
\frac{1}{n}  \sum_{i =1}^n \nabla_{\boldsymbol{\gamma}} \mathfrak{L} \left(  \boldsymbol{\theta}_i  ;  \boldsymbol{\gamma}_0  \right)    +   \left[ \frac{1}{n}   \sum_{i =1}^n  \int_0^1 \mathcal{H}_{\boldsymbol{\gamma}\boldsymbol{\gamma}} \mathfrak{L} \left[  \boldsymbol{\theta}_i  ; \boldsymbol{\gamma}_0  + t   \left( \hat{\boldsymbol{\gamma}} - \boldsymbol{\gamma}_0 \right)  \right]   dt  \right] \left( \hat{\boldsymbol{\gamma}} - \boldsymbol{\gamma}_0 \right) =   0
\end{equation*}and, consequently, asymptotic normality of $\sqrt{n} ( \hat{\boldsymbol{\gamma}} - \boldsymbol{\gamma}_0 ) $ follows by the fact that $\E [  \nabla_{\boldsymbol{\gamma}} \mathfrak{L} (  \boldsymbol{\theta}  ;  \boldsymbol{\gamma}_0  )] = 0$, by Slutsky's theorem, and from \begin{equation*}
\frac{1}{n}  \sum_{i=1}^n    \int_0^1 \mathcal{H}_{\boldsymbol{\gamma}\boldsymbol{\gamma}} \mathfrak{L} \left[  \boldsymbol{\theta}_i  ; \boldsymbol{\gamma}_0  + t   \left( \hat{\boldsymbol{\gamma}} - \boldsymbol{\gamma}_0 \right)  \right]   dt    \ \overset{p}{\rightarrow}   \  \E \left[  \mathcal{H}_{\boldsymbol{\gamma}\boldsymbol{\gamma}} \mathfrak{L} \left(  \boldsymbol{\theta}  ; \boldsymbol{\gamma}_0  \right)   \right] \ = \  - \Sigma  ;
\end{equation*}where here $\overset{p}{\rightarrow}$ follows from Lemma 4.3 in \cite{NeweyMcFadden94}. Finally, $\hat{\Sigma} \overset{p}{\rightarrow} \Sigma$ can also be obtained from this lemma.	\qed

\paragraph{Proof of Corollary \ref{corol:cost}.} By the characterization of $(\boldsymbol{\alpha}, \boldsymbol{\beta})$ in Lemma \ref{le:costfun} and by Assumption \ref{assump:inte}, $(\boldsymbol{\alpha}, \boldsymbol{\beta})$ can be regarded as a function of $\boldsymbol{\gamma}$ that is continuous at $\boldsymbol{\gamma}_0$. Then, the desired result follows immediately by consistency of $(\hat{\boldsymbol{\alpha}}, \hat{\boldsymbol{\beta}})$.   \qed

\paragraph{Proof of Lemma \ref{lem:boundest}.} Let $c_{3.1}, c_{3.1}^\prime , c_{3.2},\dots$ denote finite constants that do not depend on $n \in \mathbb{N}$, nor $m = 1,\dots, M_n^J$. We also write $\hat{\mathfrak{p}}^{(u)} ( \mathbf{q}) = \lowmathcal{s}(  \mathbf{q} - \ushort{\mathbf{q}}_{m( \mathbf{q}  )} ) \cdot  \overline{\boldsymbol{\pi}}_{m(  \mathbf{q} )}$, $\mathfrak{p}_m ( \mathbf{q} ; \mathbf{z}) = \lowmathcal{s} (  \mathbf{q}  -  \ushort{\mathbf{q}}_m  )\cdot (\mathcal{M}_n  \mathbf{z} )$ for $\mathbf{z} \in \mathbb{Z}^L$, and let $\hat{\mathbf{z}}_m$ be defined so that $\mathfrak{p}_m ( \mathbf{q} ; \hat{\mathbf{z}}_m) = \hat{\mathfrak{p}}^{(u)} ( \mathbf{q}) $ for $\mathbf{q} \in \mathcal{R}_{n,m}$.

We start with the following observation that follows from the arguments in the proof of Theorem 7.4.1 in \cite{KorostelevTsybakov1993}: there exist constants $c_{3.1}, c_{3.1}^\prime, c_{3.1}^{\prime\prime}>0$ such that, for any $n$ sufficiently large and each $m = 1,\dots, M_n^J$, there is $\mathbf{z}_m \in \mathbb{Z}^L$ that satisfies the next conditions: $\mathcal{M}_n \mathbf{z}_m  \in \Pi_n$, and for all $  \mathbf{q} \in \mathcal{R}_{n,m}$,   $\mathfrak{p}_m ( \mathbf{q}; \mathbf{z}_m )   \geq   \mathfrak{p} ( \mathbf{q}) $,  \begin{equation}
\left|    \mathfrak{p}_m ( \mathbf{q} ;  \mathbf{z}_m )  - \mathfrak{p}  ( \mathbf{q})  \right|     \leq     c_{3.1}   M_n^{-(S+1)} ,  \ \   \|   \nabla_\mathbf{q} \mathfrak{p}_m ( \mathbf{q} ;  \mathbf{z}_m )  - \nabla \mathfrak{p}  ( \mathbf{q})  \|_{\infty}     \leq     c_{3.1}^\prime   M_n^{-S} ,
\end{equation}and $ \|   \mathcal{H}_\mathbf{q} \mathfrak{p}_m ( \mathbf{q} ;  \mathbf{z}_m )  -  \mathcal{H} \mathfrak{p}  ( \mathbf{q})  \|_{\infty}     \leq     c_{3.1}^{\prime\prime}  M_n^{-S-1} $ if $S\geq2$. The rest of the proof is divided into four steps, and we culminate in proving that  \begin{equation*}
 \left\|  \hat{\mathfrak{p}}^{(u)}  -   \mathfrak{p}   \right\|_{ \mathcal{R}_n  , \infty} = O_p \left( {M}_n^{-S-1} \right) ,  \ \   \left\|  \nabla \hat{\mathfrak{p}}^{(u)} -   \nabla \mathfrak{p}   \right\|_{ \mathcal{R}_n  , \infty} = O_p \left( {M}_n^{-S} \right) ,
\end{equation*}and $\|  \mathcal{H} \hat{\mathfrak{p}}^{(u)}  -  \mathcal{H} \mathfrak{p}   \|_{ \mathcal{R}_n  , \infty}=O_p ( {M}_n^{-S+1} )$ if $S\geq2$: the desired results then follow immediately by extending symmetrically these results to the other piecewise polynomial, $\lowmathcal{s}(  \cdot  - \ushort{\mathbf{q}}_{m( \cdot  )} ) \cdot  \underline{\boldsymbol{\pi}}_{m( \cdot)}$.

In the first step, letting \begin{eqnarray*}
\mathcal{T}_m \left( \mathbf{z}  , \mathbf{z}^\prime \right)  & = & \int_{\mathcal{R}_{n,m}} \left[  \mathfrak{p}_m ( \cdot ; \mathbf{z})    - \mathfrak{p}_m ( \cdot ; \mathbf{z}^\prime)   \right] \mathds{1} \left[ \mathfrak{p}_m ( \cdot ; \mathbf{z})    > \mathfrak{p}_m ( \cdot ; \mathbf{z}^\prime )   \right] \   \  \text{and}  \\ 
 \mathcal{T}_m^\circ  ( \mathbf{z} ) &  = & \int_{\mathcal{R}_{n,m}} \left[  \mathfrak{p}  ( \cdot )  -  \mathfrak{p}_m ( \cdot ; \mathbf{z})   \right] \mathds{1} \left[   \mathfrak{p} ( \cdot )  > \mathfrak{p}_m ( \cdot ; \mathbf{z})  \right] ,
\end{eqnarray*}we prove that there exists $c_{3.2} >0$ such that \begin{equation}
 \max_{m = 1,\dots, M_n^J}  \  \Pr  \left[   M_n^{S+1+J} \times  \mathcal{T}_m \left( \mathbf{z}_m  , \hat{\mathbf{z}}_m  \right)    \geq C    \right]     \   \leq  \  2C^L \times n^{-c_{3.2} \times C}   
 \label{eq:le3step1}
\end{equation}for any $C \geq 1$ and $n  \in \mathbb{N}$ sufficiently large. To do so, for $r \in \mathbb{N}$, define the set of integers \begin{equation*}
\mathscr{Z}_m (r, C) =   \left\{ \mathbf{z} \in \mathbb{Z} :  \ (r - 1) C \leq \|  \mathbf{z} - \mathbf{z}_m \|_\infty  < r C  ,  \  M_n^{S+1+J} \times  \mathcal{T}_m \left( \mathbf{z}_m  , \mathbf{z}  \right)  >  C \right\}
\end{equation*}and note that $\# \mathscr{Z}_m (r , C) \leq  2(rC)^{L}$. Then, by the arguments in \citet[p.\ 192]{KorostelevTsybakov1993}, we can bound $
M_n^{S+1+J} \times  \mathcal{T}_m^\circ  ( \mathbf{z} )   \   \geq   \   c_{3.2}^\prime r C$, for all $\mathbf{z} \in \mathscr{Z}_m (r, C)$ and some $c_{3.2}^\prime >0$. This leads to the inequality \begin{equation*}
\forall \ \mathbf{z} \in \mathscr{Z}_m (r, C) , \     \int_{\mathcal{R}_{n,m}}  \mathds{1}  \left[   \mathfrak{p} ( \cdot )  > \mathfrak{p}_m ( \cdot ; \mathbf{z})  \right]  \  \geq   \   \frac{  c_{3.2}^{\prime\prime} r C    }  { M_n^{S+1+J}  }  ,
\end{equation*}for some $c_{3.2}^{\prime\prime} > 0$, and therefore for all $\mathbf{z} \in \mathscr{Z}_m (r, C) $, we have that \begin{eqnarray*}      \mathrm{P}_{i,m} (  \mathbf{z} ) : = \Pr \left[ P_i \leq   \mathfrak{p}_m ( \mathbf{Q}_i ; \mathbf{z} ) \  \& \ \mathbf{Q}_i \in \mathcal{R}_{n,m} \right]  & =  &  1   -   \int_{\mathcal{R}_{n,m}}  \mathds{1}  \left[   \mathfrak{p} ( \cdot )  > \mathfrak{p}_m ( \cdot ; \mathbf{z})  \right]  f_{\mathbf{Q}} (\cdot)  \\ 
 & \leq &   1   -  \ushort{f}_{\mathbf{Q}}  \int_{\mathcal{R}_{n,m}}  \mathds{1}  \left[   \mathfrak{p} ( \cdot )  > \mathfrak{p}_m ( \cdot ; \mathbf{z})  \right]   \\
 & \leq &   1   - \frac{\ushort{f}_{\mathbf{Q}}  c_{3.2}^{\prime\prime} r C}{M_n^{S+1+J}} .
\end{eqnarray*}As a result, after taking $c_{3.2} = \ushort{f}_{\mathbf{Q}} c_{3.2}^{\prime\prime}/4$, Eq.\ (\ref{eq:le3step1}) emerges from the next inequalities: \begin{eqnarray*}
\Pr  \left[   M_n^{S+1+J} \times \mathcal{T}_m \left( \mathbf{z}_m  , \hat{\mathbf{z}}_m  \right)   \geq C    \right]   &  \leq & \sum_{r \in \mathbb{N}}  \ \# \mathscr{Z}_m (r  , C )    \times  \max_{\mathbf{z} \in \mathscr{Z}_m (r  , C ) } \mathrm{P}_{i,m} (  \mathbf{z} )   \\
&  \leq & 2C^L  \sum_{r \in \mathbb{N}}   r^L \left(  1   - \frac{\ushort{f}_{\mathbf{Q}}  c_{3.2}^{\prime\prime} r C}{M_n^{S+1+J}}  \right)^n  \ \leq  \    2C^L \times n^{-c_{3.2} \times C}  ;
\end{eqnarray*}the first inequality follows from the arguments in \citet[pp.\ 192-193]{KorostelevTsybakov1993} that rely on the inequality constraints in the definition of the estimator; the second is straightforward, and the third follows from the form of $M_n$.

In the second step, we show that there exists $c_{3.3} >0$ such that \begin{equation}
\max_{m = 1,\dots, M_n^J}  \    \Pr \left(    \left\|  \hat{\mathbf{z}}_m -  \mathbf{z}_m    \right\|_{\infty} \geq  C \right) \   \leq  \  n^{-c_{3.3} \times C }
 \label{eq:le3step2}
\end{equation}for any $C \geq 1 $ and $n  \in \mathbb{N}$ sufficiently large. Note that, by the change of variable formula and the equivalence of all norms in a finite-dimensional vector space, there is $c_{3.3}^{\prime} > 0$ such that $  \left\|  \hat{\mathbf{z}}_m -  \mathbf{z}_m    \right\|_{\infty}  \  \leq \ 	c_{3.3}^{\prime} \times  M_n^{S+1+J}  \times \left\| \hat{\mathfrak{p}}^{(u)} -  \mathfrak{p}_m ( \cdot ; \mathbf{z}_m ) \right\|_{\mathcal{R}_{n,m} , 1}$; see \citet[Lemma 4.2.1]{KorostelevTsybakov1993}. Moreover, by construction of $\hat{\mathfrak{p}}$, we can also bound $ \| \hat{\mathfrak{p}}^{(u)} -  \mathfrak{p}_m ( \cdot ; \mathbf{z}_m ) \|_{\mathcal{R}_{n,m} , 1}	 \leq  2 \mathcal{T}_m \left( \mathbf{z}_m  , \hat{\mathbf{z}}_m  \right) $ and therefore Eq.\ (\ref{eq:le3step2}) follows from (\ref{eq:le3step1}).

In the third step, we establish the uniform convergence of $ \hat{\mathfrak{p}}^{(u)}$. For that purpose, observe that we can find $c_{3.4}>0$ such that, for any constant $C > c_{3.1} + L / \min\{1 , c_{3.3} \}$ sufficiently large, we have that \begin{multline*}
\Pr \left(  \left\|  \hat{\mathfrak{p}}^{(u)} -  \mathfrak{p}    \right\|_{\mathcal{R}_{n,m} , \infty}  \geq   C  \times M_n^{-(S+1)}   \right)\  \leq  \  \Pr \left(   \left\|  \hat{\mathfrak{p}}^{(u)} -  \mathfrak{p}_m   ( \cdot ; \mathbf{z}_m ) \right\|_{\mathcal{R}_{n,m} , \infty} \geq  (C - c_{3.1}) \times M_n^{-(S+1)} \right)  \\
\leq  \  \Pr \left(    \left\|  \hat{\mathbf{z}}_m -  \mathbf{z}_m    \right\|_{\infty} \geq  c_{3.4}(C - c_{3.1}) / L \right) ;
\end{multline*}the second inequality, as well as existence of $c_{3.4}>0$, follows by construction of $\Pi_n$. Consequently, Eq.\ (\ref{eq:le3step2}) implies  \begin{eqnarray*}
\Pr \left(  \left\|  \hat{\mathfrak{p}}^{(u)} -  \mathfrak{p}    \right\|_{\mathcal{R}_n , \infty}    \geq   C  \times M_n^{-(S+1)}   \right) 
& \leq &     \sum_{m=1}^{M_n}  \Pr \left(  \left\|  \hat{\mathfrak{p}}^{(u)} -  \mathfrak{p}    \right\|_{\mathcal{R}_{n,m} , \infty}  \geq  C  \times M_n^{-(S+1)}   \right) \\
&\leq  &   M_n  \times  n^{ - c_{3.3} \times c_{3.4} ( C   - c_{3.1})/L}.
\end{eqnarray*}

In the last and fourth step, we establish the uniform convergence of $\nabla_{j_1} \hat{\mathfrak{p}}^{(u)}$ and $\mathcal{H}_{j_1 , j_2}\hat{\mathfrak{p}}^{(u)}$, $j_1, j_2 = 1,\dots,J$, for which it suffices to show that $  \|  \nabla_{j_1} \hat{\mathfrak{p}}^{(u)} ( \cdot ) -   \nabla_{j_1} \mathfrak{p}_m ( \cdot ;  \mathbf{z}_m )    \|_{ \mathcal{R}_n  , \infty} = O_p ( {M}_n^{-S} ) $ and  $\|  \mathcal{H}_{j_1 , j_2} \hat{\mathfrak{p}}^{(u)} ( \cdot ) -   \mathcal{H}_{j_1 , j_2}\mathfrak{p}_m ( \cdot ;  \mathbf{z}_m )    \|_{ \mathcal{R}_n  , \infty} = O_p ( {M}_n^{-S+1} ) $, respectively. With this aim, by construction of $\Pi_n$, note that we can bound \begin{eqnarray*}
{M}_n^{S}  \|  \nabla_j \hat{\mathfrak{p}}^{(u)} ( \cdot ) -   \nabla_j \mathfrak{p}_m ( \cdot ;  \mathbf{z}_m )    \|_{ \mathcal{R}_{n,m}  , \infty}     &  \leq  & c_{3.5}  (L - 1)      \left\|  \hat{\mathbf{z}}_m - \mathbf{z}_m    \right\|_{\infty}  \  \ \text{and} \\
{M}_n^{S-1}  \|  \mathcal{H}_{j_1 , j_2} \hat{\mathfrak{p}}^{(u)} ( \cdot ) -  \mathcal{H}_{j_1 , j_2}  \mathfrak{p}_m ( \cdot ;   \mathbf{z}_m )    \|_{ \mathcal{R}_{n,m}  , \infty}     &  \leq  & c_{3.5}^\prime (L - 2)      \left\|  \hat{\mathbf{z}}_m - \mathbf{z}_m    \right\|_{\infty} \  \    \text{if $S \geq 2$} ,
\end{eqnarray*}for some $c_{3.5} , c_{3.5}^\prime >0$. Then, following arguments similar to those used in the previous step, the desired result follows from Eq.\ (\ref{eq:le3step2}). \qed

\paragraph{Proof of Theorem \ref{thm:asympapp}.}  We start by establishing consistency. First, note that  \begin{equation}
\sup_{ i , \boldsymbol{\gamma} } \left| \log \left[ f \left( \hat{\boldsymbol{\theta}}_i  ;  \boldsymbol{\gamma}  \right)   \right]  -  \log \left[ f \left( \boldsymbol{\theta}_i  ;  \boldsymbol{\gamma}  \right)   \right] \right|    \leq   c_{\ref*{thm:asympapp}.1} \left\| \nabla \hat{\mathfrak{p}} -   \nabla \mathfrak{p}   \right\|_{ \tilde{\mathcal{R}} , \infty}  \ \underset{\text{Lemma \ref{lem:boundest}}}{=} \  o_p \left(\frac{1}{\sqrt{n}}\right) 
\label{eq:proofmle1}
\end{equation}for some constant $c_{\ref*{thm:asympapp}.1} >0$, with $\sup$ being taken over $\{ i : \mathbf{Q}_i  \in \tilde{\mathcal{R}} \}\times \Gamma$. Second, by standard arguments, we can also bound \begin{multline}
\sup_{ \boldsymbol{\gamma} \in \Gamma  }  \left|   \int_{  \tilde{\mathcal{R}}}  f \left[   \nabla\hat{\mathfrak{p}} ( \mathbf{q} ) ; \boldsymbol{\gamma}    \right]  \left|  \mathcal{H} \hat{\mathfrak{p}} ( \mathbf{q} )  \right| d\mathbf{q}    -    \int_{  \tilde{\mathcal{R}}}  f \left[   \nabla{\mathfrak{p}} ( \mathbf{q} ) ; \boldsymbol{\gamma}    \right]  \left|  \mathcal{H} {\mathfrak{p}} ( \mathbf{q} )  \right| d\mathbf{q}  \right|  \\
  \leq  \  c_{\ref*{thm:asympapp}.2} \max \left\{   \|  \nabla \hat{\mathfrak{p}}  -  \nabla \mathfrak{p}   \|_{ \mathcal{R}  , \infty}  ,  \|  \mathcal{H} \hat{\mathfrak{p}}  -  \mathcal{H} \mathfrak{p}   \|_{ \mathcal{R}  , \infty}  \right\}  \ \underset{\text{Lemma \ref{lem:boundest}}}{=} \  o_p \left(\frac{1}{\sqrt{n}}\right)  
  \label{eq:proofmle2}
\end{multline}for some constant $c_{\ref*{thm:asympapp}.2} >0$. Then, combining Eqs.\ (\ref{eq:proofmle1})-(\ref{eq:proofmle2}) with the facts that $\tilde{\mathcal{S}} = \nabla \mathfrak{p} ( \tilde{\mathcal{R}} )$ and $
 \int_{  \tilde{\mathcal{R}}}   f \left[   \nabla{\mathfrak{p}} ( \mathbf{q} ) ; \boldsymbol{\gamma}    \right]  \left|  \mathcal{H} {\mathfrak{p}} ( \mathbf{q} )  \right| d\mathbf{q}    = \int_{\tilde{\mathcal{S}}}   f \left( \mathbf{t}; \boldsymbol{\gamma}  \right) d \mathbf{t}$,
 yields $
\sup_{\boldsymbol{\gamma} \in \Gamma } \frac{1}{n}  \sum_{i=1}^n   \left|   \hat{\mathfrak{L}}_i ( \boldsymbol{\gamma} )   -     \mathfrak{L} (  \boldsymbol{\theta}_i; \boldsymbol{\gamma} )     \right|  =  o_p \left(\frac{1}{\sqrt{n}}\right),$ from which $\check{\boldsymbol{\gamma}} \overset{p}{\rightarrow} \boldsymbol{\gamma}_0 $ follows by the arguments in the proof of Theorem \ref{thm:asymp}.

To establish asymptotic normality, first, note that $\sum_i  \nabla_{\boldsymbol{\gamma}}  \hat{\mathfrak{L}}_i ( \boldsymbol{\gamma} ) = 0$ hold w.p.a.1 because the empirical criterion function is differentiable on $\mathrm{int} ( \Gamma )$. Second, since $f( \cdot; \cdot)$ is twice continuously differentiable, observe that proceeding as in Eqs.\ (\ref{eq:proofmle1})-(\ref{eq:proofmle2}) yields \begin{equation}
\sup_{i=1,\dots,n}  \sup_{\boldsymbol{\gamma} \in \tilde{\Gamma}}  \left\| \nabla_{\boldsymbol{\gamma}}  \hat{\mathfrak{L}}_i ( \boldsymbol{\gamma} )    -   \nabla_{\boldsymbol{\gamma}}  \mathfrak{L} (  \boldsymbol{\theta}_i ; \boldsymbol{\gamma} ) \right\|_{\infty}   \ = \   o_p \left(\frac{1}{\sqrt{n}}\right)  ,
\label{eq:dercon}
\end{equation}and therefore  $
\frac{1}{n}  \sum_{i=1}^n  \nabla_{\boldsymbol{\gamma}} \mathfrak{L} \left(  \boldsymbol{\theta}_i; \check{ \boldsymbol{\gamma} } \right)    + o_p \left(\frac{1}{\sqrt{n}}\right) = 0$. Then asymptotic normality follows by the same reasoning as in the proof of Theorem \ref{thm:asymp}.
Finally, consistency of $\check\Sigma$ follows from Eq.\ (\ref{eq:dercon}) and the same arguments as in the proof of Theorem \ref{thm:asymp}.   \qed

\paragraph{Proof of Theorem \ref{thm:costapp}.} We start by establishing consistency. First, note that  \begin{equation}
\sup_{ i , \boldsymbol{\gamma} } \left| \log \left[ f \left( \hat{\boldsymbol{\theta}}_i  ;  \boldsymbol{\gamma}  \right)   \right]  -  \log \left[ f \left( \boldsymbol{\theta}_i  ;  \boldsymbol{\gamma}  \right)   \right] \right|    \leq   c_{\ref*{thm:asympapp}.1} \left\| \nabla \hat{\mathfrak{p}} -   \nabla \mathfrak{p}   \right\|_{ \tilde{\mathcal{R}} , \infty}  \ \underset{\text{Lemma \ref{lem:boundest}}}{=} \  o_p \left(\frac{1}{\sqrt{n}}\right) 
\label{eq:proofmle1}
\end{equation}for some constant $c_{\ref*{thm:asympapp}.1} >0$, with $\sup$ being taken over $\{ i : \mathbf{Q}_i  \in \tilde{\mathcal{R}} \}\times \Gamma$. Second, by standard arguments, we can also bound \begin{multline}
\sup_{ \boldsymbol{\gamma} \in \Gamma  }  \left|   \int_{  \tilde{\mathcal{R}}}  f \left[   \nabla\hat{\mathfrak{p}} ( \mathbf{q} ) ; \boldsymbol{\gamma}    \right]  \left|  \mathcal{H} \hat{\mathfrak{p}} ( \mathbf{q} )  \right| d\mathbf{q}    -    \int_{  \tilde{\mathcal{R}}}  f \left[   \nabla{\mathfrak{p}} ( \mathbf{q} ) ; \boldsymbol{\gamma}    \right]  \left|  \mathcal{H} {\mathfrak{p}} ( \mathbf{q} )  \right| d\mathbf{q}  \right|  \\
  \leq  \  c_{\ref*{thm:asympapp}.2} \max \left\{   \|  \nabla \hat{\mathfrak{p}}  -  \nabla \mathfrak{p}   \|_{ \mathcal{R}  , \infty}  ,  \|  \mathcal{H} \hat{\mathfrak{p}}  -  \mathcal{H} \mathfrak{p}   \|_{ \mathcal{R}  , \infty}  \right\}  \ \underset{\text{Lemma \ref{lem:boundest}}}{=} \  o_p \left(\frac{1}{\sqrt{n}}\right)  
  \label{eq:proofmle2}
\end{multline}for some constant $c_{\ref*{thm:asympapp}.2} >0$. Then, combining Eqs.\ (\ref{eq:proofmle1})-(\ref{eq:proofmle2}) with $\tilde{\mathcal{S}} = \nabla \mathfrak{p} ( \tilde{\mathcal{R}} )$ and $
 \int_{  \tilde{\mathcal{R}}}   f \left[   \nabla{\mathfrak{p}} ( \mathbf{q} ) ; \boldsymbol{\gamma}    \right]  \left|  \mathcal{H} {\mathfrak{p}} ( \mathbf{q} )  \right| d\mathbf{q}    = \int_{\tilde{\mathcal{S}}}   f \left( \mathbf{t}; \boldsymbol{\gamma}  \right) d \mathbf{t}$,
 yields $
\sup_{\boldsymbol{\gamma} \in \Gamma } \frac{1}{n}  \sum_{i=1}^n   \left|   \hat{\mathfrak{L}}_i ( \boldsymbol{\gamma} )   -     \mathfrak{L} (  \boldsymbol{\theta}_i; \boldsymbol{\gamma} )     \right|  =  o_p \left(\frac{1}{\sqrt{n}}\right),$ from which $\check{\boldsymbol{\gamma}} \overset{p}{\rightarrow} \boldsymbol{\gamma}_0 $ follows by the arguments in the proof of Theorem \ref{thm:asymp}.

To establish asymptotic normality, first, note that $\sum_i  \nabla_{\boldsymbol{\gamma}}  \hat{\mathfrak{L}}_i ( \boldsymbol{\gamma} ) = 0$ hold w.p.a.1 because the empirical criterion function is differentiable on $\mathrm{int} ( \Gamma )$. Second, since $f( \cdot; \cdot)$ is twice continuously differentiable, observe that proceeding as in Eqs.\ (\ref{eq:proofmle1})-(\ref{eq:proofmle2}) yields \begin{equation}
\sup_{i=1,\dots,n}  \sup_{\boldsymbol{\gamma} \in \tilde{\Gamma}}  \left\| \nabla_{\boldsymbol{\gamma}}  \hat{\mathfrak{L}}_i ( \boldsymbol{\gamma} )    -   \nabla_{\boldsymbol{\gamma}}  \mathfrak{L} (  \boldsymbol{\theta}_i ; \boldsymbol{\gamma} ) \right\|_{\infty}   \ = \   o_p \left(\frac{1}{\sqrt{n}}\right)  ,
\label{eq:dercon}
\end{equation}and therefore  $
\frac{1}{n}  \sum_{i=1}^n  \nabla_{\boldsymbol{\gamma}} \mathfrak{L} \left(  \boldsymbol{\theta}_i; \check{ \boldsymbol{\gamma} } \right)    + o_p \left(\frac{1}{\sqrt{n}}\right) = 0$. Then, asymptotic normality follows from the arguments in the proof of Theorem \ref{thm:asymp}.
Finally, consistency of $\check\Sigma$ follows from Eq.\ (\ref{eq:dercon}) and again from the arguments in the proof of Theorem \ref{thm:asymp}.   \qed

\vspace{1in}

\bibliographystyle{econometrica}

\bibliography{references}

\end{document}